\documentclass{mn2e}
\usepackage{amssymb, amsmath}
\usepackage{ifthen}
\usepackage{graphicx}
\usepackage{float}		

\newcommand{\comment}[1]{}              

\def\br {{\bf r}}
\def\bv {{\bf v}}

\def\bz {{\bf z}}
\def\bn {{\bf n}}
\def\bk {{\bf k}}
\def\bx {{\bf x}}
\def\bI {{\bf I}}
\def\bD {{\bf D}}
\def\bS {{\bf S}}
\def\bs {{\bf s}}

\def\bV {{\bf V}}

\def \bnabla {{\boldsymbol{\nabla}}}
\def \bchi {{\boldsymbol{\chi}}}

\def \btheta {{\boldsymbol{\theta}}}
\def \bTheta {{\boldsymbol{\Theta}}}

\def \rec {{\rm rec}}

\def \nablaperp {{\nabla_{\negthickspace\perp}}}
\def \bnablaperp {{\bnabla_{\negthickspace\perp}}}

\def \nablax {\nabla_{\bx}}
\def \bnablax {{\bnabla_{\negthickspace\bx}}}
\def \bnablaxstar {{\bnabla_{\negthickspace\bx^*}}}
\def \nablaxstar {{\nabla_{\negthickspace\bx^*}}}

\def \brdot {{\dot \br}}
\def \brddot {{\ddot \br}}
\def \bxdot {{\dot \bx}}
\def \tilden {{\tilde n}}
\def \bzhat {{\hat \bz}}
\def \bK {{\bf K}}
\def \bM {{\bf M}}
\def \bN {{\bf N}}
\def \bSigma {{\boldsymbol{\Sigma}}}
\def \bzero {{\bf 0}}

\def\Tr {{\rm Tr}}

\def \ispace {{\:}}	

\def\[{\begin{equation}}
\def\]{\end{equation}}

\def\m@th{\mathsurround=0pt }
\def\eqalign#1{\null\,\vcenter{\openup1\jot \m@th
 \ialign{\strut\hfil$\displaystyle{##}$&$\displaystyle{{}##}$\hfil
 \crcr#1\crcr}}\,}

\begin{document}

\title[Lensing bias in the distance-redshift relation]{On the bias of the distance-redshift relation from gravitational lensing
%
%
}
\author[Kaiser \& Peacock]{Nick Kaiser$^1$ \& John A. Peacock$^2$ \\
$^1$Institute for Astronomy, University of Hawaii, 2680 Woodlawn Drive, Honolulu, HI 96822-1839, USA \\ 
$^2$Institute for Astronomy, University of Edinburgh, Royal Observatory, Edinburgh EH9 3HJ, UK}
\maketitle 

\begin{abstract}
A long standing question in cosmology is whether gravitational
lensing changes the distance-redshift relation $D(z)$ or the mean flux density of sources.
Interest in this has been rekindled by recent studies in
non-linear relativistic perturbation theory that find biases in both
the area of a surface of constant redshift and in the mean distance to this surface,
with a fractional bias in both cases on the order of the mean squared convergence 
$\langle \kappa^2 \rangle$.  
Any such area bias could alter CMB cosmology, and the corresponding bias
in mean flux density could affect supernova cosmology.
Here we show that, in an ensemble averaged sense, the perturbation to
the area of a surface of constant redshift is in reality much smaller,
being on the order of the cumulative bending angle squared, 
or roughly a part-in-a-million effect.
This validates the arguments of Weinberg (1976) that the
mean magnification $\mu$ of sources is unity and of Kibble \& Lieu (2005)
that the mean direction-averaged inverse magnification is unity.
It also validates the conventional treatment of lensing in analysis of CMB anisotropies.
But the existence of a scatter in magnification will cause any non-linear function of
these conserved quantities to be statistically biased.
The distance $D$, for example, is proportional to $\mu^{-1/2}$ so
lensing will bias $\langle D\rangle$ even if $\langle \mu \rangle=1$.
The fractional bias in such quantities is generally of order 
$\langle \kappa^2 \rangle$, which is orders of magnitude larger
than the area perturbation.  Claims for large bias in area or
flux density of sources appear to have resulted from misinterpretation of such effects:
they do not represent a new non-Newtonian effect, nor do they invalidate standard
cosmological analyses.
\end{abstract}


\section{Introduction}

In homogeneous and isotropic cosmologies the ratio between the proper size of
a source and the angle subtended at the observer -- the angular diameter distance $D$ --
is solely a function of redshift.
In an inhomogeneous universe, gravitational
lensing by intervening metric fluctuations
can cause magnification of the 
angular size -- with associated change of flux density, since surface brightness is unaffected by lensing.
Thus the apparent distance to objects at a given $z$ becomes in effect a randomly
fluctuating quantity. Equivalently, the flux density measured on a sphere surrounding an
object at redshift $z$ is a random function of position on the sphere. 
The question we shall address here is whether distances
or flux densities are perturbed in the mean.

This subject has a long history with pioneering studies by Zel'dovich (1964) and
Feynman (in a colloquium at the California Institute of Technology in 1964; see Gunn 1967b)
with detailed calculations using point masses performed by Bertotti (1966), using the
`optical scalar' formalism of Sachs (1961), and by Gunn (1967a,b).
Swiss-cheese models (Einstein \& Straus, 1945) were used by Kantowski (1969) and
later by Dyer \& Roeder (1972, 1974) who generalised Kantowski's results to
include a cosmological constant.  
These works suggested that there is a non-vanishing
perturbation to the mean flux densities of distant sources caused by intervening structures,
at least for sources that are viewed along lines of sight that avoid mass concentrations.

\subsection{Flux conservation}

Weinberg (1976), however, argued via conservation of photons that 
for transparent lenses there could be no mean flux density amplification
and that the uniform universe formula for $D(z)$ remains valid.
The apparent distance $D$ of a source at a fixed $z$ is, by 
definition, proportional to $1 / \sqrt{\Omega}$
where $\Omega$ is the solid angle a `standard source' subtends (or would if resolved),
while conservation of surface brightness means that
the flux density $S$ is proportional to $\Omega$.  In terms of the
magnification $\mu \equiv S / S_0$, where $S$ is the actual flux density
and $S_0$ is the flux density a standard source
would have at the same $z$ if the structure were smoothed out,
Weinberg says that $\langle \mu \rangle_A = 1$,
where the averaging is over sources, or equivalently
over area on the source sphere (hence the subscript A).
Alternatively, one can say that $\langle D_0^2 / D^2 \rangle_A = 1$, where
$D_0$ is the angular diameter distance in the smoothed out background.
This result, however, rests on the implicit assumption that the area of the constant-$z$
surface is unaffected by lensing.

This invariance of the mean flux density, however, appears to contradict a
well-known theorem of gravitational lensing, stating that
at least one image is always magnified 
(Schneider 1984; Ehlers \& Schneider 1986; Seitz \& Schneider 1992).
Taking a somewhat different approach, 
Seitz, Schneider \& Ehlers (1994) have used the optical scalars
formalism of Sachs (1961) to show that
the square root of the proper area of a narrow bundle of rays 
$D = \sqrt{A}$ obeys the `focusing equation':
\begin{equation}
\ddot D / D = -(R + \Sigma^2) .
\label{eq:SSEfocusingequation}
\end{equation}
Here $\ddot D$ is the second derivative of $D$ with respect to affine distance along
the bundle;
$R = R_{\alpha\beta} k^\alpha k^\beta / 2$ is the
local Ricci focusing from matter in the beam, which for non-relativistic velocities is 
just proportional to the matter density;
and $\Sigma^2$ is the squared rate of shear from the integrated effect of up-beam Weyl focusing -- i.e.\ the
tidal field of matter outside the beam.  The resulting
{\em focusing theorem\/} is that the RHS of (\ref{eq:SSEfocusingequation}) is non-positive, so
that beams are always focused to smaller sizes, 
at least as compared to empty space-time, where 
beams obey $\ddot D = 0$. 
(see Schneider, Ehlers \& Falco 1992 and Narlikar 2010 for
further details and discussion).

In the cosmological context Seitz, Schneider \& Ehlers (1994) therefore state that
``a light beam cannot be less focused than a reference beam that
is unaffected by matter inhomogeneities'', at least up until caustic formation
and ``no source can appear fainter [...] than in the case that there
are no matter inhomogeneities close to the line-of-sight to the source''.
But it would be incorrect to conclude that inhomogeneities always cause
magnification: this analysis actually compares the
flux density of sources in a universe containing
a uniform density component plus localised positive density
lenses with sources in a universe containing only the uniform component.
This is not quite the same as the real question of interest, which is the
mean degree of focusing caused by perturbations about the mean
density -- i.e. lenses whose density can be negative as well as positive.

In a spatially flat FRW model, bundles of rays emanating from a source
or observer travel in straight lines at a constant speed in conformal coordinates,
so also obey $\ddot D = 0$.
For general weak-field perturbations to such a model, 
appendix \ref{sec:opticalscalars} proves an analogue of
(\ref{eq:SSEfocusingequation}) where the RHS is $- (\delta R + \Sigma^2)$.
For weakly perturbed bundles with $D$ close to $D_0$,
the unperturbed distance to redshift $z$,
we can average this equation, assuming $\langle \delta R \rangle$ 
vanishes and setting $D = D_0$ in the denominator, to obtain the
linearised {\em averaged focusing theorem\/} 
\begin{equation}
\langle \ddot D \rangle / D_0 = - \langle \Sigma^2 \rangle < 0 .
\label{eq:avgfocusingtheorem}
\end{equation}
This implies that $\langle D \rangle < D_0$ so objects viewed through
inhomogeneity have distances that are systematically decreased even when we allow
correctly for the fact that the mean mass of lenses is zero.

The transport equation for the rate of shear $\Sigma$
(see appendix \ref{sec:opticalscalars}) shows that,
in the perturbative regime at least, the resulting mean
change in the distance from this cumulative effect of
tidal shearing of beams by up-beam structure is, at leading order, 
$\langle \Delta D \rangle / D_0 \sim \langle \kappa^2 \rangle$,
where $\kappa$ is the usual first order lensing convergence
and $\Delta D \equiv D - D_0$.
The convergence for galaxies at $z \sim 1$ is on the order of 1\%
at degree scales, rising to a few percent for the cosmic microwave
background (CMB) at $z \simeq 1000$, so the mean squared value is $\langle \kappa^2 \rangle \sim 10^{-3}$
(e.g.\ Seljak 1996), which is non-negligible.
Furthermore, $\langle \kappa^2 \rangle$ 
is a strongly decreasing function of averaging scale,
so there is potentially a large effect for compact sources such as supernovae at high redshift.

While interesting and suggestive, 
one should not necessarily conclude that (\ref{eq:avgfocusingtheorem}) 
invalidates Weinberg's argument that $\langle D_0^2 / D^2 \rangle_A = 1$.
First, the focusing theorem is concerned with
$\langle D / D_0 \rangle$, which is not the same thing, and
second the focusing equation provides the apparent distance to
the far end of a ray propagated along some
chosen direction from the observer.  Averaging this, as we
shall discuss in more detail presently, is not the same as averaging over sources.

\subsection{Lensing and the CMB}

The subject has received much further attention over the years,
though with varied results, and the scope has expanded to incorporate lensing of the CMB.  
 
A significant general development came from Kibble \& Lieu (2005), who emphasised the
important distinction between averaging over sources -- which is 
appropriate for SN1a cosmology -- and averaging over directions on the
observer's sky -- which is more appropriate for CMB studies.
They went on to show that, averaged over the sky with equal weight per
unit solid angle $\Omega$, which we will denote by $\langle \ldots \rangle_\Omega$
it is the {\em inverse\/} magnification that is conserved: $\langle \mu^{-1} \rangle_\Omega = 1$, at least to the extent that multiple lensing is unimportant.  
But, as with Weinberg's argument, Kibble \& Lieu also assume that the area of the constant-$z$
surface is unperturbed.

Despite the conservation arguments,
many lensing analyses
have continued to claim large effects in the mean. Frequently, such calculations
make use of Swiss-cheese models.
Kantowski, Vaughan \& Branch (1995) and Kantowski (1998), for example,
claim to confirm Kantowski's earlier conclusions in his 1969 paper and show there
should be large effects for SN1a cosmology.
Ellis, Bassett \& Dunsby (1998) claim that Weinberg's assumption of invariance
of area may be strongly violated by strong lensing from small-scale structure
if one is considering observations of supernovae.
Clifton \& Zuntz (2009) find $\sim$ few percent bias in source
magnitudes using Swiss-cheese models.
Bolejko (2011a), also using Swiss-cheese models, finds that the
distance to the CMB last-scattering surface is strongly affected by
structure, with significant impact on cosmological parameter estimation.
Similar results are presented in Bolejko (2011b) and Bolejko \& Ferriera (2012).
Bolejko (2011a) provides a very useful and extensive review of other studies, some of which
(e.g.\ Marra et al.\ 2007) find large effects; some which
find effects at the level of a few percent (which would still be significant if correct);
while others claim that the effect is very small. An important example of the latter is  
Metcalf \& Silk (1997); they integrated the geodesic deviation equation, claiming
that the mean magnification is only a part-in-a-million effect.

Clarkson et al.\ (2012)
provide an extensive review of source amplification statistics,
focusing mostly on SN1a observations but also touching on the implications for the CMB.
They claim that the mean magnification of a source is 
\begin{equation}
\langle \mu \rangle \simeq 1 + \langle 3 \kappa^2 + \gamma^2 \rangle + \ldots
\label{eq:Clarkson++2012avgmu}
\end{equation}
where $\gamma$ is the usual first order image shear.
This is in conflict with Weinberg's result, though
it is qualitatively in line with the expectation for 
$\langle \Delta D \rangle / D_0$ from averaging the focusing equation,
both in the sign and in the order of magnitude of the effect,
but would indicate potentially serious problems for SN1a cosmology if correct.
This group has carried out a systematic analysis of the distance perturbation
in 2nd order relativistic perturbation theory (Umeh et al.\ 2014a,b).
They calculate both the perturbation to the redshift and the 
distance as a function of affine parameter (using the geodesic and optical scalar
equations respectively) and then solve the resulting pair of parametric equations.
A similar calculation has been carried out by Marozzi (2014). 

Most recently, Clarkson et al.\ (2014; hereafter CUMD14), find that there is a 
perturbation to the relation between distance and redshift  that, in our notation, is
\begin{equation}
\langle D / D_0 \rangle = 1 + \frac{3}{2} \langle \kappa^2 \rangle .
\label{eq:CUMD14Delta}
\end{equation}
They present several arguments to support this, and say that
``It implies that the total area of a sphere of constant redshift will be larger than in the background''.
They also compute the perturbation to the proper area of a surface of 
constant $z$ (the integral of $D^2$ over
the observer's sky) using the optical scalars transport equation and find this to be
the square of the integral along the ray of the first order perturbation $\Delta \theta$ to the
rate of expansion $\theta = \dot A / 2 A$ of the ray.  Here $A$ is the beam area expressed
in conformal background -- i.e.\ `co-moving' -- coordinates and dot denotes
the derivative with respect to conformal distance; $\theta$ is not to be
confused with an angle.  To zeroth order, and for a spatially
flat background, as we shall assume, $\theta$ is just the inverse of the conformal distance.
But at first order $\theta$ includes the additional rate of change of the beam area
caused by inhomogeneity.  Expressed in terms of the usual first order convergence $\kappa$ their
result for the area is
\begin{equation}
\langle D^2 / D_0^2 \rangle_\Omega = 1 + 4  \langle \kappa^2 \rangle .
\label{eq:CUMD14Deltad2}
\end{equation}
This is also in direct conflict with Weinberg.
Of this CUMD14 say ``This is a purely relativistic effect with no Newtonian counterpart --
and it is the first quantitative prediction for a significant change in
the background cosmology when averaging over structure'' 
(citing the review of dynamical backreaction by Clarkson et al.\ 2011).
They discuss how this may be thought of as arising 
because of `crumpling' of the surface of constant redshift which enhances its area.

CUMD14 applied their results to compute the mean 
perturbation to the distance of the cosmic photosphere in terms of
the matter density power spectrum; a significant advance over calculations
that use idealised spherical Swiss-cheese models.
They found the strength of the effect in conventional models
to be at the $\simeq 1$\% level. 
This, they argued, might significantly affect CMB cosmological
parameters -- in particular, resolving the tension between $H_0$ as inferred from
the CMB (Planck collaboration 2013) and via direct distance methods
(Riess et al.\ 2011; although see Efstathiou 2014).     

A puzzling feature of the calculation is the sign of the
effects (\ref{eq:CUMD14Delta}, \ref{eq:CUMD14Deltad2}): both distance
and area are {\em increased\/} by structure.  The first might seem to be opposite to 
the qualitative expectation from the averaged focusing equation.
The latter seems to be at odds with (\ref{eq:Clarkson++2012avgmu}); if 
the area of a surface of constant $z$ around a source is increased
then, following Weinberg, one would think that conservation of photons would imply that the
mean flux density seen by observers on that surface should be {\em decreased\/}.

Another surprising feature  is that much of the effect
arises from quite small scale structure.  The relevant information in the CMB
is encoded in the angular 
frequency $\ell$ of the `acoustic peaks' in 
the power spectrum of the temperature fluctuations.
These arise from perturbations of comoving scale of order 100 Mpc.
As mentioned, the mean square convergence at the photosphere on 
this scale is only $\sim 10^{-3}$
so it is hard to see how a $\sim 1\%$ effect arises.  
But the mean squared convergence is a strongly decreasing function of angular scale, scaling
roughly inversely with angle, and CUMD14 emphasise that their calculation
obtains a large contribution from lensing by structures down to $\sim$ 10 kpc scale.
Again this is hard to understand: as argued by
Ellis, Bassett \& Dunsby (1998), lensing by small scale structure should not
affect the angular size of extended objects such as the acoustic peak scale features.
However, no such objection exists with SN1a cosmology, where any lensing biases
could indeed reflect the high small-scale variance in $\kappa$. Thus the CUMD14
results can potentially induce a profound change in the inferences about the 
cosmological model that are normally drawn from high-$z$ SN1a
(e.g.\ Riess et al.\ 1998; Perlmutter et al.\ 1999).

\begin{figure}
\begin{center}
\includegraphics[width=82mm]{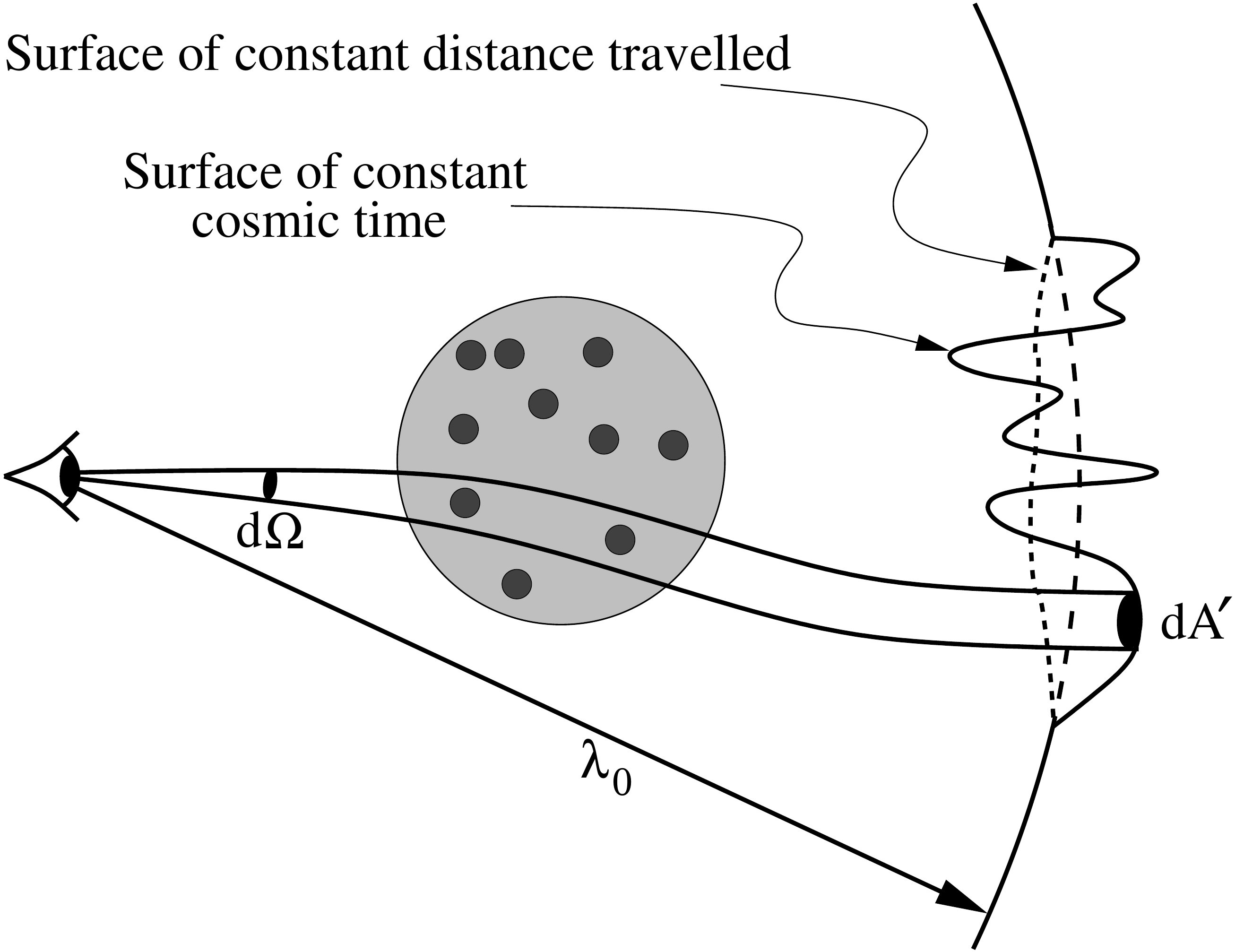}
\end{center}
\caption{In a hypothetical
universe with inhomogeneity in some finite region of space, consider the mean 
fractional change to the area of a surface of constant redshift, 
or cosmic time, which, in the absence of structure, lies at comoving distance $\lambda_0$
(note that our notation here differs from that of Weinberg 1976, who used $\lambda$ to denote
affine parameter).
We find that the area $dA'$ {\em is\/} biased, but to an extremely small extent, as a result  
of two competing effects:
(1) the radius reached by light rays is reduced because they are not straight; 
(2) the surface is `wrinkled' owing to time delays induced by the density fluctuations.
Regarding the first effect, a single lensing structure would cause
a deflection $\Theta_1 \sim \phi$ where $\phi$ is the metric 
perturbation (or the dimensionless Newtonian potential) and the 
corresponding fractional decrease in distance
reached would be $\Delta r / r \sim \Theta_1^2$.  The effect of $N \sim \lambda / L$
of these structures with metric fluctuations of random sign
-- assumed to have size $L$ and lying along a path length $\lambda$ --
would be $N$ times larger.  So $\langle \Delta r \rangle / r \sim \langle \Theta^2 \rangle \sim \phi^2 \lambda / L$
where $\langle \Theta^2 \rangle \sim N \Theta_1^2$ is the cumulative mean square deflection.
As for the second effect, one can draw an analogy with the surface of a swimming pool
perturbed by random waves of small amplitude.  These cause a fractional
increase in the area of the surface that is on the order of the mean square
tilt of the surface.  Here the surface is perpendicular to the
light rays, so we expect that the area increase is also, to order
of magnitude, $\langle \Delta A \rangle / A \sim \langle \Theta^2 \rangle$.
Both effects are caused predominantly by structures on scales of tens of Mpc, 
and these give only a part-in-a-million effect, counter to
much larger recent claims from relativistic perturbation theory.
This is the main new result of this paper, discussed at length in \S\ref{sec:areabias}.
}
\label{fig:sphere}
\end{figure}

\subsection{Overview of the present paper}

In the work presented here,
we dispute the above claims for significant flux amplification of sources,
or equivalently significant violation of conservation of area, and we attempt
to clarify the situation and explain the apparently discordant results that can be
found in the literature.  We also show that, despite its name, the
focusing theorem does not indicate any tendency for inhomogeneities to
cause magnification on average.

In the first part of the paper we show how, under the conventional assumption that the total
area of a surface of constant $z$ is unaffected by lensing, quantities such as the mean distance-redshift
relation {\em are\/} biased by lensing.  If the flux density 
$S$ is unbiased, then so is $4 \pi S / L = 1/D^2$; thus 
$\langle 1/D^2 \rangle_A = 1 / D_0^2$, when averaged over standard sources.
But the magnification is a fluctuating quantity, so there is a dispersion in
values of $1/D^2$ for different lines of sight or different source regions.
These therefore provide what are, in effect, noisy estimates of $1 / D_0^2$.
If one takes a non-linear function of $1/D^2$, such as $D$, and then
averages, the noise effectively gets rectified and inevitably
$\langle D \rangle \ne D_0$ even though $\langle 1/D^2 \rangle$ itself
is unbiased.

We show that the claims for non-zero mean source amplification or 
surface area bias in the calculations described above arise
partly from failing to make this distinction between distance bias and flux-density bias,
but mostly from ignoring the distinction between averaging over sources
and averaging over direction.
We find that the RHS of (\ref{eq:Clarkson++2012avgmu}) is the direction averaged (rather than source averaged)
amplification and (\ref{eq:CUMD14Delta}) is the bias in the source-averaged distance,
while the direction averaged distance, which is more relevant for CMB
observations, is
\begin{equation}
\langle D / D_0 \rangle_\Omega = 1 - \frac{1}{2} \langle \kappa^2 \rangle .
\label{eq:diravgdistance}
\end{equation}
The RHS of (\ref{eq:CUMD14Deltad2}) is the source averaged
inverse amplification $\langle \mu^{-1} = D^2 / D_0^2\rangle_A$ rather than
the average over the observer's sky (it also happens to be
the direction average of $\mu$) and so it does not
reflect any increase in the area of the photosphere or surface of constant $z$.

The rest of the paper consists of a calculation of the
perturbation to the area of a surface of constant redshift.
This is the net result of the competing effects of wiggling
of rays, which reduces the radius they reach, and the wrinkling
of the surface via time delays, which increases its area.
We show, using both the the geodesic equation (appendix \ref{sec:areacalculation})
and via the much more arduous route of the optical scalars formalism
(appendix \ref{sec:opticalscalars}), that the area bias is on the order of the
mean squared cumulative deflection angle, not the 
much larger mean squared convergence.
This means that, at least as far as sub-horizon scale structure is concerned,
Weinberg's flux-conservation argument is actually good to about one part in a million,
and no radical changes to SN1a cosmological inferences need to be made.
The calculation is somewhat involved, but a (only slightly over-simplified)
order-of-magnitude argument for why this should be the case is given in the caption
to Figure \ref{fig:sphere}.

The outline of the paper is as follows:
In \S\ref{sec:statisticalbias} we compute the statistical
bias in quantities such as the apparent distance under the assumption that area is unbiased by lensing.  
In \S\ref{sec:fluxconservation} we consider biases that
arise when averaging over sources.
In \S\ref{sec:diravgmagnification}, turning to the CMB, we consider
the statistics of quantities that are averaged over direction, rather
than averaging over sources.  In \S\ref{subsec:avginvmag} we consider the 
argument of Kibble \& Lieu (2005) that the direction averaged inverse magnification is conserved,
and in \S\ref{subsec:MSreview} we recall the
calculations of Metcalf \& Silk (1997).
In \S\ref{subsec:thinscreen} we calculate the mean inverse magnification caused by a thin screen
of lenses and find this is zero, consistent with Kibble \& Lieu
and we discuss the generalisation of this to a shell containing
deflectors of a finite size.
We then give the statistical bias in the
direction averaged distance and magnification and show that the
latter nicely accounts for (\ref{eq:Clarkson++2012avgmu}). 

In \S\ref{sec:areabias} we expand on the simple-minded
argument in the caption to Figure \ref{fig:sphere} and attempt to give a heuristic
explanation of the results of the detailed calculation presented in appendix \ref{sec:areacalculation}.
We note that the argument above is over-simplified in one respect, but we show that this
does not significantly alter the basic conclusion that the area bias is essentially zero.
In \S\ref{subsec:whatsizestructure} we identify the scale of structures that dominate
the ensemble effect on the area.
In \S\ref{subsec:areafluctuations} we consider fluctuations about the
ensemble average area increase that we have calculated.  We argue that
for sub-horizon scale density perturbations alone these are small, so the
area of one observer's sky will be close to the ensemble mean, and
the mean fractional change to flux densities will be close to $-\langle \Delta A \rangle / A_0$.
But for horizon scale perturbations there is a first order change to the
area that is typically on the order of the metric perturbation for these modes
and is actually larger in mean modulus than the ensemble mean from sub-horizon scale structure.
In \S\ref{subsec:CMBbias} we discuss how different ways of analysing CMB data could, in principle, result in
biased results, but argue that the conventional analysis method
(Hu 2000; Challinor \& Lewis 2005) avoids this.

Appendix \ref{sec:areacalculation} contains the detailed
calculation of the mean perturbation to the photosphere area
at second order in the metric perturbations, arising from
gravitational time delays and the associated light path deflection
(though the result is obtained entirely
as the average of the products of first order quantities).  
There, in \S\ref{subsec:weakfield}, we describe why the weak-field model for
metric fluctuations provides an adequate description and we
recall the analogy between light propagation in a weakly perturbed FRW cosmology
and light propagating in a medium with spatially varying, but locally isotropic, 
refractive index (`lumpy glass').
In \S\ref{subsec:rayBCs} we discuss the appropriate boundary conditions for the
end of the rays, and the distinction between surfaces of constant $z$ and
the cosmic photosphere (the latter being a surface of constant optical path in the
lumpy glass analogy).

The resulting ensemble mean for the fractional area perturbation
$\langle \Delta A \rangle / A_0$
emerges as a weighted integral along the line of sight of 
\begin{equation}
J \equiv - 8 \int\limits_{-\infty}^0 dy \ispace \xi'_\phi(y) / y
=2\pi \int k \, \Delta^2_\phi(k)\; d\ln k,
\label{eq:Jdefinition_maintext}
\end{equation}
where $\xi'_\phi$ is the derivative with respect to conformal (or `co-moving') 
background coordinates of the two-point spatial auto-correlation 
function of the dimensionless Newtonian gravitational potential fluctuations (divided by $c^2$); $\Delta^2_\phi$ is the dimensionless power spectrum of $\phi$
(variance per $\ln k$).
Physically, $J$ is the rate of change with respect to path length of the ensemble
mean square angular deflection of a ray.  It is 
similar to the `$J_3$' integral (Peebles 1981) and is dominated by large scale 
density fluctuations around the peak of the matter power spectrum.
This demonstrates rigorously that the effect is on the order of the mean
squared cumulative deflection angle,
and is therefore many orders of magnitude smaller than the statistical biases such as in
(\ref{eq:Clarkson++2012avgmu}), (\ref{eq:CUMD14Delta}), (\ref{eq:CUMD14Deltad2}) and (\ref{eq:diravgdistance}).

If the potential fluctuations are non-evolving
then $\langle \Delta A \rangle / A_0 = (2/3) \lambda_0 J$ where
$\lambda_0$ is the conformal distance to redshift $z$ (in units where
conformal distance has dimensions of length).
The value of $J$ in the `concordance' cosmological model is $J \simeq 9.9\times 10^{-11} h / {\rm Mpc}$
(this is the asymptotic value at high redshift when the potential is non-evolving; at low $z$ the potential decreases
with time and $J$ falls to about 60\% of this value at $z = 0$).
The overall path length is $\lambda_0 \simeq 9800 h^{-1} {\rm Mpc}$ so the 
net perturbation to the
area of the photosphere is $\langle \Delta A \rangle / A_0 \simeq 6 \times 10^{-7}$. 

We argue in \S\ref{sec:discussion} that, while the calculation is performed using perturbation
theory, this is valid even if non-linear lensing by very small
scale structure causes the shear and amplification
of most lines of sight to high redshift to be significant.

Several other technical calculations are consigned to appendices.
In appendix \ref{sec:expansion} we calculate the first-order
beam expansion rate that is used in appendix \ref{sec:areacalculation}.  
In appendix \ref{sec:geodesicdeviation} we show how the result of Metcalf \& Silk's 
calculation of the mean magnification, while qualitatively very similar
to ours, differs at a detailed level, particularly in regard to the effect from nearby lenses.
In appendix \ref{sec:opticalscalars} we show how our results 
can be obtained from the optical scalar formalism.
In appendix \ref{sec:sourceaveragedkappa} we show how the
non-vanishing inverse magnification averaged over sources
can be understood as arising because light paths to sources
tend to avoid over-dense regions.

Although some of the detail in the appendices is admittedly excessive in the face of what turns out to be a very small correction, there is value in collecting this material together. Flux conservation will probably
continue to be of great importance in gravitational lensing, and it is important to understand the issue in depth. We hope the present paper is a useful contribution to this process.
\vfill
\section{Statistical Biases}
\label{sec:statisticalbias}

In this section we show how quantities
such as distance can be statistically biased.  We consider both
averages over sources and over directions, presenting the conservation 
arguments of Weinberg (1976) and Kibble \& Lieu (2005) and showing
how powers of the distance may or may not be biased. We
illustrate these general points with the specific case of a thin
deflecting screen.

\subsection{Source averaged properties}
\label{sec:fluxconservation}

\subsubsection{Photon conservation}
\label{subsec:W76argument}

Weinberg (1976) argued that transparent lenses 
cannot change the mean flux density of sources
on the grounds of conservation of the flux of photons.  
The idea is that if a monochromatic source emits $N$ photons per period of the
emitted radiation then there must also be $N$ photons per (redshifted) period
passing through any surface of constant redshift.
Additionally, static lenses do not affect the redshift of sources. 
So, while individual sources may be
magnified or de-magnified, and some may be multiply imaged, the average 
fraction of photons from a source at redshift $z$ that 
we detect is the ratio of our telescope aperture to the proper area
of the sphere around each source on which the redshift has value $z$.
Averaged over the observers
that uniformly populate the sphere around a particular source, the flux density
is thus unbiased.

To obtain the quantity of more interest, which is the mean
flux density of sources seen by one observer, one can argue that
the average over the entire ensemble of pairs of
sources and the observers who see them to have redshift $z$ the flux density is also unbiased,
and if we are not a special observer the average over the
sources that we see with redshift $z$ should also have unbiased flux density.
Weinberg thus concluded that sources are, on average, unmagnified and that
the conventional formula for $D(z)$ remains valid. In fact, as we show below,
Weinberg's result holds for every observer, not merely in an ensemble-average sense.

This is a very powerful and general argument, which is not restricted to the weak-lensing regime --
though it does require that multiple images of sources from strong lensing are
either unresolved or that the flux densities of the multiple images have been aggregated. 
If we define the magnification of a source $\mu$ as the ratio of its flux density
to that which an identical source would have at the same redshift in an unperturbed FRW model,
or viewed along a path with no inhomogeneity, 
and imagine the source sphere at redshift $z$ to be tessellated into a very large number of 
equal area elements, each containing one standard source, then averaging over
these sources is equivalent to averaging over area and 
Weinberg's argument is that $\langle \mu \rangle_A = 1$ where the subscript indicates 
averaging $\mu$ weighted by area on the constant-$z$ surface.

The flux density is also inversely proportional to 
$dA / d\Omega$, the Jacobian of the transformation between position
on the source plane and angle on the observer's sky
(conservation of surface brightness means the flux density increases with
$d\Omega$ for given $dA$). The average of the
inverse of the Jacobian, weighted by area on the source sphere, is
$\langle d \Omega / d A \rangle_A = 
\int dA (d \Omega / d A) / \int dA = 4 \pi / A$.
We emphasise that $\langle \cdots\rangle$ is not an ensemble average, but
simply an average over the source sphere. Multiple lensing is accounted for
because the $d\Omega$ for the different images add into a single element of
total solid angle.
Invariance of mean flux density is therefore equivalent to the assertion
that the surface of constant $z$ has the same proper
area as would be the case if the matter inhomogeneity were smoothed out.

\subsubsection{Distance bias}
\label{subsec:distancebias}

Weinberg's endorsement of the conventional formula for $D(z)$
does {\em not\/} imply that the distance, averaged over sources, is unaffected by lensing.  
Rather, the mean {\em flux density\/} of standard candles uniformly
or randomly distributed over the constant-$z$ surface is unperturbed; i.e.\ the average of $1/D^2$ is
the same as its value in a uniform universe.  
Now, the distance is a non-linear function of $1/D^2$, as is the magnitude, 
and $1/D^2$ is a quantity that fluctuates between different lines of
sight (having a first order fractional perturbation $2 \kappa$ in the linear regime).
As a result, the distance- and magnitude-redshift relations are both biased
with respect to the conventional formula for $D(z)$.

Estimating this bias for point-like sources is difficult since small-scale
structure may cause large fluctuations in the magnification for narrow beams.
Consider a (possibly fictitious, though of the kind considered in perturbation
theory) universe with only small amplitude surface density
perturbations. The distance is proportional to 
$\mu^{-1/2}$ which, can be expanded, with $\Delta \mu \equiv \mu - 1$, as
$D / D_{0} \simeq 1 - \Delta \mu / 2 + 3 (\Delta \mu)^2 / 8 + \ldots$.  The average over sources
of the linear term vanishes, according to Weinberg, but the second order term does not average to zero.  
Instead, there is a statistical bias in $D$, with respect to its value in a homogeneous
universe $D_{0}$, of
\begin{equation}
\langle D / D_0 \rangle_A = 1 + \frac{3}{8} \langle (\Delta \mu)^2 \rangle_A + \ldots = 1 + \frac{3}{2} \langle \kappa^2 \rangle + \ldots
\label{eq:deltadoverd0}
\end{equation}
where the second equality, involving the mean squared weak lensing convergence $\kappa$, applies in the perturbative regime
where $\Delta \mu = 2 \kappa + \ldots$. Note that as the average distance perturbation is second order we
do not need to specify whether the average of $\kappa^2$ is weighted by area or solid angle as the
difference between these is a third order effect.

Similarly the average of $D^2 / D_0^2 = \mu^{-1}$ is readily found to be
\begin{equation}
\langle D^2 / D_0^2 \rangle_A = 1 + \langle (\Delta \mu)^2 \rangle_A + \ldots = 1 + 4 \langle \kappa^2 \rangle + \ldots
\label{eq:deltad2overd02}
\end{equation}
These are precisely the same as the distance (\ref{eq:CUMD14Delta}) and area
(\ref{eq:CUMD14Deltad2}) perturbations found by CUMD14.  But clearly (\ref{eq:deltad2overd02})
is not the perturbation to the area: that would be the average over {\em directions\/} rather than over source-plane
area, whereas (\ref{eq:deltad2overd02}) is the average over sources of $D^2 / D_0^2$ assuming that the area is actually precisely
unperturbed.

The applicability of these formulae to point-like sources in the real Universe is somewhat
questionable since galaxy
clustering observations tell us that $\langle \kappa^2 \rangle$ grows roughly inversely
with scale while the effective beam size, which introduces a cut-off, is tiny and
extrapolation is difficult.  Ellis, Bassett \& Dunsby (1998; hereafter EBD98) 
argue quite convincingly that this `ultraviolet divergence' problem for $\langle \kappa^2 \rangle$ is
potentially real and should not be ignored, though this is constrained
empirically by modelling of the scatter in supernova flux densities and,
out to $z \simeq 1$ at least, any enhancement in the scatter from lensing is small
(Sullivan et al.\ 2011; Conley et al.\ 2011).  
The bias estimated from large-scale structure alone, 
however, would apply in a hypothetical observation where measurements of
the average flux density are made on a patches of sky containing
large numbers of sources and the inverse square roots of these then averaged.
It should also correctly describe angular area magnification of structures in the CMB.

\subsubsection{Ellis, Bassett \& Dunsby's objection}
\label{subsec:EBD98}

\begin{figure}
\begin{center}
\includegraphics[width=80mm]{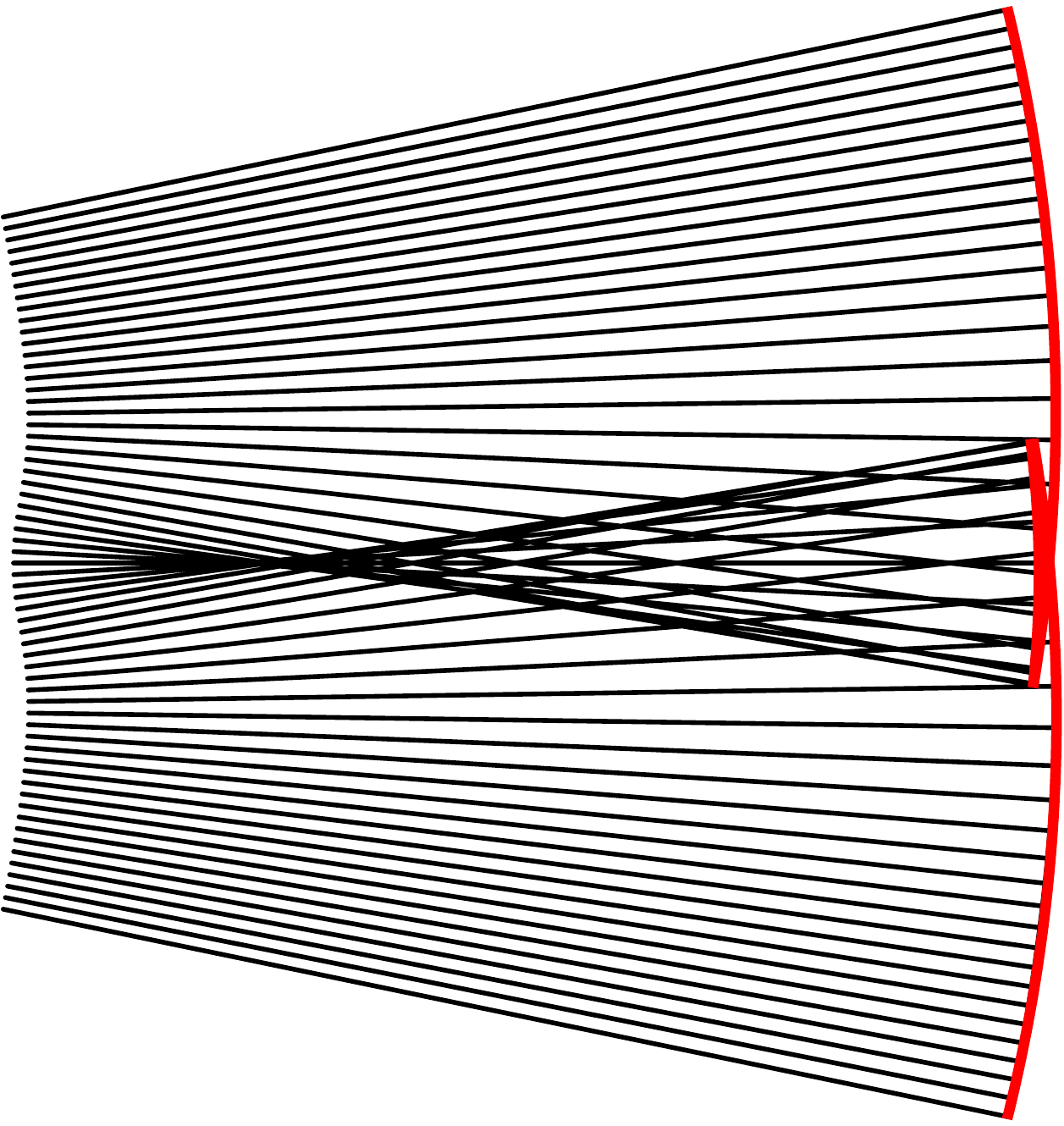}
\end{center}
\caption{Grossly exaggerated illustration of the form of the surface of constant redshift
in the case of strong lensing.
The lines are rays of light that start on, and are perpendicular to, a wavefront on the left.
This surface is distorted as a result of time delays induced by the lenses that
the light has previously encountered (not shown).
The rays are propagated to a constant redshift surface on the right.  This can either be viewed as the
surface of sources that an observer sees to have redshift $z$ at some epoch, or as the surface
around a source hosting observers who see that source to have redshift $z$.  Weinberg's flux conservation argument
relies on the assumption that e.g.\ the area of the {\em outer surface\/} here is identical to the area of
a sphere of the same constant $z$ in an unperturbed universe. If it is, the flux {\em density\/}, averaged over observers
on this surface is the same as for a homogeneous universe.  In reality, this surface is slightly
deformed, and its area is biased, so the mean flux density is not precisely unbiased.  
But as we argued in the caption to Figure \ref{fig:sphere} and discuss further in \S\ref{sec:areabias}
and in appendix \ref{sec:areacalculation}, the bias is predominantly caused by large-scale density
perturbations that are well understood, and the bias is extremely small and, for all 
practical purposes, negligible.}
\label{fig:zsurface}
\end{figure}

A weakness of Weinberg's argument, as was emphasised by EBD98,
is that he assumes that the surface of constant $z$ is a sphere and that
its area is unaffected by structures along the line of sight.
It is true that static lenses have little effect on the redshift of sources,
but in the real universe the set of observers
who see a source to have redshift $z$ at some time $t$ do not lie on a sphere, rather the surface will
in general will be slightly aspherical because of time delays associated with the inhomogeneity,
and if there are caustics it will be folded over on itself on small scales, 
so along any light path from the source there may be multiple
observers at slightly different distances who see the source to have redshift $z$ (each one of these
observers will see multiple images with very slightly different redshifts).  Similarly the set of sources 
that we perceive to have redshift $z$ at the present will lie on some aspherical and generally microscopically multi-foliated
surface, a section of which is illustrated schematically, though in grossly exaggerated form, in Figure \ref{fig:zsurface}.

EBD98's focus is on the effect of small-scale structure and its associated caustics. They emphasise 
the UV-divergence mentioned before and how this may in principle significantly increase the observed
areas corresponding to a net solid angle even when averaged over large angular scales.
This seems to us to be beside the point.  The effect of folding of the surface is already taken into
account by Weinberg in requiring that multiply imaged sources are either unresolved or their flux densities be aggregated.  
More significant is how much the area is biased, not counting the small-scale folding.
Referring again to Figure \ref{fig:zsurface} we would argue that 
the relevant question is: what is the effect of structure on the area of the outer surface?
We answer this in \S\ref{sec:areabias}.  For now we assume that there is no effect, and turn to consider
direction averages which are more relevant for CMB studies.

\subsection{Direction averaged properties}
\label{sec:diravgmagnification}

The studies mentioned above were mostly concerned with the
magnification of point sources.  Regarding the lensing of anisotropies
of the CMB, many studies have followed the pioneering work by Cole \&
Efstathiou (1989).  Here we shall focus only on the issue of the mean
magnification, reviewing the argument presented by Kibble \&
Lieu (2005): when averaging over directions on the sky, it is the 
{\it inverse\/} magnification that is conserved; we also discuss how sky-
and source-averages are related.

\subsubsection{Conservation of inverse magnification}
\label{subsec:avginvmag}

Kibble \& Lieu discussed the average magnification
using a model of uncorrelated random clumps of matter.  
But more significantly they emphasised the important and general distinction between averages over sources -- or 
equivalently over areas on the source plane -- and averages over directions on the
sky (i.e.\ averages weighted by solid angle):

\begin{quote}
``We may choose at random one of the sources at redshift z, or we may choose a random direction in the sky and look for sources there. These are not the same; the choices are differently weighted. If one part of the sky is more magnified, or at a closer angular-size distance, the corresponding area of the constant-z surface will be smaller, so fewer sources are likely to be found there. In other words, choosing a source at random will give on average a smaller magnification or larger angular-size distance.''
\end{quote}

For source averaging, Kibble \& Lieu reason that since the distance is,
by definition, $D = \sqrt{dA / d\Omega}$ and the flux density $S$ is proportional
to $1/D^2$ then, if $D_0$ is the distance for a standard
source viewed along an unperturbed path, the amplification is $\mu = D_0^2 / D^2$
and its average over area on the source (or observer) surface is
\begin{equation}
\langle \mu \rangle_A = D_0^2 \left\langle \frac{d\Omega}{dA} \right\rangle_A =
D_0^2 \frac{\int dA \ispace (d \Omega / d A)}{\int dA} = \frac{4 \pi D_0^2}{A} .
\label{eq:KLavgmu}
\end{equation}
We have already invoked this result above in saying that
Weinberg's result $\langle \mu \rangle_A = 1$ implicitly assumes that
the area is $A = 4 \pi D_0^2$ and is unaffected by lensing.  

For direction averaging, they show that a
precisely analogous statement can be made concerning $\langle \mu^{-1} \rangle_\Omega$:
\begin{equation}
\langle \mu^{-1} \rangle_\Omega = D_0^{-2} \left\langle \frac{dA}{d\Omega} \right\rangle_\Omega =
\frac{\int d \Omega \ispace (d A / d \Omega)}{D_0^2 \int d\Omega} = \frac{A}{4 \pi D_0^2} 
\label{eq:KLavginvmu}
\end{equation}
so, again if one assumes the total area $A$ is unperturbed, it is the direction average
of $\mu^{-1}$ that is conserved. 

In the absence of strong lensing both of the above results are unexceptionable.
But with multiple imaging the last step in (\ref{eq:KLavginvmu}) is questionable: if an
element of surface area can be reached via paths that start in
disjoint elements of solid angle, it would be counted multiple times --
so that one would expect $\int d \Omega \ispace (d A / d \Omega)$ to be greater than
$A$.  Kibble \& Lieu claim that (\ref{eq:KLavginvmu}) is of general validity,
but in doing so they take a very different definition of magnification
than the one employed here.  Rather than taking $D_0^2 \mu^{-1}$ to be the modulus
of $d A / d \Omega$, they include the sign of the Jacobian of the transformation
from angle to area coordinates, so that for some images $\mu^{-1}$ is formally
negative.  When there are multiple images, and in general there
are an odd number $2n + 1$ of these, then $n$ of them have odd parity
(Blandford \& Narayan 1986); these therefore have negative Jacobian, which
effectively cancels the multiple counting of areas.
In (\ref{eq:KLavgmu}) the integral over
area is understood to be over the outer surface -- which has a one-to-one
mapping to solid angle -- and the parity of the outer surface
is, as shown again by Blandford \& Narayan, always even.
Since the parity is not easily observable, (\ref{eq:KLavginvmu}) is of limited
practical utility when there are strong lenses.  
But to the extent that strong lensing can be ignored -- if the optical depth
is very low or if one is concerned with unresolved compact sources or
with the size of large structures (such as acoustic
peak scale ripples in the CMB) -- then it is the mean of the inverse of the absolute magnification
that is conserved.  

These results can also be understood in terms of the probability distribution for
amplification.  One can imagine calculating $\mu = D_0^2 d \Omega / dA$ for
an ensemble of rays fired in random directions
and propagated a path length $D_0$.  Denoting the probability distribution for $\mu$
in such an experiment by $P_\Omega(\mu)$ then $P_\Omega(\mu) d\mu$ is the fraction of
solid angle for which $\mu$ lies in a range $d\mu$ around $\mu$, so $P_\Omega(\mu) d\mu = d \Omega / 4 \pi$.
If there are no multiple images, the element $d\Omega$ 
maps to an area $dA = D_0^2 d \Omega / \mu$.
The fraction of the total area is thus $dA / A = D_0^2 d\Omega / \mu A = (4 \pi D_0^2 / A) \mu^{-1} P_\Omega(\mu) d\mu$;
but this must also be equal to $P_A(\mu) d\mu$,
where $P_A(\mu)$ is the probability distribution for $\mu$ over area, so the
two probability distribution functions are related by
$P_A(\mu) = (4 \pi D_0^2 / A) \mu^{-1} P_\Omega(\mu)$.
This gives
\begin{gather}
\langle \mu \rangle_A = 
\frac{\int d\mu \ispace \mu P_A(\mu)}{\int d\mu \ispace P_A(\mu)} = 
\frac{4 \pi D_0^2 \int d\mu \ispace P_\Omega(\mu)}{A \int d\mu \ispace P_A(\mu)} = \frac{4 \pi D_0^2}{A} \\
\langle \mu^{-1} \rangle_\Omega = 
\frac{\int d\mu \ispace \mu^{-1} P_\Omega(\mu)}{\int d\mu \ispace P_\Omega(\mu)} = 
\frac{A \int d\mu \ispace P_A(\mu)}{4 \pi D_0^2\int d\mu \ispace P_\Omega(\mu)} = \frac{A}{4 \pi D_0^2},
\end{gather}
consistent with (\ref{eq:KLavgmu}) and (\ref{eq:KLavginvmu}).
Putting these together
shows that $\langle \mu \rangle_A \langle \mu^{-1} \rangle_\Omega = 1$,
so conservation of one implies conservation of the other.  And clearly
both rest on the assumption that area is conserved.

If this assumption is correct then because $\mu^{-1}$ is
a fluctuating quantity one would expect $\langle \Delta \mu \rangle_\Omega \ne 0$.
Writing $\Delta \mu = (1 + \Delta \mu^{-1})^{-1} - 1$ (where $\Delta \mu^{-1} \equiv \mu^{-1} - 1$)
and expanding gives
$\langle \Delta \mu \rangle_\Omega = - \langle \Delta \mu^{-1} \rangle_\Omega 
+ \langle (\Delta \mu^{-1})^2 \rangle_\Omega + \ldots$.  But $\langle \mu^{-1} \rangle_\Omega = 1$
means the first term is zero and, since $\Delta \mu^{-1} = -2 \kappa + \ldots$, we
would have, in the perturbative regime, $\langle \mu \rangle_\Omega = 1 + 4 \langle \kappa^2 \rangle + \ldots$ (which,
we note, is the same as $\langle \mu^{-1} \rangle_A$ obtained in \S\ref{subsec:distancebias}).

An alternative way to get this result is to note that conservation of area $\int dA = 4 \pi D_0^2$
and the definition of magnification  $\mu = D_0^2 d\Omega / dA$ 
imply that for any function of the magnification $F(\mu)$ the different averages are related by
\begin{equation}
\begin{split}
\langle F \rangle_A & = \frac{\int dA \ispace F}{\int dA} = \frac{\int d\Omega \ispace (dA / d \Omega) F}{D_0^2 \int d\Omega} \\
& \quad\quad= \frac{\int d\Omega \ispace \mu^{-1} F}{\int d\Omega} = \langle \mu^{-1} F \rangle_\Omega 
\end{split}
\label{eq:avgFrelations1}
\end{equation}
and similarly
\begin{equation}
\langle F \rangle_\Omega = \langle \mu F \rangle_A,
\label{eq:avgFrelations2}
\end{equation}
these relations being exact to the extent that the optical depth for 
multiple images is small.  With $F = \mu$ in (\ref{eq:avgFrelations2}) we have
$\langle \mu \rangle_\Omega = \langle \mu^2 \rangle_A = \langle 1 + 2 \Delta \mu + (\Delta \mu)^2 \rangle_A$
so $\langle \Delta \mu \rangle_\Omega = \langle (\Delta \mu)^2 \rangle$ is exact (though
$\langle \Delta \mu \rangle_\Omega = 4 \langle \kappa^2 \rangle$ is only true to 2nd order precision).

\subsection{Geodesic deviation calculations}
\label{subsec:MSreview}

Metcalf \& Silk (1997; hereafter MS97) used geodesic deviation (rather than using
the optical scalar formalism as in most other studies) to calculate the
magnification of the cosmic photosphere to second order precision.
With a COBE normalised power spectrum of density perturbations (Bennett et al.\ 1996)
they found that lensing produces a non-zero mean magnification of structures on
surfaces of constant redshift if weighted by solid angle on the sky, but it is only at the
$\sim 10^{-6}$ level (it is on the order of $\langle \Theta^2 \rangle$)
and so is, for all practical purposes, observationally negligible.  This would seem to say
$\langle \mu \rangle_\Omega = 1$, where the subscript
denotes an average over direction (i.e.\ averaging with equal weight per unit solid angle).

That, however, would be at odds with Kibble \& Lieu whose conservation of 
inverse amplification implies, as we have seen, a relatively large ${\cal O}(\langle \kappa^2 \rangle)$
bias in $\langle \mu \rangle_\Omega$.
The resolution, however, is straightforward; the quantity that MS97 calculate is
actually the mean inverse magnification, as we now show.

MS97 calculate
the trace of the distortion tensor $\bD = \partial \delta \bTheta / \partial \bTheta$
(where $\bTheta$ denotes the 2D angular position vector on the flat sky), and
the expectation value of its integral that gives the parallel component of the change in
separation on the source plane of pairs of beams on the sky of separation $\bs$:
\begin{equation}
\beta_\parallel(|\bs|) = \int\limits_{-\bs/2}^{\bs/2} {\hat \bs} \cdot \langle \bD \rangle \cdot d\bTheta .
\end{equation}  
For $s = |\bs|$ smaller than the angle subtended by the coherence length, and dividing by $s$
to get a fractional quantity, this
is $\beta_\parallel / s = \langle \Tr(\bD) \rangle$, i.e.\ the {\em trace\/} of the
distortion.  This is not the magnification.  Nor, in general, is it the
inverse magnification, which is the {\em determinant\/} $|\bD|$.
However, if we write the distortion as $\bD = \bI + \bS_1 + \bS_2 + \ldots$,
where $\bI$ is the identity matrix and subscripts denote terms that are
of 1st order, 2nd order etc.\ in the potential, then 
$|\bD| = 1 + \Tr(\bS_1) + (|\bS_1| + \Tr(\bS_2)) + \ldots$ up
to second order.  The trace of the first order term vanishes for
a random fluctuating potential with zero mean.  It turns out (see below) that if the
potential fluctuations are statistically spatially homogeneous the ensemble average of the determinant of the
first order distortion vanishes also, $\langle |\bS_1| \rangle = 0$, so, for random lenses, the mean 
inverse amplification perturbation is just the trace of $\langle \bD \rangle$.  

We show, in appendix \ref{sec:geodesicdeviation}, that MS97's result can be expressed as
\begin{equation}
\langle \mu^{-1} \rangle = 1 - \frac{4}{\lambda_0} 
\int\limits_0^{\lambda_0} d\lambda \ispace (\lambda_0 - \lambda) J
\label{eq:MS97inversemag}
\end{equation}
where $\lambda$ is conformal distance along the path and $J$ is as defined
in (\ref{eq:Jdefinition_maintext}).  If the lensing structures 
have `coherence length' $L$ then $J \sim \langle \phi^2 \rangle / L$ 
so $\langle \mu^{-1} \rangle - 1 \sim \phi^2 \lambda / L$,
consistent with the hand-waving argument in the caption to Figure 1 that
this is on the order of the mean square cumulative deflection angle.
As we shall see, however, the actual effect differs from
(\ref{eq:MS97inversemag}), particularly for lenses close to the observer, 
but (\ref{eq:MS97inversemag}) has the correct order of magnitude.
In any case, MS97's result is not in conflict with Kibble \& Lieu, which is
the main point of this section.

\subsection{Effect of a thin lensing screen}
\label{subsec:thinscreen}

Further evidence against there being any
${\cal O}(\langle \kappa^2 \rangle)$ perturbation to $\langle \mu^{-1} \rangle_\Omega = 1$ comes from
considering lensing by a single deflecting screen or shell at conformal distance $\lambda_d$; this
is similar in principle to, but much simpler than, the full 3D calculation of MS97.  
As we shall see in the following sub-section, this also sheds light on claims
for significant source-averaged flux amplification.

In this model rays travel along straight paths in conformal
coordinates, receiving a small transverse
deflection at the lensing screen.  
As we are primarily concerned with small structures,
a reasonable first approximation is
to work in the `flat-sky' limit where both the screen and
the source surface are assumed to be planar, using a 2-D Cartesian
coordinate system to describe deflections and displacement of
rays for a beam that propagates along the $z$ axis.
The matrix relating positions $\bx'$ on the source plane 
(scaled by $\lambda_d / \lambda_s$) 
to positions on the deflector plane $\bx$  is 
\begin{equation}
\frac{d\bx'}{d\bx} = \left[\begin{array}{cc}
1 - \kappa + \gamma_1 & \gamma_2 \\
\gamma_2 & 1 - \kappa - \gamma_1 \\
\end{array}\right]\; ,
\end{equation}
where $\kappa = \nablaperp^2 \Phi$ and $\{\gamma_1,\gamma_2\} = -\{ \Phi_{11} - \Phi_{22}, 2 \Phi_{12} \}$
with $\Phi_{ij} \equiv \partial^2 \Phi / \partial x_i \partial x_j$ and 
$\Phi = [\lambda_d (\lambda_s - \lambda_d) / \lambda_s] \int d \lambda \; \phi$
where $\phi$ is the Newtonian potential and the integration is through the
deflecting shell.  Thus $\kappa$ is the usual weak lensing convergence (the surface density in units
of critical value) and $\gamma$ the image shear.  It follows from the 
definition of $\kappa$ and $\gamma$ that $\kappa^2 - \gamma^2 = 4(\Phi_{11} \Phi_{22} - \Phi_{12}^2) = 4 |\bnablaperp \bnablaperp \Phi|$.

The determinant of this matrix is the inverse magnification:
\begin{equation}
\mu^{-1} = \left| \frac{d\bx'}{d\bx} \right| =
(1 - \kappa)^2 - \gamma^2
= 1 - 2 \kappa + \kappa^2 - \gamma^2 .
\label{eq:determinant}
\end{equation}
At linear order this is just $1-2 \kappa$, and the average of $\kappa$
over directions from the observer -- which is equivalent to an average over
the deflecting screen -- vanishes, so the net averaged non-linear
inverse magnification is $\langle \mu^{-1} \rangle_\Omega = 1 + \langle \kappa^2 - \gamma^2 \rangle$.

In the `flat-sky' limit we can write the 2D lensing potential $\Phi(\bx)$ as a Fourier sum
$\Phi(\bx) = \sum_\bk {\tilde \Phi}_\bk \exp(i \bk \cdot \bx)$.
For a statistically homogeneous random deflection
screen, the expectation value of the product of the potential coefficients
for distinct Fourier modes vanishes, so $\langle {\tilde \Phi}_\bk {\tilde \Phi}^*_{\bk'} \rangle = P_\Phi(\bk) \delta_{\bk \bk'}$
with $P_\Phi(\bk)$ the power spectrum.
This follows directly from the assumed translational invariance of 
the statistical properties of the random deflector screen.
An immediate consequence is that 
$\langle \kappa^2 - \gamma^2 \rangle =
4 \langle |\bnablaperp \bnablaperp \Phi| \rangle = 
4 \langle \Phi_{11} \Phi_{22} - \Phi_{12}^2 \rangle 
= 4 \sum_\bk P_\Phi(\bk) (k_x^2 k_y^2 - (k_x k_y)^2) = 0$.
Thus the 2nd order contributions to
the inverse magnification, $\kappa^2 - \gamma^2$,
average to zero when we average over positions on the deflection
screen, or equivalently over direction at the observer.
The direction averaged inverse amplification in this model is therefore unity,
consistent with Kibble \& Lieu.  
Note that we have not imposed any restriction on the strength of the lensing
screen; however, as with (\ref{eq:KLavginvmu}), our approach is 
of questionable utility for strong lensing as the inverse magnification
is the Jacobian, which will be negative for some directions, rather than the modulus
of $D_0^2 dA / d\Omega$. 

To obtain the actual effect -- which does not vanish -- one needs to allow for
the finite size of the deflecting structures and compute the deflection
with post-Born corrections allowing for the non-flatness of the sky etc.  
If we imagine gravitational 
potential fluctuations of size $L$ -- the `coherence scale' --
and consider a shell of such objects around us then e.g.\ the convergence is 
$\kappa = [\lambda_d (\lambda_s - \lambda_d) / \lambda_s] \int d \lambda \; \nablaperp^2 \phi 
\sim \lambda \int d \lambda \; \nablaperp^2 \phi$ (where we are assuming a `typical' distance
to the screen; i.e.\ not very close to the observer or to the sources).
At first order this integral can be taken along the unperturbed path.  At next order
one must allow for the 1st order deviation of
the ray from the unperturbed path by 
a perpendicular displacement $\Delta \bx_\perp \sim \bnablaperp \phi L^2 \sim \phi L$.
Allowing for this might suggest a second order contribution to $\kappa$ whose ensemble average is non-zero:
$\langle \kappa \rangle \sim \lambda \int d \lambda \; \Delta \bx_\perp \cdot \bnablaperp \nablaperp^2 \phi
\sim  (\lambda / L) \phi^2$.
Comparing this to $\langle \kappa^2 \rangle \sim  (\lambda / L)^2 \phi^2$
we see that this is much smaller.  In fact,
as we shall see, and as is suggested by (\ref{eq:MS97inversemag}), there is no effect 
that is of first order in $\lambda / L$ as there are other corrections that cancel.
The leading order effect of a single thick screen is $\langle \Delta \mu^{-1} \rangle \sim \phi^2$, independent of $\lambda / L$.

\subsubsection{Direction averaged distance and magnification for a thin screen}
\label{subsec:diravgdist}

While the mean of the determinant (\ref{eq:determinant}) is unity, the same is not
true for its square root, or equivalently the apparent distance $D / D_0 = \sqrt{\mu^{-1}}$.  
Expanding this for small $\kappa$, $\gamma^2$ and keeping only up to second order contributions gives
\begin{equation}
D / D_0 = 1 - \kappa - \gamma^2 / 2 + \ldots
\label{eq:distancefromdeterminant}
\end{equation}
Taking the ensemble average of this, the first term vanishes and since $\langle \gamma^2 \rangle = \langle \kappa^2 \rangle$
we have
\begin{equation}
\langle \Delta D / D_0 \rangle_\Omega =  - \langle \kappa^2 \rangle / 2 + \ldots
\label{eq:deltadoverd0diravg}
\end{equation}
similar to (\ref{eq:deltadoverd0}) but with $-1/2$ in place of $+3/2$.
We can also obtain this from $\langle \mu^{-1} \rangle_\Omega = 1$ much
as we did for the source averaged distance, since 
$D / D_0 = (1 + \Delta\mu^{-1})^{1/2} = 1 + \Delta\mu^{-1} / 2 - (\Delta\mu^{-1})^2 / 8 + \ldots$
where again averaging over directions the first order term vanishes
and we can use $\Delta\mu^{-1} = - 2 \kappa + \ldots$.

Similarly if we take the inverse of (\ref{eq:determinant}) and expand we have, up to 2nd order,
\begin{equation}
\mu = 1 + 2 \kappa + (3 \kappa^2 + \gamma^2) + \ldots
\label{eq:mutosecondorder}
\end{equation}
and taking the average over the deflector surface the linear term goes away and we have
$\langle \mu \rangle = 1 + \langle 3 \kappa^2 + \gamma^2 \rangle + \ldots$.
This would seem to be the origin of the result (\ref{eq:Clarkson++2012avgmu}) of
Clarkson et al.\ 2012 for the mean amplification of sources.
But averaging over the deflector surface is an average over directions on the
sky, not an average over sources.  We saw in the previous section that
$\langle \kappa^2 - \gamma^2 \rangle = 0$ for a statistically homogeneous
screen, so $\langle \kappa^2 \rangle = \langle \gamma^2 \rangle$ so the above
is $\langle \mu \rangle_\Omega = 1+ 4 \langle \kappa^2 \rangle + \ldots$ consistent with
the result given at the end of \S\ref{subsec:avginvmag}.

Finally, we can note the further consequence that the mean convergence to sources is biased
low. Averaging (\ref{eq:mutosecondorder}) over area, and using 
$\langle \gamma^2 \rangle = \langle \kappa^2 \rangle$, we see that
\begin{equation}
\langle \kappa \rangle_A = -2\langle \kappa^2\rangle\; .
\end{equation}
The interpretation of this result is discussed further in 
appendix \ref{sec:sourceaveragedkappa}.
\section{Area bias}
\label{sec:areabias}

As we have emphasised, the above conservation theorems depend
on the assumption that the source surface has an area that is unaffected by metric
fluctuations.:
\begin{equation}
\eqalign{
\langle \mu \rangle_A &= {4\pi D_0^2\over A}\, ;\cr
\langle \mu^{-1} \rangle_\Omega &= {A\over 4\pi D_0^2} \; .
}
\label{eq:muavgpair}
\end{equation}
Weinberg's flux-conservation argument is that the mean flux density is
the ratio of the telescope
aperture to the area of the outer surface of constant $z$; if this area is increased, 
then observers will measure a decreased mean flux density and vice versa.
But as indicated in Figure \ref{fig:zsurface}, there are good grounds
to expect that the outer surface is indeed not
precisely spherical in the presence of foreground inhomogeneities.
We now elaborate on this issue and calculate the corrections to (\ref{eq:muavgpair}), which
turns out to be very small -- on the order of the squared cumulative deflection angle.

Equations (\ref{eq:muavgpair}) indicate a reciprocal relationship $\langle \mu \rangle_A = 1 / \langle \mu^{-1} \rangle_\Omega$.
If we relax the assumption that the area is unperturbed by lensing we can generalise this as follows:
If we consider a solid angle  $d\Omega$, the number of unit areas $N$ that
fall within this beam on the surface at redshift $z$ is proportional to $dA / d\Omega$, 
while the flux density of standard sources (or $1/D^2$) is inversely proportional to this.
The mean flux density -- or equivalently the mean inverse apparent distance squared -- is therefore
\begin{equation}
\left\langle \frac{1}{D^2} \right\rangle_A = \frac{\int d\Omega \ispace N d \Omega / dA}{\int d\Omega \ispace N}
= \frac{\int d\Omega}{\int d\Omega \ispace dA / d\Omega} =
\left\langle \frac{dA}{d\Omega} \right\rangle^{-1}_\Omega.
\end{equation}
The solid angle here can be considered to be at the observer, in which 
case the area weighted average is an average over the sources seen by that
observer, or it may be considered to be at the source, 
in which case the area weighted average is an average over observers
as considered in Weinberg's argument.
Regardless of which interpretation one adopts,
the above formula says that, as before, we have $\langle \mu^{-1} \rangle_\Omega  = 1 / \langle \mu \rangle_A$
so the reciprocity of these averages is valid in general.
This relation means that one can calculate the mean of $1/D^2$ for the sources seen by
an observer by calculating the average over that observer's sky of $dA / d\Omega$ and then
taking the inverse.

But if the area is biased, this cannot be to the same extent 
for all observers (there will always be
some rare observers who inhabit spheres containing negligible fluctuations).
Ideally, therefore, one would want to know the probability distribution
for the sky average of $dA / d\Omega$, but calculating that
is very difficult.  What {\em is\/} amenable to calculation, however,
is to calculate the {\em ensemble\/} average of $\langle dA / d\Omega \rangle_\Omega$
(averaged over an ensemble of randomly placed observers).
What makes this tractable is the fact that we wish, naturally, to
assume that the metric perturbations take the form of a statistically 
homogeneous and isotropic random field.  Under that assumption
the ensemble average of $\langle dA / d\Omega \rangle_\Omega$
is precisely the same as the average of $dA / d\Omega$ over an
ensemble of realisations for the metric perturbation field for
a single ray fired from the origin along a random direction (or along the 
$z$-axis say).  We will denote this average by $\langle dA / d\Omega \rangle_{\rm ens}$.

This does not provide the full probability distribution for $\langle dA / d\Omega \rangle_\Omega$,
for which one would also need to know, at least, the RMS fluctuation about the
ensemble mean.
But if we assume that the average over any one observer's sky
of $dA / d \Omega$ comes from a large number of effectively statistically independent 
regions then it would seem reasonable to assume that sky average
for any observer will be given, to a good approximation, by the ensemble average
of the sky average.
And if so it should also be
valid to approximate the mean flux-density amplification of sources
for one observer by the inverse of $D_0^{-2} \langle dA / d \Omega \rangle_{\rm ens}$.
We shall therefore calculate, in the first instance, the ensemble average of
$dA / d \Omega$ (which, when multiplied by $4 \pi$, gives the ensemble mean of the
area and hence the mean fractional perturbation to the area $\langle A \rangle_{\rm ens} / A_0 - 1$
with $A_0$ the unperturbed area), although we shall return
to the question of fluctuations shortly.

The calculation of $\langle dA / d \Omega \rangle_{\rm ens}$ is presented in appendix \ref{sec:areacalculation}.
Here we give an overview of the essential points.
As before, we motivate this via a simple model of random over- or under-densities of scale $L$
and density contrast $\Delta$ for which the Newtonian gravitational potential --
cast in dimensionless form by dividing by $c^2$ -- is $\phi \sim H^2 L^2 \Delta / c^2$.
Along the way we provide the more quantitative key results from appendix \ref{sec:areacalculation}
which are valid for arbitrary random perturbations; the mean area perturbation 
being expressed purely in
terms of the 2-point function of the metric perturbations, independent
of higher-order statistics.  
Following this, in \S\ref{subsec:whatsizestructure} we show that the mean
bias is dominated by structures of scale of order tens of Mpc.
In \S\ref{subsec:areafluctuations} we return to the question of how large
are the fluctuations in the area for any particular observer compared to the
ensemble average.  In the interest of clarity henceforth all averages $\langle \ldots \rangle$
will be understood to be ensemble averages unless otherwise explicitly indicated.

\subsection{Surface of constant distance travelled}

We first consider a geometrical effect: owing to the wiggly nature of the light
paths, the radius reached by a path of total length $\lambda_0$
will be less than $\lambda_0$. As usual, we carry out this calculation viewing
the rays as propagating backwards in time from the observer.
The light deflection angle by a single localised structure is the integral of the
transverse potential gradient through the structure so this is $\Theta_1 \sim L \nablaperp \phi \sim \phi$.
As we are assuming a spatially flat background, simple geometry tells us
that the deflector must be displaced from the straight line from observer
to the surface by $d_\perp = \Theta_1 \lambda_{\rm od} \lambda_{\rm ds} / \lambda_{\rm os}$,
where the subscripts denote observer, deflector and (source) surface
and where $\lambda$ is background conformal coordinate distance along the path.
Pythagoras tells us that the change in distance reached (as compared to 
the sum of the hypotenuses $\lambda_{\rm od} + \lambda_{\rm ds}$)
is $\Delta \lambda = - (1/2) d_\perp^2 \lambda_{\rm os} / (\lambda_{\rm od} \lambda_{\rm ds})$ working to
second order precision.
Combining these gives the change in distance reached 
$\Delta \lambda = - (1/2) \Theta_1^2 \lambda_{\rm od} \lambda_{\rm ds} / \lambda_{\rm os}$.
We see here the usual `lensing kernel' that suppresses the effect of deflectors close to either
end of the path. Thus there is no scope for anomalously
large effect from deflectors near the end point.

We next assume that the effect (on the distance reached) of the $N \sim \lambda / L$ multiple
deflectors along a line of sight is simply the sum of the effects of individual deflectors.  
In this regard, we note that in the above paragraph we are not `solving the lens equation' for
some given configuration of observer, source and deflector.  We simply
fire off a ray in an arbitrary direction that happens to meet a deflector,  and ask how far away in background coordinates will the end of the ray 
be after it travels a net path length $\lambda_{\rm od} + \lambda_{\rm ds}$.
Thus the reduction in background conformal distance from $N$ independent deflectors
is just the sum of the (second order) effects from individual small angular deflections,
to give $\langle \Delta r \rangle / r \sim \langle \Theta^2 \rangle$ and
a corresponding change in area of twice this.
This is confirmed in the perturbative regime in appendix \ref{sec:areacalculation}, where
we find that the perturbation to the area of the constant distance travelled surface is
\begin{equation}
\langle \Delta A \rangle / A_0 = 2 \langle \Delta r \rangle / r = 
- \frac{2}{\lambda_0^2}\int\limits_0^{\lambda_0} d\lambda\; \lambda (\lambda_0 - \lambda) J(\lambda) \, ,
\label{eq:distancetravelledDeltaA}
\end{equation}
with $J$ as defined in (\ref{eq:Jdefinition_maintext}) and where we
see, as expected from the consideration of a single deflector, the
presence of the lensing kernel $\lambda (\lambda_0 - \lambda) / \lambda_0$.

The mean perturbation to the area of the constant distance travelled surface is thus
determined solely by the power spectrum, or equivalently by the 
2-point correlation function, of the metric fluctuations.
The perturbation to the distance reached for any individual line of
sight is also of second order in the metric perturbations, there being no
first order perturbation.  
This means that if the size of the perturbations $L$ (or the correlation length) 
is much less than the path length -- which is a good
approximation for the structures that are relevant here -- the
variation in $\Delta r / r$ between between different paths will be very small.  More
precisely, we would expect 
$\langle (\Delta r / r - \langle \Delta r \rangle / r)^2 \rangle^{1/2} \sim N^{-1/2} \langle \Delta r \rangle / r \ll \langle \Delta r \rangle / r$,
with the numerical coefficient being determined by higher than 2-point statistical properties of the
metric fluctuations.  Thus the surface of constant (conformal background coordinate) distance travelled
should be visualised as almost exactly spherical and with radius in background coordinates $r = \lambda_0 - \Delta r$.

\subsection{Surfaces of constant redshift or cosmic time}

We now discuss the countervailing increase of area from wrinkling 
of the surface.  In the Introduction we made an analogy with the
surface of a swimming pool perturbed by small amplitude -- i.e.\ height $\ll$ wavelength -- random
waves, which yield a fractional change in area of $\langle \Theta^2/2 \rangle$.
We now explore this in more detail, and draw attention to
one short-coming of the analogy; but we show that this does
not significantly change the basic conclusion that the 
cosmological effect is also of order $\langle \Theta^2 \rangle$.  
 
Despite the conceptual simplicity, this calculation is rather more subtle in detail
than the radial bias from the distance-covered effect. A number of terms arise,
whose origin is as follows. We start with a specific beam of solid angle $d\Omega$ at
the observer, which would correspond to an area $dA_0 = \lambda_0^2d\Omega$ at the
photosphere in the absence of structure.
As a result of lensing magnification, this beam passes through a different
area $dA$ at the constant distance travelled surface, which we can write 
as $dA = r^2 d\Omega'$, where we are defining the fictitious
solid angle $d\Omega'$ as that which the area $dA$ (which is 
perpendicular to the outward normal, though not perpendicular to the
beam direction) would subtend if there were no light deflection.
It follows then that 
\begin{equation}
\frac{dA}{dA_0} = \frac{d\Omega'}{d\Omega} \left(1 + \frac{2 \Delta r}{r} \right) .
\label{eq:dAdA0}
\end{equation}

What we actually want is the expectation value of the area of the intersection
of the beam with the actual (i.e.\ perturbed) photosphere, which we
will denote by $dA'$ (see Figure 1).  This differs from $dA$ by two further multiplicative
factors:
\begin{equation}
\frac{dA'}{dA} = (1 - \Theta^2/2) \times (1 + 2 \theta \Delta \lambda) .
\label{eq:dAprimedA}
\end{equation}
These arise as follows:  The beam is not, in general, perpendicular to 
the surface of constant distance travelled but has some tilt, which we denote
here by $\Theta$. This is a first order quantity that we compute using the
geodesic equation.  The first factor (times $dA$) is therefore the cross-sectional area
of the beam at that point.
The second factor in (\ref{eq:dAprimedA}) is the amount by which the 
beam expands or contracts in passing from the surface of constant
distance travelled to the actual photosphere.
Here $\theta \equiv \dot A / 2 A$ is the expansion rate, 
with $A$ the cross-sectional beam area in conformal background coordinate units
and $\dot A$ its rate of change with path length,
and $\Delta \lambda$ is the extra path length -- which may be positive or
negative -- caused by the gravitational time delay:
\begin{equation}
\Delta \lambda = 2 \int d\lambda \ispace \phi .
\label{eq:Deltalambda}
\end{equation}  
As we are considering the effect of intervening lenses
we can ignore the effect of the perturbations at the end of the
rays, so the photosphere is the intersection of our past light cone
with the surface of constant cosmic time $t = t_{\rec}$.  That means
that it is a surface of constant optical path or, equivalently, 
a wave-front or location of a backward propagating pulse of radiation.  
It is therefore perpendicular to the ray direction at the end of the beam,
so there is no additional angular correction factor needed in (\ref{eq:dAprimedA}).

In the absence of perturbations the beam expansion rate is just $\theta = 1/\lambda$,
so we can write $\theta = 1/\lambda + \Delta \theta$, where
$\Delta \theta$ is the perturbation to the expansion,
and combine (\ref{eq:dAdA0}) and (\ref{eq:dAprimedA}) to obtain
\begin{equation}
\begin{split}
\frac{dA'}{dA_0} & = \frac{d\Omega'}{d\Omega} 
\left(1 + \frac{2 \Delta r}{r} \right)
\left(1 - \frac{\Theta^2}{2} \right) \\
& \quad \times (1 + 2 (1/\lambda + \Delta \theta) \Delta \lambda) . \\
\end{split}
\label{eq:dAprimedA0}
\end{equation}
We need to ensemble average this equation, retaining all terms up to 2nd order.
Parts of this are straightforward: $\Delta r$ is a 2nd order quantity 
so we do not need to worry about correlations between it and any other
factors.  The same is true of $\Theta^2$.
In appendix \ref{sec:areacalculation}, we also show that 
$\langle d\Omega'/d\Omega \rangle =1$. For the
present calculation, we can therefore replace
$d\Omega'/d\Omega$ by $1-2\kappa$, and the only complication
is to allow for the correlation between $\kappa$ and other first-order
terms.  We now discuss the various factors here in terms of the
`random blobs' model.

The first order (i.e.\ Born approximation) time-delay -- or perturbation
to the path length to the photosphere -- 
for a single perturber is $\Delta \lambda_1 = 2 \int d\lambda \ispace \phi$,
where the integral is through the structure, so
$\Delta \lambda_1 \sim \phi L$.  The cumulative
effect is a random sum of $N$ of these with RMS value 
$\Delta \lambda \sim \sqrt{N} \Delta \lambda_1 \sim \phi \sqrt{\lambda L}$
where now $\phi$ is the RMS potential fluctuation.  This averages to
zero when multiplied by the zeroth order expansion $1/\lambda$
but it correlates with the first order expansion $\Delta \theta$.
At the end of a path that happens to be over-dense, both $\Delta \lambda$ 
and $\Delta \theta$ will be negative, and vice versa for an under-dense
path.  The result is a systematic positive bias to the area at 2nd order. 
Now $\Delta \theta$ is the rate of change
of the first order convergence $\kappa$ so 
$|\Delta \theta| \sim |\kappa| / \lambda$.  Since $|\kappa| \sim \phi (\lambda / L)^{3/2}$ this
means that $\langle \Delta \lambda \times \Delta \theta \rangle \sim \phi^2 \lambda / L$
or just the same, to order of magnitude,
as the effect of the light path wiggling reducing the distance.
In fact, $2 \langle \Delta \lambda \times \Delta \theta \rangle = \langle \Theta^2 \rangle$
so this, combined with the third factor in (\ref{eq:dAprimedA0}) results in an
area increase $1 + \langle \Theta^2 \rangle / 2$ exactly as in the swimming pool analogy.

The increase of area described so far depends only on the variance of the
ray directions at the surface.  It does not seem to depend on where along the path the deflections
were imposed.  Also, and interestingly, we find that for the case that
there is no evolution of the metric fluctuations (as is the case for linear
perturbations of an Einstein-de Sitter model) this increase in area
cancels the decrease (\ref{eq:distancetravelledDeltaA}) from 
distance reached being less than distance travelled
and the ensemble average effect would be zero.

But there are two more factors we have not considered.
One is the possibility of a significant 2nd order (i.e.\ post-Born approximation) contribution from
$\Delta \lambda$ itself, as this multiplies the zeroth order expansion rate.
But in fact this turns out to be sub-dominant and can be ignored.
Finally, we need to consider the fact that $\kappa$ in $d\Omega'/d\Omega$ is correlated
with the path length perturbation $\Delta \lambda$. 
This gives a 2nd order term
$- 4 \langle \kappa \Delta \lambda \rangle / \lambda$.
With $\kappa \sim \phi (\lambda / L)^{3/2}$ and $\Delta \lambda \sim \phi \sqrt{\lambda L}$
this is yet another contribution to $\Delta A / A \sim \phi^2 \lambda / L$
so this is also of order
$\langle \Theta^2 \rangle$ so this does not change the
conclusion regarding the order of magnitude strength of the effect
(but it does mean that the net effect is not zero for non-evolving metric fluctuations).

\comment{
We also considered an alternative approach, calculating the area increase as an ensemble average of the
fractional end-of-beam area change for a specific element of area on the 
unperturbed source surface (i.e. averaging over fictitious rays that travel from
the observer without undergoing deflection). Initially, this seems more
straightforward, as the element of area on the true wrinkled photosphere is
larger than on the unperturbed photosphere by the foreshortening factor
$1+\Theta^2/2$. However, in this approach one is fixing
both ends of the ray, so it is necessary to allow for the fact that light paths
between two given points follow perturbed trajectories, which 
tend to avoid over-densities (see appendix \ref{sec:sourceaveragedkappa}).
As a result of integration over the perturbed trajectory, an additional
second-order contribution to $\Theta^2$ will arise, complicating the calculation.
We believe that this approach should be capable of reproducing the results of our
direct averaging over true light rays, but we have not verified this in detail.
}

The final result for the fractional change in area, 
combining the reduced distance travelled and the
area enhancement from wrinkling, is obtained in appendix \ref{sec:areacalculation}:
\begin{equation}
\langle \Delta A \rangle / A_0 
= \frac{1}{\lambda_0^2} \int\limits_0^{\lambda_0} d\lambda \; (2 \lambda (\lambda_0 - \lambda) + \lambda^2) J(\lambda) .
\label{eq:finalDeltaAoverA}
\end{equation}
This result is of second order in the metric fluctuations and is valid at leading order
in the assumed small parameter $L / \lambda$.  For constant $J$ this is
$\langle \Delta A \rangle / A_0 = + (2/3) \lambda_0 J$, which is positive
-- so the competing effects of paths wiggling and surface crinkling
do not cancel.
However, as anticipated in the order-of-magnitude argument presented
in the Introduction, the change is extremely small: roughly a part-in-a-million effect.
Appendix \ref{sec:areacalculation} shows that $J$ may also be interpreted
as the rate of change of the squared transverse deflection with path length,
so quite generally the perturbation to the area is on the order of
of the cumulative deflection angle squared.  

If one is concerned with discrete sources, rather than the CMB, then
the observationally relevant area is not a surface of constant cosmic time, but a surface
of constant redshift.  For linear density perturbations -- and we will
shortly see that the effect is dominated by such perturbations -- the surface
of constant cosmic time is not at constant observed redshift because of the
ISW effect.  One result of this, as we show in \S\ref{subsec:rayBCs},
is to change the first order perturbation to the path $\Delta \lambda$ 
-- to sources at distance $\lambda_0$ as caused by structure at distance $\lambda$ --
introducing  a factor $1 + (\phi' / \phi)_\lambda (a'/a)_{\lambda_0}$ 
in the integral in (\ref{eq:Deltalambda}).
Here $\phi' \equiv \partial \phi / \partial \eta$ and  $a' \equiv d a / d \eta$.  
Another is that, unlike the photosphere, this surface is not normal to the beam direction,
so there is an extra factor $1 + \Theta'^2/2$ -- where $\Theta'^2$ is the
squared angle between the normals of the constant-$z$ and constant cosmic time surfaces --
to convert from cross-section to area at constant $z$.   
These effects, however, are only significant for sources at low redshift
and do not qualitatively change our conclusions regarding the size of the effects.

\subsection{What size of structures are important?}
\label{subsec:whatsizestructure}

\begin{figure}
\begin{center}
\includegraphics[width=83mm]{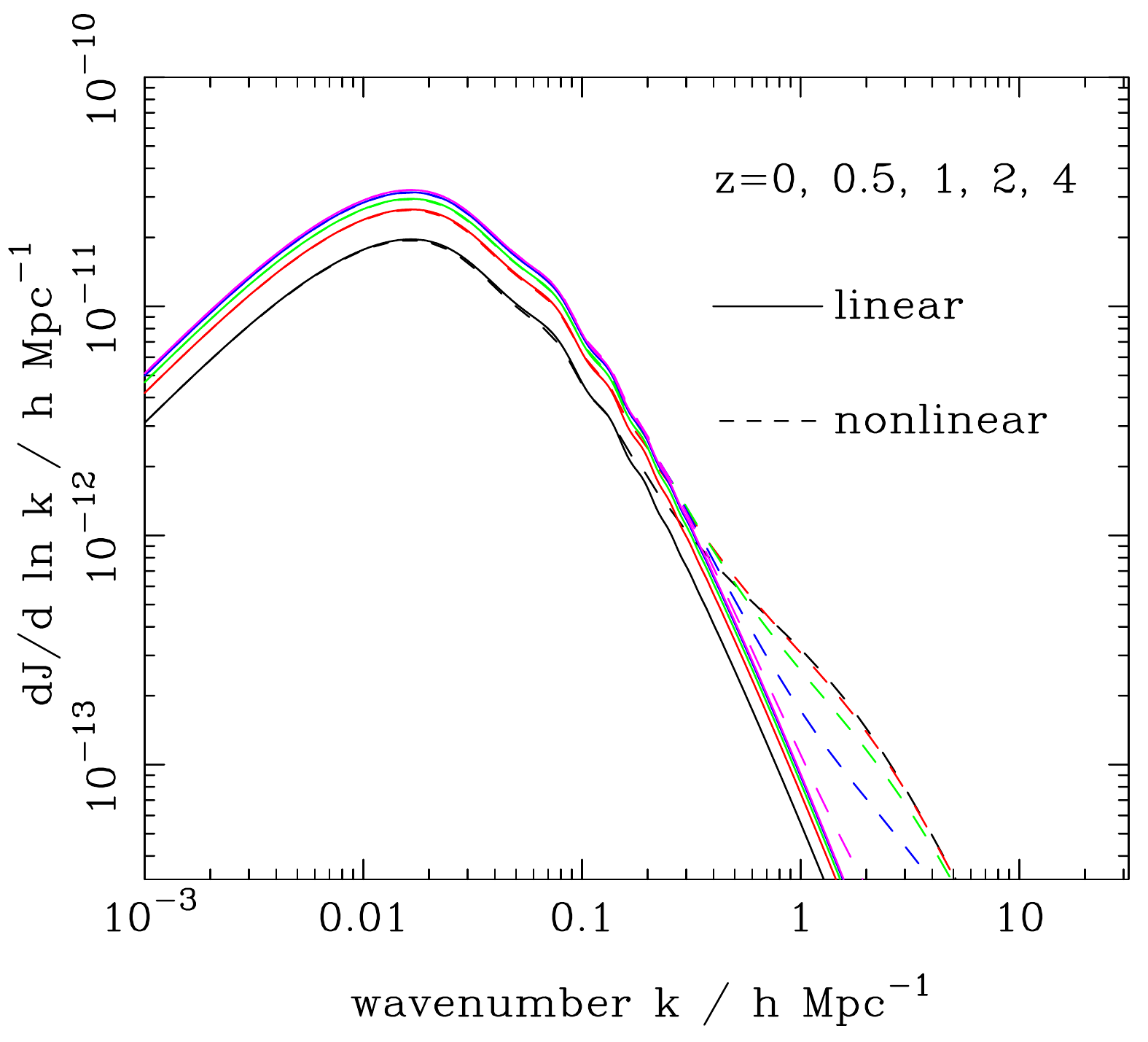}
\end{center}
\caption{Contribution to $J$ for the the concordance model as
a function of wave-number.  This quantity, when multiplied by the
path length gives the fractional perturbation to the area,
which we see here is dominated by modes of scale $k^{-1} \simeq 50 h^{-1}$Mpc.
See \S\ref{subsubsec:deflectionvariance} for details.}
\label{fig:dJdlnk}
\end{figure}

Unlike $\langle \kappa^2 \rangle$, one can argue
that $\langle \Theta^2 \rangle$ is
dominated by large-scale structure, so that uncertainty from
highly non-linear small-scale structure is negligible, and the
overall effect is definitely extremely small.
The evidence from galaxy clustering -- in the quasi-linear and linear regime -- is
that $\xi \propto 1 / r^2$ or thereabouts. This measures the density variance, so the density contrast of structures of some scale
$L$ is $\Delta \sim \sqrt{\xi} \propto 1/L$.
As we have seen, the mean squared deflection is $\langle \Theta^2 \rangle \sim N \Theta_1^2 \sim (H L / c)^3 \Delta^2$.  With $\Delta \propto 1/L$ this
is an increasing function  of scale.  This increase does not continue to indefinitely large
scales in conventional models.  As the spectral index increases the total variance converges, with most of the
variance coming from the logarithmic interval where $n \simeq 0$ or scales of tens of Mpc.
This is quantified in Figure \ref{fig:dJdlnk} which shows the contribution to $J$ per
logarithmic interval of wave-number from equation (\ref{eq:JfromP}):
$dJ/d\ln k = 2\pi\,  k\, \Delta^2_\phi$.
As can be seen, the modes that contribute most
strongly have inverse wave-numbers $k^{-1} \sim 50 h^{-1} {\rm Mpc}$, while 
non-linear structures have very little effect.

The shear $\gamma$ and the
convergence $\kappa$ from sub-horizon scale structures
are much larger, being on the order of $\kappa \sim \lambda \Theta / L \sim (H L / c)^{1/2} \Delta$.
In contrast to the deflection angle this is a decreasing function of scale.
For $\sim 100$Mpc scale structures with $\Delta \sim 15$\% the convergence is a
few percent (e.g.\ Seljak 1996) while the deflection is $\sim 30$ times smaller (about a few arc-minutes or $\sim 10^{-3}$ in radians), 
and $\langle \kappa^2 \rangle \sim 10^{3} \langle \Theta^2 \rangle$.
More quantitatively, equation (\ref{eq:finalDeltaAoverA}) indicates that the
ensemble average of the fractional change in area caused by
lensing by large-scale structure along the line of sight
is very small, being slightly less than a part-in-a-million effect.

\subsection{Fluctuations in the area}
\label{subsec:areafluctuations}

We have calculated the ensemble average of the area
of a surface of some redshift $z$, but it is also relevant to ask
if there could be large fluctuations around this figure.
Regarding the second order effects, we have already
shown that there is very little variation in the distance reached
for constant distance travelled.  As for the increase in area from the
wrinkling of the surface, this depends on the square of the angular tilt of
the surface. This will certainly vary
between different directions, but for the
scale of perturbations that are significant for the mean bias there
are a large number of coherence areas over the sky ($N \sim \lambda^2 / L^2$) so there should be small ${\cal O}(1 / \sqrt{N})$ fluctuations in the integral over the
sky.  This would suggest that it is very safe to assume that the change in 
total photosphere area shows negligible fluctuations between observers.

But there is the first order contribution to the fluctuation in the
area in (\ref{eq:dAprimedA0}): $\Delta A / A = 2 \Delta \lambda / \lambda$.
For any one ray, this will be of order $\sim \phi (L / \lambda)^{1/2}$,
and with $N \sim \lambda^2 / L^2$ independent regions on the sky we
would expect this effect to give rise to fluctuations in the net area
$\Delta A / A_0 \sim \phi (L / \lambda)^{3/2}$.  For sub-horizon
wavelength perturbations this is once again a tiny effect, but for horizon-scale
density perturbations ($L \sim \lambda$) this would be of order the RMS potential fluctuation
on those scales.  This would be similar in magnitude to the fluctuations in
temperature of the CMB on these scales, or $\sim 10^{-4.5}$.
In an ensemble average sense this vanishes as it is a first order effect,
but it does mean that, in all likelihood, the area of our photosphere 
or a surface of constant redshift differs
from the ensemble mean at this level, which is actually larger than the
ensemble mean perturbation itself.  But this is still a very small
effect and is, for all practical purposes, negligible.
\section{Summary and Discussion}
\label{sec:discussion}

\subsection{The area of the cosmic photosphere}

The main new result in this paper is to show that gravitational
lensing causes a non-vanishing
perturbation to the area of a surface of constant redshift
or of the CMB photosphere.
The result (\ref{eq:finalDeltaAoverA}) is valid at second order in
weak-field metric fluctuations, 
and was obtained under the assumption that the scale
of the perturbations that are responsible for the effect is much less than the path length
(i.e.\ that we are dealing with sub-horizon scale structures).

Under these assumptions, the problem is isomorphic to
optics in a refractive medium with random spatial variations of the
refractive index. 
The effects here are non-linear, but are not in any way associated with the
non-linearity of Einstein's equations.  The structures involved may have
$\delta \rho / \rho \gg1$, but the metric fluctuations are small.
We see no scope for additional intrinsically relativistic effects beyond the usual treatment of
light deflection in terms of the Newtonian potential
and the curvature of the spatial hypersurfaces.

Our result was derived in perturbation theory, formally assuming that
the image shear and magnification along all rays are small.
In this approximation the dominant contribution to $\langle \Delta A \rangle / A_0$
comes from structures on scales of tens of Mpc with height-to-wavelength
ratio -- and therefore surface tilt -- on the order of 
$\langle \Theta^2 \rangle^{1/2} \sim \phi \sqrt{\lambda / L} \simeq 10^{-3}$.
The mean of the change of area and hence the mean flux amplification is
much smaller, being on the order $\langle \Theta^2 \rangle$. This quantity
converges to a well-defined limit, with little contribution from the smaller-scale structures
responsible for strong lensing.

Ray tracing through the Millennium simulation (Hilbert et al.\ 2007) shows that the 
high-$z$ asymptotic optical depth for strong lensing is only $\tau \simeq 10^{-3}$, which is dominated
by clusters of mass $M \sim 10^{14} M_\odot$, with a similar optical depth 
probably arising from galaxy-scale
haloes if baryonic effects are taken into account (Hilbert et al.\ 2008).
It is possible that much smaller-scale structures 
cause most rays to high $z$ to be significantly sheared and amplified,
and the constant $z$ surface may be fractal on small scales, as argued by EBD98 and
discussed in \S\ref{sec:fluxconservation}.
But we believe that our result remains valid for the following reason.
A high degree of
small-scale folding of the surface might conceivably
result in a large decrease in the mean flux density, {\em but only if
the multiple images are resolved.\/}  For unresolved, or flux density aggregated,
sources any further change to the mean flux density is negligible compared
to the (already tiny) effect from large-scale structure, simply because the
bending angles associated with small-scale strong lenses is so small.

In particular, we may ask whether the neglect of small-scale strong lensing
could have a significant impact on the CMB.
Like Ellis, Bassett \& Dunsby (1998) we do not see how arcminute-scale 
strong lensing can affect the observed CMB sky at degree scales,
in contrast to CUMD14's claim of percent level effects
by including structures down to scales of order 10 kpc. 
This is because the area of the photosphere
mapped to by a disk of solid angle $\Delta \Omega$ is determined, at the linear level, only by the
mass density excess within the tube that the boundary of $\Delta \Omega$ traces out.
This is a consequence of the 2-dimensional version of Gauss's law.  Unlike paths
to sources, which tend to avoid over-densities
(see appendix \ref{sec:sourceaveragedkappa}), beams of randomly
chosen direction sample a density that is unbiased.  The increase in $\Delta A / \Delta \Omega$
for those paths that pass between clusters is compensated for by the decrease for
those beams that happen to encompass a cluster.

We noted the minor distinction between a surface of constant-$z$
and the photosphere.  These are not
precisely the same, as the Rees-Sciama and related effects cause slight
perturbation to the redshift of the photosphere.  
This changes the area perturbation but does not qualitatively 
change our essential conclusion.

\subsection{Lensing conservation theorems}

The fact that the area of constant-redshift surfaces is in practice invariant
justifies Weinberg's (1976) claim that the mean flux density, or equivalently the
mean inverse square distance, is unchanged by lensing when averaged over sources.
It also confirms the complementary result of Kibble \& Lieu (2005), that
the inverse amplification averaged over directions is also unperturbed.

Nevertheless, a major thrust of this paper has been to emphasise the importance of
statistical bias in any non-conserved quantities -- anything
that is a non-linear function of magnification (or its inverse if averaging
over directions).  This includes distances and distance moduli.
We have provided formulae (equations \ref{eq:deltadoverd0}, \ref{eq:deltad2overd02}, \ref{eq:deltadoverd0diravg})
for various examples of these biases in the perturbative regime and have shown
that these are on the order of $\langle \kappa^2 \rangle$.
Recent claims in the literature that find large 
results from non-linear relativistic perturbation theory 
for e.g.\ the perturbation to the area of the photosphere
seem to have resulted from a confusion of these effects and between source and direction averaging.

We have shown in appendix \ref{sec:opticalscalars} that these effects may
also be derived, although with considerable difficulty, from the focusing equation obtained from the optical scalar formalism.  
We have also described how the non-vanishing mean inverse magnification
of sources can be understood as arising because light paths to sources
tend to avoid over-dense regions and therefore sample paths that
have a convergence that is, on average, negative.  We find that
the fractional bias in column density is $ \langle\kappa\rangle_A=- 2 \langle \kappa^2 \rangle$,
but this was obtained in the perturbative regime and may only be a 
crude model for real absorption line studies.

The virtue of the optical scalar analysis is that it is more explicitly
relativistic in form. Agreement with our more simple-minded
discussion in terms of lengths of rays and wrinkling of surfaces
therefore provides some reassurance that this viewpoint is
not lacking some
subtle non-Newtonian relativistic effect.

This analysis also helps clarify the meaning of the focusing
equation (\ref{eq:SSEfocusingequation}).
We have emphasised that despite the RHS of this 
being on average greater than in a structure-free universe, it does not
indicate any tendency for structure to focus beams in the sense of changing their
average area.  The perturbation to the distance 
$\langle D \rangle / D_0 = 1 - \langle \kappa^2 \rangle / 2$
obtained from averaging the focusing equation in appendix \ref{sec:opticalscalars} just what is obtained
under the assumption the mean beam area is precisely unperturbed. 
Thus, despite its name, the focusing theorem 
does not reflect any particular tendency for
cosmic inhomogeneity to cause any systematic gravitational amplification of source flux densities.
There is a real systematic change to the square root of the
beam area, which is potentially large, being on the order of $\sim \langle \kappa^2 \rangle$
or $\sim \phi^2 \lambda^3 / L^3$, but once again this simply reflects the
statistical bias in $\sqrt{A}$ owing to $d A / d \Omega$ being a fluctuating
quantity.  The effect on the {\em area\/} in perturbation theory is 
suppressed relative to the mean distance perturbation by two powers of 
$L / \lambda$ to give the {\em averaged un-focusing theorem\/}
\begin{equation}
\langle \Delta A \rangle / A_0 = 0 + {\cal O}(\phi^2 \lambda / L)
\end{equation}
or, for all practical purposes, $\langle \Delta A \rangle / A_0 = 0$.
Contrary to what
the focusing theorem might na\"{\i}vely be taken to suggest, beams
of light tracked back in time from the observer
actually wind their way through an inhomogeneous universe
with barely any change to their average area.

\subsection{Possible statistical biases}
\label{subsec:CMBbias}

Evidently, for these $\sim \langle \kappa^2 \rangle$ effects,
the distinction between sky-plane and source-plane averaging is
important, as is the choice of variable used as the diagnostic.
Depending on the latter issue in particular, there may or may not
be a bias. We therefore need to look at how analyses are
actually performed in the critical cosmological cases of 
SN1a and CMB analyses.

For the case of SN1a, lensing is routinely included in modern analyses.
For example, Sullivan et al.\ (2011); Conley et al.\ (2011) account for a magnitude scatter of
$\sigma_m=0.055\, z$ in their fitting. Interestingly, however, the
magnitudes are implicitly taken to be unbiased in the regression
procedure, and it is not clear that this is correct. Denoting flux density
by $S$, this is affected by the magnification as $S\propto \mu$, so that
$\langle S/S_0 \rangle = 1$ (under area averaging, as is appropriate
in this case). But 
\begin{equation}
\eqalign{
\langle \ln (S/S_0) \rangle &= \langle \ln(1 + \Delta\mu) \rangle
\simeq -{1\over 2}\langle (\Delta\mu)^2\rangle \cr
&= -2\langle\kappa^2\rangle = -{1\over 2}\sigma_{\ln S}^2\, .
}
\end{equation}
This relation is not entirely straightforward, since we need to worry about what 
happens at $S=0$ in performing the averaging. It is safer to work in reverse
and ask if, for a Gaussian distribution of $\ln S$ centred at zero, the flux is
unbiased; it is not, by the same offset given above. Thus by fitting to
magnitudes, high-$z$ supernovae are in effect treated as being fainter than
they should be for their redshift. The effect is most marked at high $z$, where 
the dispersion is largest. For example, at $z=2$ the nominal $\sigma_m=0.11$
yields a 0.3\% increase in distance, which is equivalent to a shift of about
$\Delta w=0.01$ in the dark-energy equation of state. With the precision
of present data, this effect is therefore unimportant -- but a more careful
incorporation of the constraint of flux conservation may be necessary
in future generations of experiment, with a target of sub-percent precision in $w$.

In the case of CMB anisotropy measurements, it might seem that
the appropriate average is sky-plane.  The observers decide {\it a priori\/} 
where to look and measure some property such as the angular harmonic $\ell$
of the first peak of the angular power spectrum.  One could 
imagine averaging this quantity over different patches of the sky.  That measurement
{\em would\/} be
biased by lensing, since $\ell_{\rm peak}\propto \mu^{-1/2}$.  But if one were
to average $\ell_{\rm peak}^2$, which is proportional to 
the inverse magnification, this would not
be biased.  But one could also, in principle, detect peaks on the CMB sky
and use their curvature as a cosmological diagnostic, and then average over peaks (which are equivalent
to sources).  Clearly in a region with positive (negative) lensing amplification
both the number and curvature of the peaks will be biased low (high) so
the result would be biased.  But if one were to peak-average the {\em inverse\/}
curvature this is like $\langle \mu \rangle_A$ and the result would be unbiased.  

While it is therefore possible to analyse CMB data in a biased manner, 
the standard analysis method is not susceptible to such a bias.
What is done (Hu 2000; Challinor \& Lewis 2005) is to calculate the
angular power spectrum by modelling the observed sky as the primary
fluctuations on an unperturbed sphere being distorted by the transverse deflections
from foreground structures, and keeping terms up to
second order in the Newtonian potential (e.g.\ equation 15 of Challinor \& Lewis 2005).  
Cosmological parameters are obtained by computing the likelihood as the probability of
the actual measured spectrum given this lensed prediction (as a function of the
parameters of interest).
Thus there is no scope to obtain a bias in the inferred parameters
since the relevant quadratic effects arising from lensing are already
properly accounted for. Any apparent tensions between the CMB and astrophysical
estimates of parameters such as $H_0$ cannot be explained by lensing.

\section{Acknowledgements}

We thank Cifar and the members of the C\&G programme for stimulating
discussions over the years.  NK thanks Shaun Cole and Henk Hoekstra for
interesting discussions on the subject and is grateful for the hospitality
of the Higgs Centre for Theoretical Physics.

\appendix
\section{The Perturbation to the Source Surface Area}
\label{sec:areacalculation}

We now compute the second order correction to the area of the cosmic photosphere
or of a surface of constant redshift.
We first justify the use of weak-field metric perturbations and we
note the analogy between light propagation in
a weakly perturbed FRW cosmology and in a medium with a non-uniform
refractive index.  We next
discuss the boundary conditions for the end of the rays (which for the
photosphere corresponds to a surface of constant optical path in the
lumpy glass analogy).  We then perform the calculation; we do this
in two steps.  We first calculate the distance reached after
propagating a given path length $\lambda_0$ 
(i.e. a fixed path length in background coordinates) and the
mean area of the intersection of a narrow bundle of rays with given solid angle at the observer
with this surface.  We then compute the extra contribution to the
mean area that arises by propagating the extra (positive or negative)
distance to the surface of constant optical path, allowing for the
correlations between the the various first order effects 
(e.g.\ the expansion rate of the bundle and the extra displacement).

\subsection{Light deflection in weak field gravity}
\label{subsec:weakfield}

We are interested in very weak field perturbations to FRW cosmology -- metric fluctuations
of order $h_{\alpha\beta} \sim 10^{-4.5}$ or smaller -- associated with very nearly Newtonian perturbations of scale
(mostly) much less than the horizon size.  For simplicity we will consider a flat background, as
this seems to be a very good approximation to reality.  We first consider the very weak
field limit in which the metric has only one degree of freedom (in GR).  Then we 
generalise to a metric that includes the off-diagonal terms associated with
bulk motion of matter and show that this has an extremely small effect on
the deflection of light rays (much less than one might imagine from the
size of the metric perturbation).

\subsubsection{Light deflection by a static source}
\label{subsubsec:veryweakfields}

The weak field 
metric is usually taken to be
\begin{equation}
ds^2 = a^2(\eta) \left[-(1 + 2 \psi) d\eta^2 + (1 - 2 \phi) (dx^2 + dy^2 + dz^2)\right]
\label{eq:metric}
\end{equation}
where the potentials $\psi$ and $\phi$ are some functions of the coordinates.
In GR the two potentials are equal for nonrelativistic matter perturbations, 
$\psi = \phi$, and we assume that henceforth.
The linearised versions of Einstein's equations show that this is the metric
generated by non-relativistic matter with density fluctuations related
to $\phi$ by $\nabla^2 \phi = 4 \pi G \delta \rho / c^2$ where the Laplacian is in
proper coordinates and $\delta \rho$ is the matter density perturbation.  
As mentioned in the Introduction, we take the scale factor 
to be dimensionless: $a = 1 / (1 + z(\eta))$,
so our conformal coordinates have dimensions of length.

Writing this as
\begin{equation}
ds^2 = a^2(\eta) (1 + 2 \phi) \left[- d\eta^2 + n^2 (dx^2 + dy^2 + dz^2)\right]
\label{eq:metric2}
\end{equation}
with
\begin{equation}
n = \sqrt{\frac{(1 - 2 \phi)}{(1 + 2 \phi)}} \simeq 1 - 2 \phi
\label{eq:nfrommetric}
\end{equation}
we see that null rays have coordinate speed $d |\br | / d \eta = 1 / n$
and extremise the coordinate time:
\begin{equation}
\delta \int d |\br| \ispace n(\br) = \delta \int d \lambda \ispace | {\dot \br} | n(\br) = 0,
\end{equation}
where ${\dot \br} = d \br / d\lambda$ and where the parameterisation of the path $\br(\lambda)$ is arbitrary. 
The Euler-Lagrange equation from this is
\begin{equation}
n {\ddot \br} + \brdot (\brdot \cdot \bnabla n) - \frac{1}{2} \overset{.}{|\brdot|} n = |\brdot|^2 \bnabla n 
\end{equation}
which is not particularly useful, but if we fix
$\lambda$ to be equal to the coordinate distance along the path (so $|\brdot| = 1$), the geodesic equation is
\begin{equation}
{\ddot \br} = \bnablaperp \tilden
\label{eq:geodesicequation}
\end{equation}
where $\tilden \equiv \ln n$ and $\bnablaperp \equiv \bnabla - \brdot (\brdot \cdot \bnabla)$
is the derivative in the direction perpendicular to $\brdot$.
This is exactly the same as Snell's law for rays propagating in a
refractive medium with refractive index $n$ (Born \& Wolf 1965).  Optics in an expanding universe
with metric (\ref{eq:metric}) is the same as in lumpy glass
with conformal coordinates playing the same role as ordinary physical spatial
coordinates in a refractive medium.  

Additionally, if the potential is only a function of conformal position,
and not changing with time, then the lenses induce no change in the redshift
of the photons.  But if the potential is changing -- either because one is dealing
with perturbations in a non Einstein--de Sitter background or because the perturbations
are non-linear or moving or evolving internally -- then there will be
redshift perturbations (Sachs \& Wolfe 1967; Rees \& Sciama 1968; Birkinshaw \& Gull 1983).
These again are also the same as would apply in a medium with a time varying
refractive index.  If the optical path length -- i.e.\ the number of waves along the path -- 
is changing with time then $\nu_{\rm rec}$, the number of waves per unit time at the receiver, will be
the number of waves per unit time at the emitter $\nu_{\rm em}$ minus the rate of change
of the optical path, so 
$\nu_{\rm rec} = \nu_{\rm em} (1 + \int d\lambda \ispace \partial \tilden / \partial \eta)$.
This provides a novel way of thinking about the Integrated Sachs--Wolfe effect, and it
is a phenomenon that is routinely used to tune frequencies in optoelectronics.

Expressed in terms of the potential $\phi$, the geodesic equation is,
from (\ref{eq:nfrommetric}),
\begin{equation}
\ddot \br = \frac{- 2 \bnablaperp \phi}{1 - 4 \phi^2} .
\label{eq:geodesicfromphi}
\end{equation}
This seems to suggest that the linear formula $\ddot \br = - 2 \bnablaperp \phi$
would be accurate up to ${\cal O}(\phi^2 \bnablaperp \phi)$.
In fact, as we now show, if we allow for the non-relativistic matter motion
we find additional terms in the geodesic equation, but these
are smaller than the linear term by a factor $\sim \phi$.

\subsubsection{Weak fields sourced by moving matter}
\label{subsubsec:weakfields+framedragging}

The metric (\ref{eq:metric2}) is obtained using linearised gravity (i.e.\ working
only to first order in the metric perturbations 
$h_{\alpha\beta} = g_{\alpha\beta} - \eta_{\alpha\beta}$, with $\eta_{\alpha\beta}$ the
Minkowski metric) and incurring errors on the order $|h|^2$ from e.g.\ using
$\eta_{\alpha\beta}$ to raise and lower indices.  It also assumes that the
source of gravity is $T^{\alpha\beta} = {\rm Diag}\{\rho, 0, 0, 0\}$ with $\rho$ the
mass density.  A more accurate model for the metric, still within the
context of a linearised relation between the metric and the Einstein tensor, 
comes from including the momentum
density source.  This is $ds^2 = g_{\alpha \beta}dx^\alpha dx^\beta$ with 
\begin{equation}
g_{\alpha\beta} = a^2(\eta)
\begin{bmatrix}
-(1 + 2 \phi)  & \bV  \\
\bV & (1-2\phi) \bI 
\end{bmatrix}
\label{eq:4dofmetric}
\end{equation}
where $\bI$ is the 3D identity matrix and $\bV$
is on the order of the potential $\phi$ times the peculiar velocity of the matter $\bv$
(we note that in general non-vanishing 3-stress $T^{ij}$ also introduces
differences between the diagonal terms but these are of order $\phi v^2$, and 
therefore much smaller).

We now show how these extra `frame-dragging' terms affect the 
geodesic equation.  The result is very small; the corrections to the
linear term in (\ref{eq:geodesicfromphi}) being smaller by a factor $\phi$.

To calculate deflection of light in the space-time (\ref{eq:4dofmetric})
we can use the geodesic equation
\begin{equation}
\frac{d^2 x^ \alpha}{d s^2} = - \Gamma ^\alpha_{\beta\gamma} 
\frac{dx^\beta}{ds} \frac{dx^\gamma}{ds}
\end{equation} 
where $s$ is the affine parameter (unique up to a constant and a scale factor) and 
where 
\begin{equation}
\Gamma ^\alpha_{\beta\gamma} = 
\frac{1}{2} g^{\alpha\nu} (g_{\nu \beta, \gamma} + g_{\nu \gamma, \beta} - g_{\beta \gamma, \nu})
\end{equation}
is the Christoffel symbol (e.g.\ Weinberg 1972).  At zeroth order in the matter
fluctuations $g_{\alpha \beta} = a^2(\eta) \eta_{\alpha \beta}$ and we have $d^2 x^i / d s^2 = 0$ and
$d^2 \eta / d s^2 = - 2 (a' / a) (d \eta / ds)^2$, where $a' \equiv da/d\eta$, 
with solution $d \eta / ds = a^{-2}(\eta)$.
Since $d\eta = dt / a$ this means that the energy $p^0 \propto dt/ds = a d \eta / ds \propto 1 / a$
as usual.

Here we wish to compute properties of rays and wavefronts given some statistical
prescription for the metric fluctuations as a function of background
coordinates.
It is therefore more useful to let the independent variable
in the geodesic equation be the path distance in background spatial coordinates.
Applying the chain rule, the geodesic equation with path variable
being the $z = x^3$ coordinate, for instance, is
\begin{equation}
\frac{d^2 x^ \alpha}{d z^2} = 
- \Gamma ^\alpha_{\beta\gamma} \frac{dx^\beta}{dz} \frac{dx^\gamma}{dz} 
+ \Gamma ^z_{\beta\gamma} \frac{dx^\beta}{dz} \frac{dx^\gamma}{dz} \frac{dx^\alpha}{dz} \, .
\end{equation} 

If we consider a ray that happens to be travelling along the $z$ direction
(i.e.\ with $dx/dz = dy/dz = 0$), or rotate our spatial coordinate
system to align the $z$-axis with the instantaneous ray vector,
then the curvature of the path in the $x-z$ plane is
\begin{equation}
\frac{d^2 x}{d z^2} = 
- \Gamma ^x_{\beta\gamma} \frac{dx^\beta}{dz} \frac{dx^\gamma}{dz} = 
- \Gamma ^x_{zz} - n^2 \Gamma ^x_{\eta\eta} 
\label{eq:ddotxfromGammas}
\end{equation} 
and similarly for $d^2 y / dz^2$ and where
now $n^2 = (d\eta/dz)^2$.  From $g_{\alpha\beta}dx^\alpha dx^\alpha = 0$
we find, working to second order, 
\begin{equation}
n^2 = (d\eta/dz)^2 = (g_{zz} / - g_{\eta\eta})(1 - g_{z\eta})
\end{equation}
but since the Christoffel symbols are of first order in the potential $\phi$
while $g_{z\eta}$ is smaller by a factor $\bv$ then working to 
second order in $\phi$ we can ignore the last factor and take $n^2 = - g_{zz} / g_{\eta\eta}$
in (\ref{eq:ddotxfromGammas}).  This yields
\begin{equation}
\begin{split}
\frac{d^2 x}{d z^2} & = \frac{1}{2}
[g^{xx}g_{zz,x} - g^{x\eta} ( 2 g_{z\eta,z} - g_{zz,\eta})  \\
 & - n^2 (g^{xx}(2 g_{x\eta,\eta} - g_{\eta\eta,x}) + g^{x\eta} g_{\eta\eta,\eta})] \, .
\end{split}
\end{equation}   
Clearly we can drop the term involving two off-diagonal elements
as this is smaller than second order in $\phi$, as are the terms involving $g^{x\eta} \simeq - g_{x\eta}$ 
and a time derivative of one of the diagonal elements since these are
both $\sim (\phi v) \times \phi / \lambda$, with $\lambda$ being overall path length.  
Dropping these, and using the first order approximation for the inverse metric element $g^{xx} = g_{xx}^{-1}$ 
as this multiplies a first order quantity, gives
\begin{equation}
\frac{d^2 x}{d z^2} = \frac{1}{2} 
\left(\frac{g_{zz,x}}{g_{zz}} - \frac{g_{\eta\eta,x}}{g_{\eta\eta}} - \frac{g_{x\eta,\eta}}{g_{\eta\eta}}\right)
= \frac{\partial \tilden}{\partial x} - \frac{g_{x\eta,\eta}}{2g_{\eta\eta}} \, .
\end{equation}

The first term on the RHS is what we obtained in the previous section, and the second arises
from matter motion.  But this new term has no spatial derivative, so ends up producing very little effect.

For linear perturbations, if we integrate the last expression through a single structure
of size $L$ and potential fluctuation $\phi$, we get the $x$-component of the deflection angle $dx/dz$.
The first term yields $dx/dz \sim \phi$ while the velocity $\bv$ is on the
order of $t_{\rm U} \bnabla \phi \sim \lambda \phi / L$, since the age of the universe $t_{\rm U}$ is
on the order of the path length $\lambda$, so $\int dz \ispace g_{\eta x,\eta} \sim L g_{\eta x} / \lambda \sim \phi^2$.
The effect of the off-diagonal 
terms is thus smaller by a factor $L/\lambda$ than the naive expectation from the
fact that
$h_{x\eta}$ is smaller than $h_{xx}$ by a factor $\sim v$.

For a non-linear structure such as a cluster, group or galaxy that
is stable but has some bulk peculiar velocity $\bv$ the 
partial derivative with respect to time will be, to order of magnitude,
$g_{x\eta, \eta} \sim \bv \cdot \bnabla g_{x\eta} \sim v^2 \phi / L$ which, when integrated
through the object, gives a contribution to $dx/dz$ that is again on the
order of $\phi^2$ since, for virialised systems, $v^2 \sim \phi$.

In both linear and non-linear regimes the light deflection is smaller
than the (twice) Newtonian value for a test particle moving with $v = c$
by a factor $\sim \phi$ (i.e.\ considerably smaller than one might
perhaps have guessed from the relative size of the diagonal and off-diagonal
terms in (\ref{eq:4dofmetric})).

The metric (\ref{eq:4dofmetric}) is not the most general metric as it only has 
four spatial degrees of freedom.  The missing
ingredient is the two degrees of freedom in the gravitational
waves, but  
these are not effective for lensing
(Kaiser \& Jaffe 1997).  This is easily understood
in the Fourier space version of Limber's equation (Kaiser 1998)
where, for lensing by scalar perturbations, the modes that
are effective have wave-vector perpendicular to
the line of sight so that the light ray stays in
phase with the wave -- much like a rapidly moving
surfer surfing a slowly moving wave -- so the deflection
builds up systematically.  For gravitational waves,
which propagate with $|\bv| = c$, this cannot happen.
We conclude from this that to an extremely good approximation we can ignore the
additional effect on light deflection from the non-relativistic motions
associated with structure (as well as gravitational radiation)
and use the metric (\ref{eq:metric2}), with only scalar Newtonian fluctuations.

We note that owing to the non-linearity of Einstein's
equations the mean local curvature and stress-energy tensor
implied by this fluctuating metric
will not be the same as for an unperturbed cosmology with
the same expansion rate etc.  The Riemann curvature, for instance,
contains a term that is quadratic in the connection and the latter contains derivatives
of the metric so one would expect there to be a non-zero mean curvature
involving e.g.\ the products of derivatives of $\phi$ and this
carries over into the Einstein tensor and hence the stress-energy tensor also.
An alternative would be to adopt a model in which the stress-energy tensor
is unperturbed in the mean.  This simply requires adding an appropriate constant
Laplacian to the metric perturbations (i.e.\ making the spatial sections 
globally curved).
This, however, would not 
be appropriate in the context of inflationary fluctuogenesis where the large-scale
spatial flatness is a consequence of the assumed large initial value
for the inflaton field and the slowness of its roll down the assumed
potential, while the smaller scale fluctuations -- that give rise to the structures
we can actually observe -- transition from Planck to horizon scale
later and develop metric fluctuations that must be
accommodated within a globally spatially flat background.

\subsection{Boundary conditions at the end of the ray}
\label{subsec:rayBCs}

We are interested in the integrated effect of lensing by structures along
the line of sight.  So we can take the density perturbation on the actual
photosphere to vanish and consider the observed temperature fluctuations generated
by the combination of spatial variation of temperature, Doppler shift and
gravitational redshifts to be a pattern that is `painted on'.

\begin{figure}
\setlength{\unitlength}{2.25cm}
\begin{center}
\begin{picture}(3.57,5.2)(-0.27,-0.3)
\put(0.0,0.0){\vector(1,0){3.300000}}
\put(3.150000,-0.150000){\makebox(0,0){$\lambda$}}
\put(0.0,0.0){\vector(0,1){4.900000}}
\put(-0.150000,4.750000){\makebox(0,0){$\eta$}}
\put(-0.150000,0.500000){\makebox(0,0){$\eta_{\rm rec}$}}
\put(0.000000,0.500000){\line(1,0){3.300000}}
\thinlines
\qbezier(1.926250,0.000000)(1.963125,0.036875)(2.000000,0.073750)
\qbezier(1.778750,0.000000)(1.889375,0.110625)(2.000000,0.221250)
\qbezier(1.631250,0.000000)(1.815625,0.184375)(2.000000,0.368750)
\qbezier(1.483750,0.000000)(1.741875,0.258125)(2.000000,0.516250)
\qbezier(1.336250,0.000000)(1.668125,0.331875)(2.000000,0.663750)
\qbezier(1.188750,0.000000)(1.594375,0.405625)(2.000000,0.811250)
\qbezier(1.041250,0.000000)(1.520625,0.479375)(2.000000,0.958750)
\qbezier(1.000000,0.106250)(1.500000,0.606250)(2.000000,1.106250)
\qbezier(1.000000,0.253750)(1.500000,0.753750)(2.000000,1.253750)
\qbezier(1.000000,0.401250)(1.500000,0.901250)(2.000000,1.401250)
\qbezier(1.000000,0.548750)(1.500000,1.048750)(2.000000,1.548750)
\qbezier(1.000000,0.696250)(1.500000,1.196250)(2.000000,1.696250)
\qbezier(1.000000,0.843750)(1.500000,1.343750)(2.000000,1.843750)
\qbezier(1.000000,0.991250)(1.500000,1.491250)(2.000000,1.991250)
\qbezier(1.000000,1.138750)(1.500000,1.638750)(2.000000,2.138750)
\qbezier(1.000000,1.286250)(1.500000,1.786250)(2.000000,2.286250)
\qbezier(1.000000,1.433750)(1.500000,1.933750)(2.000000,2.433750)
\qbezier(1.000000,1.581250)(1.500000,2.081250)(2.000000,2.581250)
\qbezier(1.000000,1.728750)(1.500000,2.228750)(2.000000,2.728750)
\qbezier(1.000000,1.876250)(1.500000,2.376250)(2.000000,2.876250)
\qbezier(1.000000,2.023750)(1.500000,2.523750)(2.000000,3.023750)
\qbezier(1.000000,2.171250)(1.500000,2.671250)(2.000000,3.171250)
\qbezier(1.000000,2.318750)(1.500000,2.818750)(2.000000,3.318750)
\qbezier(1.000000,2.466250)(1.500000,2.966250)(2.000000,3.466250)
\qbezier(1.000000,2.613750)(1.500000,3.113750)(2.000000,3.613750)
\qbezier(1.000000,2.761250)(1.500000,3.261250)(2.000000,3.761250)
\qbezier(1.000000,2.908750)(1.500000,3.408750)(2.000000,3.908750)
\qbezier(1.000000,3.056250)(1.500000,3.556250)(2.000000,4.056250)
\qbezier(1.000000,3.203750)(1.500000,3.703750)(2.000000,4.203750)
\qbezier(1.000000,3.351250)(1.500000,3.851250)(2.000000,4.351250)
\qbezier(1.000000,3.498750)(1.500000,3.998750)(2.000000,4.498750)
\qbezier(1.000000,3.646250)(1.500000,4.146250)(2.000000,4.646250)
\qbezier(1.000000,3.793750)(1.500000,4.293750)(2.000000,4.793750)
\qbezier(1.000000,3.941250)(1.479375,4.420625)(1.958750,4.900000)
\qbezier(1.000000,4.088750)(1.405625,4.494375)(1.811250,4.900000)
\qbezier(1.000000,4.236250)(1.331875,4.568125)(1.663750,4.900000)
\qbezier(1.000000,4.383750)(1.258125,4.641875)(1.516250,4.900000)
\qbezier(1.000000,4.531250)(1.184375,4.715625)(1.368750,4.900000)
\qbezier(1.000000,4.678750)(1.110625,4.789375)(1.221250,4.900000)
\qbezier(1.000000,4.826250)(1.036875,4.863125)(1.073750,4.900000)
\thicklines
\put(0.000000,3.500000){\line(1,-1){1.000000}}
\qbezier(1.000000,2.500000)(1.500000,1.900000)(2.000000,1.300000)
\put(2.000000,1.300000){\line(1,-1){0.800000}}
\thinlines
\thicklines
\put(0.000000,4.500000){\line(1,-1){1.000000}}
\qbezier(1.000000,3.500000)(1.500000,2.950000)(2.000000,2.400000)
\put(2.000000,2.400000){\line(1,-1){0.800000}}
\thinlines
\put(2.800000,0.900000){\vector(0,-1){0.400000}}
\put(2.800000,1.200000){\vector(0,1){0.400000}}
\put(2.800000,1.050000){\makebox(0,0){$\Delta \eta_i$}}
\put(0.200000,3.650000){\vector(0,-1){0.350000}}
\put(0.200000,3.950000){\vector(0,1){0.350000}}
\put(0.200000,3.800000){\makebox(0,0){$\Delta \eta_f$}}
\multiput(3.000000,0.500000)(-0.333333,0.333333){6}{\line(-1,1){0.200000}}
\put(3.000000,0.500000){\line(0,-1){0.250000}}
\multiput(2.900000,1.600000)(-0.333333,0.333333){6}{\line(-1,1){0.200000}}
\put(2.800000,0.500000){\line(0,-1){0.250000}}
\put(3.200000,0.300000){\vector(-1,0){0.200000}}
\put(2.600000,0.300000){\vector(1,0){0.200000}}
\put(2.900000,0.200000){\makebox(0,0){$\Delta \lambda$}}
\end{picture}
\end{center}
\caption{Illustration of the first order change in the the conformal 
spatial path length $\Delta \lambda$ to a surface of constant cosmic
time (like the CMB photosphere).
Coordinates are conformal background position and time ($\lambda$, $\eta$).
The hatched lines show an over-density 
where the coordinate velocity $d \lambda / d \eta$ is changed by a factor
$1 + 2 \phi$ at linear order, so
$\Delta \lambda = 2 \int d \lambda \ispace \phi$ (which is negative for
an over-dense path).  But the surface of constant cosmic time $\eta=$ constant
is not a surface of constant redshift as it will be affected by
the ISW effect caused by the change of the potential with time, which
occurs at low redshift $z \lesssim 1$.  A decaying over-density causes
a negative perturbation to the redshift (i.e.\ a temperature enhancement
$\Delta T / T = \Delta \eta_i / \Delta \eta_f$
for the CMB). For a ray to reach the surface of constant redshift requires,
for an over-dense path, an extra path to annul the ISW effect so the
net path length perturbation is reduced. The difference for high-redshift sources is 
on the order of a few percent even for low-redshift lenses. 
For lenses {\em and\/} sources at $z \lesssim 1$, the $\Delta \lambda$ for constant
redshift is reduced, as compared to the fictitious case where the
potential does not decay, by about 50\%.  This approximately nulls the
perturbation to the area at very low $z$, as described in the text. 
}
\label{fig:Dlambda}
\end{figure}
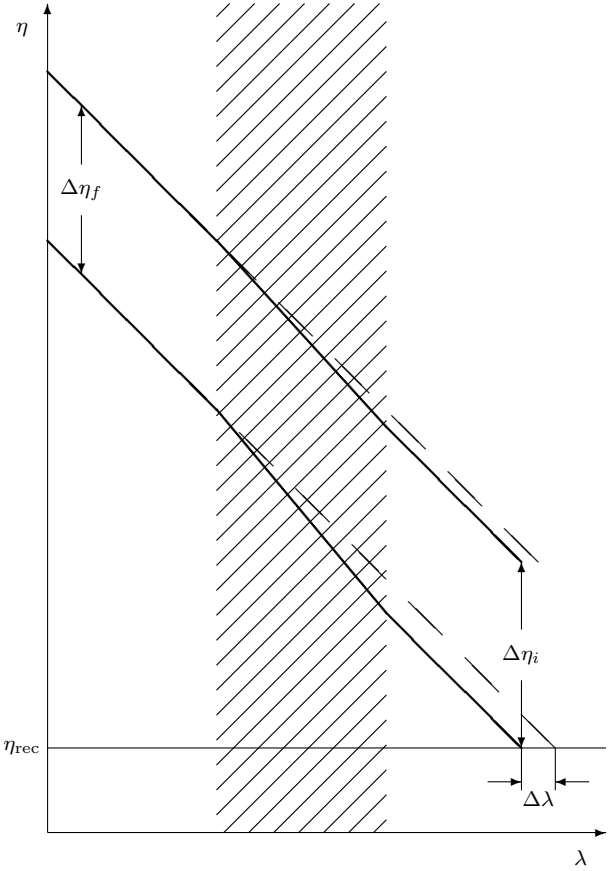

This point of view is valid even though
the photosphere -- or surface of last scattering -- is not a real surface.
It is defined as the part of our 3D past light cone where the 
cosmic time is that of recombination $t_{\rm rec}$. This decoupling
time is set by atomic and cosmological parameters, which also set the
acoustic scale of the structures that we can subsequently view in the CMB.
If recombination occurs at a fixed temperature, it may be wondered why there
are any fluctuations in the CMB at all. One answer is that the effect of fluctuations
is that the recombination temperature is reached at different times and hence
different redshifts -- which modifies the observed temperature. Thus the surface
of last scattering is in reality a surface of constant temperature but varying redshift.
Nevertheless, we can ask what temperature fluctuations would be observed if we
were able to see a surface of constant redshift, and the answer is that the observed
CMB would be the same. 

Ignoring the fluctuations at the end of the rays, the cosmic photosphere is perpendicular
to the direction of the light rays.  In the lumpy glass analogy, this is a surface
of constant optical path length, $\int n\; d \lambda$, which differs
from the physical path length at first order because of time delays (which 
in the cosmological context can be positive
or negative). 
We now ask how the area of the photosphere differs
for rays propagating backward from some observer at some time, compared
to the case of a universe -- or a line of sight -- that has no metric perturbations.
This involves computing the area of the surface with 1st order
path length perturbation
\begin{equation}
\Delta \lambda = 2 \int d\lambda \ispace \phi .
\label{eq:photosphereDeltalambda}
\end{equation}
This first order time delay gives rise, as we shall see, to a 2nd order
increase in the area of the photosphere through the `wrinkling' effect.

A surface of constant source redshift is not exactly the same as a surface of
constant cosmic time because intervening
perturbations, particularly those at low redshift, can cause perturbation to the observed
CMB temperature $T_{\rm obs}$ via the integrated Sachs \& Wolfe (1967) (ISW) or Rees \& Sciama (1968) effects, 
but do not affect $T_{\rm em}$ so $1 + z = T_{\rm em} / T_{\rm obs} \ne$ constant.
Similar effects come from moving or
dynamic lenses (Birkinshaw \& Gull 1983).  
In the perturbative regime the ISW effect produces a temperature
perturbation for the surface of constant cosmic time 
$\Delta T / T = \Delta \eta_i / \Delta \eta_f$ (see Figure \ref{fig:Dlambda}) or
$\Delta T / T  = 2 \int d\lambda \ispace \phi' $
where $\phi' \equiv \partial \phi / \partial \eta$
(which becomes non-zero when the onset of dark energy domination damps
the initial metric fluctuations).
In the background, the temperature is
decreasing as $T \propto 1/a$, so to reach the surface of constant observed
temperature (or redshift) requires an additional path length $\Delta \lambda = \Delta \eta$ where
$\Delta a / a = a' \Delta \eta / a = - \Delta T / T$ or, equivalently, $\Delta \lambda = -2  (a'/a)^{-1}_{\lambda_0} \int d\lambda \; \phi'$.
Consequently the net 1st order perturbation to the path length to constant redshift is
\begin{equation}
\Delta \lambda = 2 \int d\lambda \ispace \phi \times (1 + (\phi' / \phi)_\lambda / (a'/a)_{\lambda_0}) .
\label{eq:constantzDeltalambda}
\end{equation}
This has a very small effect for high redshift sources since for these
$(a'/a)_{\lambda_0} \gg (\phi' / \phi)_\lambda$ regardless of $\lambda$
and, as we shall see, does not qualitatively change the outcome for any source redshift.

\subsection{Distance reached vs.\ distance travelled}

Consider a ray that arrives at the observer by moving along the $-z$ axis.  
For rays propagating close to and nearly parallel to the
the $z$-axis we can set up 2D perpendicular comoving coordinates $\bx$
and take $\bnablaperp$ to also be the 2D derivative with respect to $\bx$
at linear order.
Again, we use $\lambda$ for distance along the ray.
Assuming small
displacements, the transverse velocity ${\dot \bx} = d \bx / d \lambda$ 
(equal to the deflection angle) of this ray is, at first order in $\phi$,
the  integral of the geodesic equation:
\begin{equation}
{\dot \bx}(\lambda) = -2 \int\limits_0^\lambda d\lambda' \; \bnablaperp \phi(\lambda')
\label{eq:transversexdot}
\end{equation}
where the transverse displacement is
\begin{equation}
\bx(\lambda) = - 2 \int\limits_0^\lambda d\lambda' \; (\lambda - \lambda') \bnablaperp \phi(\lambda')
\label{eq:transversex}
\end{equation}
and we have integrated by parts.

We now use this to calculate the mean distance from the observer of the end of
a ray of physical path length $\lambda_0$.  After propagating a partial path length $\lambda < \lambda_0$
the end of the ray will lie at a direction from the observer that, to first
order, is $\bn = {\hat \bz} + \bx / \lambda$ .  The amount by which
the instantaneous ray vector ${\hat \bz} + {\dot \bx}$ differs from this direction is 
\begin{equation}
\Delta \dot {\bx}(\lambda) \equiv {\dot \bx}(\lambda) - \bx / \lambda = - \frac{2}{\lambda}
\int\limits_0^\lambda d\lambda' \; \lambda' \bnablaperp \phi.
\end{equation}

In propagating a further path length $\delta \lambda$ the end of the ray will
advance a distance measured from the observer 
$\delta r = \delta \lambda \times (1 - |\Delta {\dot \bx}|^2 / 2) + \ldots$.
The distance reached after propagating a path length $\lambda_0$ is
therefore $r(\lambda_0) = \lambda_0 + \Delta r$ where
\begin{equation}
\begin{split}
\Delta r & = - \frac{1}{2} \int\limits_0^{\lambda_0} d \lambda \; |\Delta {\dot \bx}(\lambda)|^2 \\
& =  - 2 \int\limits_0^{\lambda_0} \frac{d \lambda}{\lambda^2} 
\int\limits_0^\lambda d \lambda' \; \lambda' \bnablaperp \phi' \cdot 
\int\limits_0^\lambda d \lambda'' \; \lambda'' \bnablaperp \phi'' \\
&= - \frac{4}{\lambda_0} \int\limits_0^{\lambda_0} d \lambda \; (\lambda_0 - \lambda) \bnablaperp \phi \cdot
\int\limits_0^\lambda d \lambda' \; \lambda' \bnablaperp \phi' \\
\end{split}
\label{eq:Deltar}
\end{equation}
where $\phi' \equiv \phi(\lambda')$ etc.\ and where,
in passing to the final line, we have integrated by parts and used
\begin{equation}
\int\limits_0^{\lambda_0} d\lambda' \int\limits_0^{\lambda_0} d\lambda'' \; \ldots
= 2  \int\limits_0^{\lambda_0} d\lambda' \int\limits_0^{\lambda'} d\lambda'' \; \ldots
\end{equation}
which is valid if the integrand $[\ldots]$ is a symmetric function of its arguments.
Regarding notation, here and in what follows, all gradient operators
act only on the function of position that follows them and $\phi'$ is shorthand for $\phi(\lambda')$ etc..  

Note that we only required the first order deflection here.  Higher order corrections to
(\ref{eq:transversexdot}) \& (\ref{eq:transversex}) are irrelevant.

Evidently the perturbation to the distance reached in propagating a
fixed physical path length $\lambda_0$ is a quantity that is of second
order in the potential or refractive index fluctuations.  It follows from this that the
surface of constant physical path length from the observer  has
no 1st order tilt; its outward normal is, up to first order, parallel to the local
direction from the observer. 
As discussed in \S\ref{sec:areabias}, the ray-to-ray variations in 
distance reached are expected to be small compared to the systematic offset.

\subsection{Mean distance for constant physical path}
\label{subsec:meandistancereached}

We now express the ensemble mean distance for a constant physical path length in terms of
the auto-correlation function of the potential $\xi_\phi$.
The model we shall adopt is that, at least locally, the potential is a statistically
homogeneous and isotropic random field.  The quantities that we will calculate
are of second order in the potential and so may be obtained in terms of $\xi_\phi$
without any further assumptions about higher order statistics (i.e.\ we do not
need to invoke Gaussianity, so the results are applicable for non-linear
density fluctuations).  By `local' above, we are allowing for the possibility
that the potential fluctuations may be statistically homogeneous at any
instant of cosmic time but may depend on look-back time.  If, as is the case
in conventional models, the effects of interest here are dominated by fluctuations
which are much smaller than the Hubble scale it should be a good approximation
to calculate effects by summing the effect from different
shells within which strict homogeneity is assumed to obtain. 

Writing $\phi$ as a Fourier synthesis
\begin{equation}
\phi(\br) = \int \frac{d^3k}{(2\pi)3} \; {\tilde \phi}(\bk) e^{i \bk \cdot \br} 
= \int \frac{d^3k}{(2\pi)3} \; {\tilde \phi}^*(\bk) e^{-i \bk \cdot \br}
\end{equation}
and invoking local statistical homogeneity, 
$\langle {\tilde \phi}(\bk) {\tilde \phi}^*(\bk') \rangle = (2 \pi)^3 \delta(\bk - \bk') P_\phi(|\bk|)$,
the required ensemble average of the transverse gradients at two points is
\begin{equation}
\begin{split}
\langle \bnablaperp \phi(\br) & \cdot \bnablaperp \phi(\br + \br') \rangle \\
& = \int \frac{d^3k}{(2 \pi)^3} \bk_\perp \cdot \bk_\perp P_\phi(|\bk|) e^{i \bk \cdot \br'} \\
& = - \nablaperp^2 \xi_\phi(\br') \\
\end{split} 
\end{equation}
where the auto-correlation function of the potential is
\begin{equation}
\xi_\phi(\br') \equiv \langle \phi(\br) \phi(\br + \br') \rangle
= \int \frac{d^3k}{(2 \pi)^3} P_\phi(|\bk|) e^{i \bk \cdot \br'}.
\end{equation}

The quantity that appears in the expression (\ref{eq:Deltar}) for $\Delta r$ above,
when averaged, is the
two-point function of the transverse gradients of the potential at
two points with separation parallel to the path.  Taking the potential
auto-correlation function to be locally isotropic, $\xi_\phi(\br) = \xi_\phi(r)$,
and using the standard expression for an isotropic function,
$\nabla^2 \xi = \xi'' + 2 \xi'/r$, we have
\begin{equation}
\langle  \bnablaperp \phi(\lambda) \cdot \bnablaperp\phi(\lambda') \rangle
= - \nablaperp^2 \xi_\phi(\br = {\hat \bz} y) = - 2 \xi_\phi'(y) / y
\label{eq:avgphiphi}
\end{equation} 
where $y \equiv \lambda' - \lambda$ and $\xi_\phi'(y) \equiv d \xi_\phi(y) / dy$.

It follows that the perturbation to the mean distance reached after propagating a fixed physical
path length $\lambda_0$ is
\begin{equation}
\begin{split}
\langle \Delta r \rangle &= - \frac{4}{\lambda_0} \int\limits_0^{\lambda_0} d \lambda \; (\lambda_0 - \lambda) 
\int\limits_0^\lambda d \lambda' \; \lambda' \langle \bnablaperp \phi \cdot \bnablaperp \phi' \rangle \\
&= \frac{8}{\lambda_0} \int\limits_0^{\lambda_0} d \lambda \; (\lambda_0 - \lambda)
\left[
\int\limits_{-\lambda}^0 dy\; \xi_\phi' + \lambda \int\limits_{-\lambda}^0 dy\; \xi_\phi' / y
\right] \\
& \simeq \frac{8}{\lambda_0} \int\limits_0^{\lambda_0} d \lambda \; (\lambda_0 - \lambda) 
\left[
\xi_\phi(0) + \lambda \int\limits_{-\infty}^0 dy\; \xi_\phi' / y
\right]  \; .\\
\end{split}
\label{eq:Deltaravg}
\end{equation} 
In the last step, we are invoking the idea that the 
range of correlations of the potential is limited, so for any $\lambda$ substantially greater than the correlation
length, the integrals will have converged so we can take the lower limit in the integration
over separation $y$ to be minus infinity.

Finally, in the same spirit, the second term in parentheses $[ \ldots ]$ will be much greater than
the first.  For example, if one were to consider a simple model of `blobs' of some characteristic size $L$
and randomly chosen potential with root mean squared value $\phi$ one will have
$\xi_\phi(0) \sim \phi^2$ and $\lambda \int dy\; \xi_\phi' / y \sim (\lambda / L) \phi^2 \gg \phi^2$.
Dropping the smaller term then gives
\begin{equation}
\langle r \rangle =  \lambda - \frac{1}{\lambda_0} 
\int\limits_0^{\lambda_0} d \lambda \; \lambda (\lambda_0 - \lambda) J(\lambda) \; ,
\label{eq:avgr}
\end{equation}
where we have defined
\begin{equation}
J(\lambda) \equiv - 8 \int\limits_{-\infty}^0 dy \; \xi_\phi'(y; \lambda) / y \; .
\label{eq:Jdefinition}
\end{equation}
The minus sign makes $J(\lambda)$ a positive quantity, and 
the notation $\xi_\phi(y; \lambda)$ is meant to indicate
that the two-point function has a strong dependence on separation $y$ but may
also have a weaker secular trend with conformal look-back time $\lambda$. 

Equation \ref{eq:avgr} gives the ensemble mean of the distance reached at
second order in the metric perturbation $\phi$ and is also obtained
assuming a coherence length $L \ll \lambda$, so higher order corrections to
this are smaller by at least one power of $L / \lambda$.

\subsubsection{Rate of increase of deflection variance}
\label{subsubsec:deflectionvariance}

The potential here is dimensionless, so $J$ has units of
inverse length. In the `random blobs' model $J \sim \phi^2 / L$ while the
deflection angle for a ray passing through a single blob is $\Delta \Theta_1 \sim \phi$.
For random blobs the cumulative deflection performs a random walk, and $J$
is the rate at which the cumulative deflection squared grows with path
length.  This can be made more precise: From the definition (\ref{eq:transversexdot})
of the deflection angle $\dot \bx$ it follows that
\begin{equation}
\frac{d|{\dot \bx}|^2}{d\lambda} = 8 \bnablaperp \phi(\lambda) \cdot \int\limits_0^\lambda d\lambda'\; \bnablaperp \phi(\lambda') .
\label{eq:dotThetaTheta}
\end{equation}
Taking the ensemble average using (\ref{eq:avgphiphi}) gives
\begin{equation}
\frac{d\langle |{\dot \bx}|^2 \rangle}{d\lambda} = - 16 \int\limits_{-\lambda}^0 dy\; \frac{\xi_\phi'(y)}{y}
\simeq - 16 \int\limits_{-\infty}^0 dy\; \frac{\xi_\phi'(y)}{y} = 2J
\end{equation}
where the approximation is good for any distance from the observer 
much larger than the assumed small correlation length.
Thus $J$ is the rate of increase with path length of the mean squared deflection (per component).

The above formulae also provide a useful way to express $J$ in terms of
the power spectrum of the potential fluctuations rather than in terms of the
two-point correlation function. 
With
\begin{equation}
\begin{split}
J & = -4 \int\limits_{-\infty}^0  dy \nabla_\perp^2 \xi_\phi(y) \\
& \quad = \lim_{\lambda \rightarrow \infty} 
4 \langle \bnablaperp \phi(\lambda) \cdot \int\limits_0^\lambda d\lambda' \; \bnablaperp \phi(\lambda') \rangle
\end{split}
\end{equation}
and expressing the potentials here in terms of their Fourier components we find
\begin{equation}
\begin{split}
J  = \lim_{\lambda \rightarrow \infty} 4  &\int \frac{d^3k}{(2\pi)^3} P_\phi(k) k_\perp^2 
\int\limits_{-\lambda}^0 dy\; e^{-ik_z y} \\
 = \lim_{\lambda \rightarrow \infty} 4 &\int \frac{d^3k}{(2\pi)^3} P_\phi(k) k_\perp^2 \frac{\sin k_z \lambda}{k_z} \\
 = \lim_{\lambda \rightarrow \infty} 4 &\int d \ln k \; k^2 \Delta^2_\phi(k) \\
& \quad\quad \times \frac{\lambda}{2} \int\limits_{-1}^1 d\mu \; (1 - \mu^2) \frac{\sin \mu k \lambda}{\mu k \lambda}
\end{split}
\end{equation}
where we have defined the contribution to the potential variance per log-interval
of wave-number as $\Delta^2_\phi(k) \equiv k^3 P_\phi(k) / 2 \pi^2$.
For $\lambda \rightarrow \infty$, and for finite $k$, the `sinc' function here has a 
narrow central lobe with of width $\delta \mu = 1 / (k \lambda) \ll 1$ and the
integral has very little contribution from the oscillating wings, so we can
approximate the factor $1 - \mu^2$ by unity and change the integration variable to obtain
\begin{equation}
\begin{split}
J & = 2 \int d \ln k \; k \Delta^2_\phi(k) \int\limits_{-\infty}^\infty dy \; \frac{\sin y}{y} \\
& \quad = 2 \pi \int d \ln k \; k \Delta^2_\phi(k)
\end{split}
\label{eq:JfromP}
\end{equation}
The integrand here was plotted in Figure \ref{fig:dJdlnk}.

\subsubsection{Perturbation to the area of constant distance travelled}

The average $\langle r \rangle$ in (\ref{eq:avgr}) is the ensemble average
for the distance reached by a ray fired off in a fixed direction 
from an observer (i.e.\ we are averaging over an ensemble of realisations of the potential field).  
What we are primarily interested in here is the
ensemble average of the area per unit solid angle $\langle dA / d \Omega \rangle$ or,
dividing by the constant unperturbed distance squared, we wish
to determine $\langle dA / dA_0 \rangle$ where $dA_0 \equiv \lambda_0^2 d \Omega$.

The area $dA$ that is the intersection of the bundle with the surface $\lambda = \lambda_0$
lies at a distance $r = \lambda_0 + \Delta r$ from the observer
and, as we have discussed, has a normal with no first order deviation from the
direction away from the observer.  Writing $dA = r^2 d\Omega'$ -- i.e.\ defining $d\Omega'$ to
be the solid angle that this area would subtend at this distance if there were no light
deflection -- we have $dA / dA_0 = (d \Omega' / d\Omega) (r^2 / \lambda_0^2)$.

Since the perturbation to $r$ is already second order, the ensemble average of $dA / dA_0$,
accurate to second order is
\begin{equation}
\left\langle \frac{dA}{dA_0} \right\rangle
= \left\langle \frac{d\Omega'}{d\Omega}\left( 1 + 2 \frac{\Delta r}{\lambda_0} \right) \right\rangle .
\end{equation}

Now the factor $d\Omega' / d \Omega$ here is, at linear order in the potential, just $1 - 2 \kappa$
with $\kappa$ the convergence.  This has an (ensemble) expectation average that vanishes at first order.
But we are working to second order precision here and one
might imagine that there would be a significant second order
contribution to $\langle d\Omega' / d \Omega \rangle$.

But in fact -- and this is critical in what follows -- $\langle d\Omega' / d \Omega \rangle$ is precisely unity.  
This is because the process generating realisations of the potential field is symmetric with respect to
the observer; there is no preferred direction.  So an equivalent to generating realisations of $\phi(\br)$
and averaging quantities for a single direction from the observer is to generate realisations and then,
for each of these, average over all directions from the observer.  But in doing so it is guaranteed that,
in the absence of multiple imaging, the sum of $d\Omega'$ will be $4 \pi$ since there is a
one-to-one mapping with lensing simply rearranging the sky without duplication and without missing
any regions.  Thus $\langle d\Omega' / d \Omega \rangle = 1$ and the desired expectation is
\begin{equation}
\left\langle \frac{dA}{dA_0} \right\rangle
= 1 + 2 \left\langle \frac{d\Omega'}{d\Omega}\frac{\Delta r}{\lambda_0} \right\rangle
= 1 + 2 \left\langle \frac{\Delta r}{\lambda_0} \right\rangle + \ldots
\end{equation}
where $\ldots$ denotes terms of higher than second order in the potential.

This then yields the fractional perturbation to the area of the
sphere of constant physical path length $\lambda_0$: 
\begin{equation}
\left\langle \frac{\Delta A}{A_0} \right\rangle 
= 2 \left\langle \frac{\Delta r}{\lambda_0} \right\rangle =
- \frac{2}{\lambda_0^2}\int\limits_0^{\lambda_0} d\lambda\; \lambda (\lambda_0 - \lambda) J(\lambda) \; .
\label{eq:distancereached}
\end{equation}
This is on the order of $\sim \lambda_0 J$ (for constant $J$ it is $- \lambda_0 J / 3$) or roughly equal to the
mean square deflection angle, 
though the presence of the factor $\lambda (\lambda_0 - \lambda)$
means that lenses close to the observer or the source
have relatively small effect on the distance.

\subsection{The area of the CMB photosphere}
\label{subsec:photospherearea}

We have calculated above how the wiggling of rays decreases the area of the
surface of constant path length as compared to its value in an unperturbed universe 
or uniform refractive medium.
As we have discussed, the CMB photosphere is not a surface of constant physical
path length from the observer;  it is the surface of constant optical
path length.  To first order in the potential the photosphere is the surface
of conformal path length
$\lambda = \lambda_0 + 2 \int d \lambda \, \phi$.

To calculate the ensemble average of $dA' / dA_0$, where $dA'$ is the intersection of
the bundle of rays with the photosphere, we proceed as follows:  The area $dA$ at $\lambda = \lambda_0$
considered above is not perpendicular to the ray direction; the corresponding area perpendicular
to the ray at $\lambda = \lambda_0$ is $dA \times (1 - |\Delta {\dot \bx}|^2 / 2)$ (correct
to second order).  The area of the intersection of the photosphere with the bundle (which
is perpendicular to the beam direction) is given by that perpendicular area times
$1 + 2 \theta \Delta \lambda$, where $\theta \equiv \dot A / 2 A$ is the expansion rate of the bundle.  To zeroth
order $\theta = 1 / \lambda$, but as $\Delta \lambda = 2 \int d \lambda \; \phi$ is first order
in the potential we need to consider the first order perturbation to the expansion rate
$\Delta \theta = - \lambda_0^{-2} \int d \lambda \; \lambda^2 \nablaperp^2 \phi$, as shown in Appendix \ref{sec:expansion}.
And as $\Delta \lambda$ multiplies
the zeroth order expansion we need to compute this to second order.  This is done by
writing the potential along the perturbed path as a Taylor expansion about the
unperturbed path with lowest order correction $\Delta \phi = \Delta \bx \cdot \bnablaperp \phi + \Delta \lambda \partial \phi / \partial \lambda$ with
$\Delta \bx = - 2 \int d \lambda' \; (\lambda - \lambda') \bnablaperp \phi$ and with $\Delta \lambda = 2 \int d\lambda'\; \phi(\lambda')$.
This gives for $\Delta \lambda$ correct to second order
\begin{equation}
\begin{split}
\Delta \lambda & = 2 \int\limits_0^{\lambda_0} d \lambda \; \phi - 
4 \int\limits_0^{\lambda_0} d\lambda \; \bnablaperp \phi \cdot
\int\limits_0^{\lambda} d\lambda' \; (\lambda - \lambda') \bnablaperp \phi' \\
& \quad\quad\quad + 4 \int\limits_0^{\lambda_0} d\lambda \; \frac{\partial \phi}{\partial\lambda} 
\int\limits_0^{\lambda} d\lambda' \; \phi(\lambda') .
\end{split}
\end{equation}  

But on evaluating the ensemble average of the 2nd order terms here -- making the usual assumption in the first that the range of correlations
is small compared to $\lambda_0$ and integrating the last one by parts -- one finds that these 
both involve $\xi_\phi(0)$, and give contribution to
$\Delta \lambda / \lambda_0$ only on the order of $\phi^2$ in the blob model.  They are 
therefore like the first term in the
$[\ldots]$ in the last line of (\ref{eq:Deltaravg}) and  in the same way we ignore
such sub-dominant contributions.  The upshot is
that we can just use the first order expression for $\Delta \lambda$.

Multiplying these factors, the ratio of the area of the
intersection of the bundle with the photosphere to the unperturbed area is then, at second order,
\begin{equation}
\begin{split}
\frac{dA'}{dA_0} & = \frac{d\Omega'}{d\Omega} 
\left(1 + \frac{2 \Delta r}{r} \right)
\left(1 - \frac{|\Delta {\dot \bx}|^2}{2} \right) \\
& \quad \times (1 + 2 (1/\lambda + \Delta \theta) \Delta \lambda) . \\
\end{split}
\end{equation}

Taking the ensemble expectation value we will obtain four second order contributions :
The first, $2 \langle \Delta r / r \rangle$ we have already calculated.  The second is
\begin{equation}
- \frac{\langle | \Delta {\dot \bx} |^2\rangle}{2} = - \frac{2}{\lambda_0^2} \int\limits_0^{\lambda_0} d\lambda \; \lambda
\int\limits_0^{\lambda_0} d\lambda' \; \lambda' \langle \bnablaperp \phi \cdot \bnablaperp \phi' \rangle .
\end{equation}
We can evaluate this, in the limit that the correlation length is much less than
the path length, much as we did in the calculation of $\langle \Delta r \rangle$.  The
leading order term is obtained by replacing $\lambda'$ in the second integral by
$\lambda$ and taking it outside of the integral.  Unlike the expression for $\langle \Delta r \rangle$
this involves $\int^{\lambda_0} d \lambda \int^{\lambda_0} d\lambda' \ldots$ rather
than $\int^{\lambda_0} d \lambda \int^{\lambda} d\lambda' \ldots$ so we end up with 
the complete integral $\int dy \; \xi'_\phi(y) / y$ from $-\infty$ to $\infty$.
The result is 
\begin{equation}
- \frac{\langle | \Delta {\dot \bx} |^2\rangle}{2} = - \frac{1}{\lambda_0^2} \int\limits_0^{\lambda_0} d\lambda \; \lambda^2 J(\lambda)
\label{eq:deltaxdotterm}
\end{equation}
where we have replaced $\lambda_s$ by $\lambda_0$ since the difference introduces only higher
order corrections.

The sum of (\ref{eq:distancereached}) and (\ref{eq:deltaxdotterm}) gives the
mean of the perturbation to the area perpendicular to a beam that has
propagated a path length $\lambda_0$.  We have calculated this using the
optical scalar equations in appendix \ref{sec:opticalscalars} for the case
of constant $J$.  The result is $\langle \Delta A \rangle / A_0 = - (2/3) \lambda_0 J$
which agrees with what we find here. 

Next there is the cross term
\begin{equation}
\begin{split}
2 \langle \Delta \theta \Delta \lambda \rangle & = - \frac{4}{\lambda_s^2} 
\int\limits_0^{\lambda_0} d\lambda \; \lambda^2 
\int\limits_0^{\lambda_0} d\lambda' \; \langle \phi' \nablaperp^2 \phi \rangle  \\
&= \frac{2}{\lambda_0^2} \int\limits_0^{\lambda_0} d\lambda \; \lambda^2 J(\lambda) . \\
\end{split}
\label{eq:thetalambdacrossterm}
\end{equation}
This is just twice $\langle | \Delta {\dot \bx} |^2 \rangle / 2$.
Including this we find, for the case that $J$ is non-evolving, that the sum of the effects so far vanishes.

Finally we have the cross-term from $d\Omega' / d\Omega = 1 - 2 \kappa$ and the
first order time-delay term $2 \Delta \lambda / \lambda$. This is
\begin{equation}
\begin{split}
- 4 \left\langle \kappa \frac{\Delta \lambda}{\lambda} \right\rangle 
& = - \frac{8}{\lambda_0^2} 
\int\limits_0^{\lambda_0} d\lambda \; \lambda (\lambda_0 - \lambda) 
\int\limits_0^{\lambda_0} d\lambda' \; \langle \phi'  \nablaperp^2 \phi \rangle  \\
& = \frac{4}{\lambda_0^2} \int\limits_0^{\lambda_0} d\lambda \; \lambda (\lambda_0 - \lambda) J(\lambda) . \\
\end{split}
\label{eq:kappalambdacrossterm}
\end{equation}
This is just (minus) twice $2 \langle \Delta r / r \rangle$.

Combining (\ref{eq:distancereached}), (\ref{eq:deltaxdotterm}), (\ref{eq:thetalambdacrossterm}) \& (\ref{eq:kappalambdacrossterm}) we obtain the final result
\begin{equation}
\langle \Delta A \rangle / A_0 
= \frac{1}{\lambda_0^2} \int\limits_0^{\lambda_0} d\lambda \; (2 \lambda (\lambda_0 - \lambda) + \lambda^2) J(\lambda) .
\label{eq:finalDeltaAoverA_appendix}
\end{equation}
This result is of second order in the metric fluctuations and is valid at leading order
in the assumed small parameter $L / \lambda$.  For constant $J$ this is
$\langle \Delta A \rangle / A_0 = + (2/3) \lambda_0 J$.

We can see from this that the fractional change in area of the photosphere
depends only on $J$; that it is non-zero; and that it is generally
positive -- so the effect of surface wrinkling wins out over the 
competing effect of paths wiggling.  
But, as anticipated in the order-of-magnitude argument presented
in the Introduction, it is extremely small being only on the order
of the cumulative deflection angle squared.

\subsection{The area of surfaces of constant redshift}
\label{subsec:constantzarea}

As discussed in \S\ref{subsec:rayBCs}, a surface of constant
redshift differs from a surface of constant cosmic time in that the
1st order path length perturbation $\Delta \lambda$ that appears
in (\ref{eq:thetalambdacrossterm}) and (\ref{eq:kappalambdacrossterm})
is given by (\ref{eq:constantzDeltalambda}) 
rather than (\ref{eq:photosphereDeltalambda})
and that this surface is not perpendicular to the ray direction.

The angle $\bTheta$ between the surface normal and the ray direction is
just the transverse gradient, at the end of the ray, of the
differential time delay in (\ref{eq:constantzDeltalambda}) associated with the ISW effect: 
\begin{equation}
\bTheta = \frac{2}{(a'/a)_{\lambda_0}} \int d \lambda \ispace 
\left( \frac{\phi'}{\phi} \right)_\lambda
\frac{\lambda}{\lambda_0}  \bnablaperp \phi
\end{equation}
which has mean squared expectation value
\begin{equation}
\langle \Theta^2 \rangle = \frac{2}{\lambda_0^2 (a'/a)^2_{\lambda_0}} 
\int d \lambda \ispace 
\lambda^2 \left( \frac{\phi'}{\phi} \right)^2_\lambda J(\lambda) .
\end{equation}

The upshot of this is that the fractional perturbation to the
area of a surface of constant redshift is given by an integral along
the line of sight identical to (\ref{eq:finalDeltaAoverA_appendix}),
but including a factor $1 + 2 (\phi' / \phi)_\lambda / (a'/a)_{\lambda_0}$
in the integrand, plus $\langle \Theta^2 \rangle / 2$.
This results in a substantial reduction in the perturbation to
the area, as compared to that for a surface of constant cosmic time,
for sources at low redshift.  But the conclusion that the effect
is of the same order of magnitude as the mean squared deflection is
unaltered.

All the above calculations have concentrated on the
surface around the observer where sources have redshift $z$. But one
could also 
consider the surface surrounding a single source, on which all observers
see the source to have redshift $z$. It might be expected that the
properties of these surfaces would be equivalent, but this is not so.
Consider equation (\ref{eq:finalDeltaAoverA_appendix}): with $\lambda$ the distance
from the observer, it provides the ensemble 
average of the source-surface area per unit solid angle at the observer.
But with $\lambda$ interpreted as the distance from the source it gives
the ensemble average of the observer-surface area per unit solid angle at the 
source and these are not the same, 
since (\ref{eq:finalDeltaAoverA_appendix}) is not symmetric under
$\lambda \rightarrow \lambda_0 - \lambda$.

Why this should be so may be understood in the  hypothetical situation
where the lenses only develop very recently.  In that case the observer experiences
very little perturbation to the source-surface area as both the wiggling and
wrinkling effects are suppressed (as compared to similar lensing structures
situated roughly mid-way between the sources and the observer).
The surface of a pulse of radiation from a source, on the other hand, passes
through a shell of inhomogeneity just before it reaches the surface containing
the observers who see it to have redshift $z$.  This induces little distance-reached
perturbation, but does cause the surface to be wrinkled, thus increasing the
area and thereby decreasing the mean flux density.

So the average of flux densities of the sources at redshift $z$ seen by one observer
is, in the limit that the structure appeared very recently, 
exactly unperturbed.  
But the flux densities averaged over an ensemble of sources 
and the observers who see those sources to have redshift $z$ {\em is\/} biased.
This may sound paradoxical, but is not.  The locations of the observers
in the latter case is not random; where they lie is correlated with the
location of the source and the potential fluctuations.

The distinction is of course largely academic since the effect is 
so small.  But we would argue that what is relevant observationally
is the average over sources for one observer (us) rather than the
average over an ensemble of extra-terrestrial observers.

\section{The rate of expansion of a bundle of rays}
\label{sec:expansion}

This appendix provides the first order expansion of a
bundle of rays that was used in the previous appendix.

Consider a narrow cone of rays that leave the observer, propagating
backwards in time, with central ray initially along the $z$-axis, and
label these rays by their initial direction $\bTheta$.
After propagating a distance $\lambda$ from the observer through
a refractive medium
with refractive index $n(\br) = 1 - 2 \phi(\br)$, the transverse
displacement of the ray with initial direction $\bTheta$, relative
to the location of the central ray, will be, to first order in the
potential
\begin{equation}
\Delta \bx = \bTheta \cdot \left[
\lambda \bI -2 \int\limits_0^\lambda d\lambda' \; \lambda' (\lambda - \lambda') \bnablaperp \bnablaperp \phi(\lambda') 
\right].
\end{equation}

The transformation from solid angle to areas perpendicular to the central
beam is the Jacobian: $A = |d \Delta \bx / d \bTheta | \Omega$, so the area of the
beam bundle is proportional to the determinant of the matrix $[ \ldots ]$ above
which, working to linear order, says
\begin{equation}
A(\lambda) = \lambda^2 \Omega \left( 1 - \frac{2}{\lambda} 
\int\limits_0^\lambda d\lambda' \; \lambda' (\lambda - \lambda') \nablaperp^2 \phi(\lambda') \right).
\label{eq:dA1}
\end{equation}
But this can also be expressed as
\begin{equation}
A(\lambda) = \lambda^2 \Omega \left( 1 + 2 \int\limits_0^\lambda d\lambda' \Delta \theta(\lambda') \right)
\label{eq:dA2}
\end{equation}
where what we shall call the `linearised perturbation to the rate of expansion' is
\begin{equation}
\Delta \theta(\lambda) = - \frac{1}{\lambda^2} \int\limits_0^\lambda d \lambda' \lambda'^2 \nablaperp^2 \phi(\lambda')).
\label{eq:1storderexpansion}
\end{equation}
The equivalence of (\ref{eq:dA1}) and (\ref{eq:dA2}) being easily established by integrating
the double integral obtained by substituting (\ref{eq:1storderexpansion}) in (\ref{eq:dA2})
by parts.

The meaning for the terminology is that if we define the `expansion rate' for the beam as 
$\theta \equiv \dot A / 2 A$ with $\dot A \equiv d A / d \lambda$ (analogous to the Hubble expansion rate) then
from (\ref{eq:dA2}) at linear order $\theta = \theta_0 + \Delta \theta + \ldots$
with zeroth order expansion rate $\theta_0 = 1 / \lambda$.

In CUMD14, the expansion is defined as $- d \ln A / d \lambda$ which is minus twice our definition.
In their appendix D they find that the perturbation to the area of the constant-$z$ surface
is given by $(2 \int d\lambda \ispace \Delta \theta)^2$ (in our terminology) or equivalently
$\langle \Delta A / A \rangle = 4 \langle \kappa^2 \rangle$.
We have already reached a very different conclusion, which
we confirm below in \S\ref{sec:opticalscalars} using an independent approach that
is closer to that of CUMD14.
\section{The Geodesic Deviation Approach}
\label{sec:geodesicdeviation}

The mean (inverse) magnification calculated by MS97 is qualitatively similar to our result (\ref{eq:finalDeltaAoverA})
but differs in detail and predicts much stronger mean inverse amplification for very nearby lenses.
To try to resolve this discrepancy, we first cast the MS97 analysis in the notation used here, 
where e.g.\ we work with the spatial
auto-correlation function of the potential rather than the power spectrum.

\subsection{Metcalf \& Silk's analysis}
\label{subsec:geodesicdeviation_MSanalysis}

As above, we consider rays close to a guiding ray that propagates initially along the
$z$-axis (here we will use 3-dimensional comoving coordinates $\br = \{x_1,x_2,\lambda\}$ with $|\br| = \chi$)
then the geodesic equation for the transverse displacement of the guiding ray is 
$\ddot \bx_0 = -2 \bnablaperp \phi$. 
Integrating the geodesic equation gives the transverse velocity of a
ray with initial direction $\bn = \{\Theta_1, \Theta_2, 1\}$ at $\lambda=0$
\begin{equation}
{\dot \bx}(\lambda) = \bTheta -2 \int\limits_0^{\lambda} d\lambda' \; \bnablaperp \phi'
\end{equation}
and integrating once more by parts gives a displacement
\begin{equation}
\bx(\lambda) = \bTheta \lambda - 2 \int\limits_0^\lambda d\lambda' \; (\lambda - \lambda') \; . \bnablaperp \phi'
\end{equation}

The integration is taken along the path, which to obtain $\bx(\lambda)$ to 2nd order in the potential can be taken to be the first order perturbed path, i.e.\ $\phi$ must be evaluated
at $\br = \lambda {\hat \bz} + \bx$. The location of the end of the ray after
propagating a path length $\lambda_0$ is therefore
\begin{equation}
\begin{split}
\br(\lambda_0) & = ({\hat \bz} + \bTheta) \lambda_0 - 2 \int\limits_0^{\lambda_0} d\lambda \; (\lambda_0 - \lambda) \bnablaperp \phi\biggl(({\hat \bz} + \bTheta) \lambda  \\
& \quad  - 2 \int\limits_0^\lambda d\lambda' \; (\lambda - \lambda') \bnablaperp \phi(({\hat \bz} + \bTheta) \lambda')\biggr) \\
& = ({\hat \bz} + \bTheta) \lambda_0 -2\int\limits_0^{\lambda_0} d\lambda \ispace (\lambda_0 - \lambda) \bnablaperp \phi \\
& +4 \int\limits_0^{\lambda_0} d\lambda \; (\lambda_0 - \lambda) \int \limits_0^\lambda d\lambda' \ispace (\lambda - \lambda') 
\bnablaperp \bnablaperp \phi \cdot \bnablaperp \phi'  \\
\end{split}
\label{eq:M+Sr3D}
\end{equation} 
where, in the last expression, all of the potentials are to be
understood as being evaluated along the undeflected path with
initial direction ${\hat \bz} + \bTheta$.

Differentiating with respect to the assumed infinitesimal $\bTheta$ gives the distortion tensor (the derivative of
2-D deflection $\bx = (1 - {\hat \bz} {\hat \bz} \cdot) \br$):
\begin{equation}
\begin{split}
\bD & = \frac{1}{\lambda_0} \frac{d\bx}{d\bTheta} = \bI 
- \frac{2}{\lambda_0} \int\limits_0^{\lambda_0} d\lambda \ispace (\lambda_0 - \lambda) \lambda \bnablaperp \bnablaperp \phi \\
& + \frac{4}{\lambda_0} \int\limits_0^{\lambda_0} d\lambda \; (\lambda_0 - \lambda) \int \limits_0^\lambda d\lambda' \ispace (\lambda - \lambda') \\
& \quad \times (\lambda \bnablaperp \bnablaperp \bnablaperp \phi \cdot \bnablaperp \phi' + \lambda' \bnablaperp \bnablaperp \phi \cdot \bnablaperp \bnablaperp \phi' ) \\
\end{split}
\label{eq:M+SDtensor}
\end{equation} 
as obtained by MS97 and where the potential is now evaluated along the $z$-axis.

They then proceeded to write $\phi(\br)$ as a Fourier
synthesis to obtain the mean of the trace of the distortion tensor in terms of the power spectrum (the mean
of the first order term on the first line here assumed to be vanishing).  In doing so
they take $\bnablaperp$ to be the derivatives with respect to the transverse Cartesian
coordinates $\bx = \{x_1, x_2\}$.

Here what we actually want is the inverse magnification, which is $|\bD|$ but, as discussed earlier
with $\bD = \bI + \bS_1 + \bS_2 + \ldots$ where the
subscripts `1' and `2' denote the first and second order (in $\phi$) terms appearing in
(\ref{eq:M+SDtensor}). The inverse amplification is 
$\mu^{-1} = |\bD| = 1 + \Tr(\bS) + |\bS| = 1 + \Tr(\bS_1 + \bS_2) + |\bS_1|$,
plus terms of cubic and higher order in $\phi$.

It would seem that the expectation value of the trace of $\bS_1$ vanishes as it is a first order quantity.  And the expectation
value of $|\bS_1|$ is
\begin{equation}
\begin{split}
\langle |\bS_1| \rangle & = \frac{4}{\lambda_0^2}
\int\limits_0^{\lambda_0} d\lambda\; (\lambda_0 - \lambda) \lambda \\
& \quad\quad \times \int\limits_0^{\lambda_0} d\lambda'\; (\lambda_0 - \lambda') \lambda' 
\langle \phi_{11} \phi'_{22} - \phi_{12} \phi'_{21} \rangle .
\end{split}
\end{equation}
But it is easy to show that, for a statistically homogeneous potential
this vanishes as, in Fourier space, the derivatives become multiplication by the transverse components of $\bk$.

Expressing the correlation of third and first derivatives appearing in $\bS_2$ in terms of
the power spectrum also shows that
\begin{equation}
\eqalign{
&\langle \bnablaperp \bnablaperp \bnablaperp \phi(\br) \cdot \bnablaperp \phi(\br + \br') \rangle \cr
&= - 
\langle \bnablaperp \bnablaperp \phi(\br) \cdot \bnablaperp \bnablaperp \phi(\br + \br') \rangle \cr
&= - \langle \phi(\br) \bnablaperp \bnablaperp  \nablaperp^2 \phi(\br + \br') \rangle \cr
&= - \bnablaperp \bnablaperp  \nablaperp^2 \xi_\phi(\br') \; .
}
\label{eq:geodesicnablas}
\end{equation}
We can now see that, when we take the expectation value of 
the final line in (\ref{eq:M+SDtensor}),
there will be almost complete cancellation if the range of correlations is limited (since for correlated
pairs of points $\lambda \simeq \lambda'$). 

We also see in (\ref{eq:M+SDtensor}) that there are two `post-Born' effects.  One comes from the beam being displaced
laterally.  The other comes from the change in the area of the beam.  
But from (\ref{eq:geodesicnablas}) these are almost exactly the same
but of opposite sign so the net effect is strongly suppressed.   

The trace of the mean distortion is therefore
\begin{equation}
\langle \Tr(\bS_2) \rangle = - \frac{4}{\lambda_0} \int\limits_0^{\lambda_0} d\lambda\; (\lambda_0 - \lambda) 
\int \limits_{-\lambda}^0 dy\; y^2 \nablaperp^4 \xi_\phi(\bn y)
\end{equation}
with $\bn$ a unit vector along the line of sight and where we have changed the second integration
variable from $\lambda'$ to $y = \lambda' -\lambda$.  The Laplacian here is with respect to the transverse
coordinates.  For a spherically symmetric function $F(\br) = F(r)$ we have $\nablaperp^2 F(\br) = 2 F' / r$,
where $F'(r) \equiv d F(r) / dr$, and $\nablaperp^2 (\nablaperp^2 F(\br)) = 8(F' / r)'/r$
so $\langle \Tr(\bS_2) \rangle$ can also be expressed as
\begin{equation}
\langle \Tr(\bS_2) \rangle = - \frac{32}{\lambda_0} \int\limits_0^{\lambda_0} d\lambda\; (\lambda_0 - \lambda) 
\int \limits_{-\lambda}^0 dy \ispace y (\xi_\phi' / y)'.
\end{equation}
And as above, if we assume that the potential fluctuations have a limited range of correlations,
and as we are considering sources at great distances many times the correlation length,
the mean inverse magnification will be well approximated by taking the lower limit
on the $y$-integral to be $-\infty$, with the result
\begin{equation}
\begin{split}
\langle 1/\mu \rangle_\Omega & = 1 + \langle \Tr(\bS_2) \rangle \\
& = 1 + \frac{32}{\lambda_0} \int\limits_0^{\lambda_0} d\lambda \; (\lambda_0 - \lambda) \int\limits_{-\infty}^0 dy\; \xi_\phi' / y \\
& = 1 - \frac{4}{\lambda_0} \int\limits_0^{\lambda_0} d\lambda \; (\lambda_0 - \lambda) J\; .
\end{split}
\label{eq:MS97inversemu}
\end{equation}
This is equivalent to MS97's equations 8 \& 9 but in a possibly slightly more transparent form.

Their analysis is very elegant, and seems straightforward in principle.  And their result is, at least qualitatively very 
similar to (\ref{eq:finalDeltaAoverA}) in that the inverse magnification is a weighted line integral
of the rate of increase of the mean squared deflection $J$ and is therefore clearly on the
order of the mean squared cumulative deflection which we know to be tiny.
But on closer inspection the formulae differ in the details of the weighting.  
In particular (\ref{eq:MS97inversemu}) gives much larger effect than (\ref{eq:finalDeltaAoverA}) for nearby lenses;
our (\ref{eq:finalDeltaAoverA}) is relatively suppressed for nearby lenses at distance $\lambda_d$ by a factor
$\sim \lambda_d / \lambda_0$.  
For nearby lenses, (\ref{eq:MS97inversemu}) starts growing behind
the deflection region but saturates at a constant value that is independent of $\lambda_0$,
whereas (\ref{eq:finalDeltaAoverA}) predicts an effect that decays asymptotically as $\sim 1 / \lambda_0$ for large $\lambda$.

But in the intuitive picture that the key ingredients are the change in the
distance reached because rays are wiggly and the angular deflection at the
end causing the surface to be aspherical, it seems inevitable
that the effect of structures close to the observer
should be suppressed, at least qualitatively, as in (\ref{eq:finalDeltaAoverA}).

There is also something rather strange about the MS97 result for the location
of the end point of the central ray -- the first two lines of (\ref{eq:M+Sr3D}). 
If we write this as $\br(\lambda_0) = \br_0 + \br_1 + \br_2$ where the subscripts denote the
order then the squared distance reached is 
$|\br(\lambda_0)|^2 = \br_0 \cdot \br_0 + 2 \br_0 \cdot (\br_1 + \br_2) + \br_1 \cdot \br_1 + \ldots$.
But both $\br_1$ and $\br_2$ are perpendicular to $\br_0 = {\hat \bz} \lambda_0$ so,
up to 2nd order, $|\br|^2 = |\br_0|^2 + |\br_1|^2 = \lambda_0^2 + |\br_1|^2$;
i.e.\ the distance reached is always greater than $\lambda_0$ whereas we would have expected the distance
reached to have a {\em negative\/} second order perturbation because of the wiggliness of the rays.

\subsection{A partial resolution}
\label{subsec:geodesicdeviation_resolution}

The last puzzle, at least, has a simple resolution.  The gradient operators $\bnablaperp$ in the last section 
were taken to be the derivative with respect to $\bx$.  But what appears in the
geodesic equation are the gradients in
the direction perpendicular to the instantaneous ray directions.  As the rays have a first order
deflection, this gradient is not perpendicular to the $z$-axis, so when applied to $\phi$ there
will be a second order correction.  

In general $\bnablaperp = \bnabla - \bn (\bn \cdot \bnabla)$  with $\bn$ the ray direction.
If we consider a ray that arrives at the observer with $\bn = {\hat \bz}$ then the
deviation of the direction after propagating some path length will,
to first order, be $\bn = {\hat \bz} + {\dot \bx}$, so
\begin{equation}
\begin{split}
\bnablaperp & = \bnabla - ({\hat \bz} + {\dot \bx}) ({\hat \bz} + {\dot \bx}) \cdot \bnabla \\
& = \bnablax - {\dot \bx} ({\hat \bz} \cdot \bnabla) - {\hat \bz} ({\dot \bx} \cdot \bnabla) + \ldots \\
& = \bnablax - {\dot \bx} \partial_\lambda - {\hat \bz} ({\dot \bx} \cdot \bnablax) + \ldots
\end{split}
\end{equation}
where $\bnablax = \bnabla - {\hat \bz} ({\hat \bz} \cdot \bnabla)$ is the 2D gradient with respect
to the Cartesian transverse coordinates $\bx = \{x_1, x_2\}$, and where we have used ${\hat \bz} \cdot \bnabla = \partial / \partial z = \partial_\lambda + \ldots$ to
first order. 
These are only correct to first order, but as they get applied to the potential that is all that we need.
In the second order terms in (\ref{eq:M+Sr3D}) and (\ref{eq:M+SDtensor}) we can ignore the distinction
between $\bnablaperp$ and $\bnablax$.  But working to 2nd order precision we need to
keep track of the correction to the first order terms.  

In terms of Cartesian coordinate derivatives, the geodesic equation is
\begin{equation}
{\ddot \br} = - 2 (\bnablax - {\dot \bx} \partial_\lambda - {\hat \bz} ({\dot \bx} \cdot \bnablax)) \phi
\end{equation}
with integral, for initial direction ${\dot \br}(0) = {\hat\bz}$
\begin{equation}
{\dot \br} = {\hat\bz}   - 2 \int\limits_0^{\lambda} d \lambda' (\bnablax - {\dot \bx} \partial_\lambda - {\hat \bz} ({\dot \bx} \cdot \bnablax)) \phi'
\label{geodesic_resolution_rdot1}
\end{equation}
so the transverse `velocity' ${\dot \bx} = {\dot \br} - {\hat \bz} ({\hat \bz} \cdot {\dot \br})$ is
\begin{equation}
{\dot \bx} =  - 2 \int\limits_0^{\lambda} d \lambda' (\bnablax - {\dot \bx} \partial_\lambda) \phi'
\end{equation}
which in (\ref{geodesic_resolution_rdot1}), and keeping only terms up to 2nd order in $\phi$, gives
\begin{equation}
\begin{split}
{\dot \br} = {\hat\bz} & - 2 \int\limits_0^{\lambda} d \lambda' \Biggl\{\bnablax \phi' \\
& + 2 (\partial_\lambda \phi' + {\hat \bz} \bnablax) \phi' \cdot) \int\limits_0^{\lambda'} d \lambda'' \ispace \bnablax \phi'' \Biggr\}
\end{split}
\label{geodesic_resolution_rdot2}
\end{equation}
with integral
\begin{equation}
\begin{split}
\br(\lambda_0) & = {\hat \bz} \lambda_0 - 2 \int\limits_0^{\lambda_0} d\lambda \ispace (\lambda_0 - \lambda) \Biggl\{ \bnablax \phi \\
& + 2 (\partial_\lambda \phi + {\hat \bz} \bnablax \phi \cdot) \int\limits_0^{\lambda} d\lambda' \ispace \bnablax \phi' \Biggr\} \\
\end{split}
\label{eq:geodesic_resolution_r1}
\end{equation}
where all of the potentials are to be understood as being evaluated along the 
actual perturbed path.  In order to compute expectation values we need to work in terms of the
potential along the unperturbed path $\br = \lambda {\hat \bz}$ which is obtained by
making a Taylor expansion of the first order term above (the correction to the second order
term being of cubic order).  The result is
\begin{equation}
\begin{split}
\br(\lambda_0) & = {\hat \bz} \lambda_0 - 2 \int\limits_0^{\lambda_0} d\lambda \ispace (\lambda_0 - \lambda) \Biggl\{ \bnablax \phi \\
& - 2 \bnablax \bnablax \phi \int\limits_0^{\lambda} d\lambda' \ispace (\lambda - \lambda') \bnablax \phi' \\
& + 2 (\partial_\lambda \phi + {\hat \bz} \bnablax \phi \cdot) \int\limits_0^{\lambda} d\lambda' \ispace \bnablax \phi' \Biggr\} \\
\end{split}
\label{eq:geodesic_resolution_r2}
\end{equation}
where now all the potentials are to be evaluated on the unperturbed path.

If we compare with (\ref{eq:M+Sr3D}) -- specialising to the case $\bTheta = 0$ as we are assuming here --
we see that the terms on the last line are new and, in particular, there is now a 2nd order component
of the displacement of the end of the ray parallel to the $z$-axis:
\begin{equation}
\delta \br_2 = - 4 {\hat \bz} \int\limits_0^{\lambda_0} d \lambda \ispace (\lambda_0 - \lambda) \bnablax \phi \cdot 
\int\limits_0^\lambda d\lambda' \ispace \bnablax \phi'
\end{equation}
which has a non-vanishing dot product with the unperturbed direction, so the squared distance reached is
\begin{equation}
\begin{split}
|\br|^2 & = \lambda_0^2 
- 8 \lambda_0 \int\limits_0^{\lambda_0} d \lambda \ispace (\lambda_0 - \lambda) \bnablax \phi \cdot \int\limits_0^\lambda d\lambda' \ispace \bnablax \phi' \\
& + 4 \int\limits_0^{\lambda_0} d \lambda \ispace (\lambda_0 - \lambda) \bnablax \phi \cdot \int\limits_0^{\lambda_0} d \lambda' \ispace (\lambda_0 - \lambda') \bnablax \phi'
\end{split}
\end{equation}
with expectation value
\begin{equation}
\langle |\br|^2 \rangle = \lambda_0^2 - 2 \int\limits_0^{\lambda_0} d \lambda \ispace \lambda (\lambda_0 - \lambda) J
\end{equation}
which agrees with (\ref{eq:avgr}).  Note that this contains the lensing kernel, so nearby lenses do not contribute.

Resolving the difference between the inverse amplification of MS97 and that obtained here
is much more complicated.  What one has to do is develop the 2nd order expression for
the end-point of a ray with direction at the observer ${\hat \bz} + \bTheta$ and then
differentiate with respect to $\bTheta$.  We shall not pursue that analysis here.
\section{Optical scalars and the focusing theorem}
\label{sec:opticalscalars}

Here we consider the mean inverse magnification from the perspective
of optical scalars -- the rates of expansion, shear and possibly rotation of
a bundle of light rays that appear in Raychaudhuri's equation.
This formalism was originally developed
by Sachs (1961) in the context of propagation of
gravitational radiation, but it applies for any massless field
in the geometric optics limit.  
The optical scalar transport equations (see Schneider, Ehlers \& Falco 1992, 
Narlikar 2010 for derivations) are particularly important
in the present context since, as we have discussed, they are the basis for the
`focusing theorem' 
(Seitz, Schneider \& Ehlers 1994), which appears to show that
inhomogeneities cause
systematic focusing of beams of light, and which underlies
the claims of Clarkson et al.\ 2012 and CUMD14.
Our goals here are to provide a check
on the analysis in the main text; to show that there is 
no subtle relativistic effect hidden in these equations;
and to elucidate the meaning of the focusing equation.
 
We first develop the optical scalar transport equations in the form
appropriate for calculating distances and beam areas given some
statistical prescription for the metric fluctuations as a function
of background coordinates.
We then solve these perturbatively, up to second order in the amplitude of 
the metric fluctuations and compare with the results obtained
in the main text.

\subsection{The optical scalar equations in the weak field limit}
\label{sec:os_lumpy_glass}

As discussed in \S\ref{subsec:weakfield}, light rays propagating through a perturbed FRW background with statistically isotropic metric fluctuations are exactly equivalent to optics in a
medium with refractive index $n(\br)$ and obey
\begin{equation}
{\ddot \br} = \bnablaperp \tilden
\label{eq:os_geodesicequation}
\end{equation}
where $\tilden \equiv \ln n$ and $\bnablaperp \equiv \bnabla - \brdot (\brdot \cdot \bnabla)$
is the derivative in the direction perpendicular to $\brdot$.
In terms of the metric (\ref{eq:metric}) 
$n = [(1 - 2 \phi(\br) / c^2)/(1 + 2 \phi(\br) / c^2)]^{1/2}$ 
with $\br$ being conformal background coordinates,
and dot being derivative with respect to path length in these coordinates so $|\brdot| = 1$.

The optical scalar equations are a set of coupled non-linear differential 
equations that describe the evolution of the rate of expansion, the vorticity and the rate of shear
of a bundle of rays (here we are interested
here in a bundle of rays that left the observer, propagating backward in time, within
a circular cone of infinitesimal solid angle $d\Omega$).
These equations are of interest here because
the rate of expansion can be integrated to give the area of the beam.

\begin{figure}
\setlength{\unitlength}{2.700000cm}
\begin{center}
\begin{picture}(3,3)(-1.5,-1)
\thicklines
\qbezier(0.000000,0.000000)(-0.750000,0.000000)(-1.500000,-0.135000)
\qbezier(0.000000,0.000000)(0.750000,0.000000)(1.500000,-0.135000)
\thinlines
\put(-1.500000,0.000000){\vector(1,0){3.000000}}
\put(1.400000,0.050000){\makebox(0,0){x}}
\put(0.000000,-1.000000){\vector(0,1){3.000000}}
\put(-0.050000,1.900000){\makebox(0,0){z}}
\qbezier(-1.000000,0.000000)(-0.940000,-0.500000)(-0.820000,-1.000000)
\qbezier(-1.000000,0.000000)(-1.120000,1.000000)(-1.000000,2.000000)
\qbezier(0.000000,0.000000)(0.000000,-0.500000)(0.060000,-1.000000)
\qbezier(0.000000,0.000000)(0.000000,1.000000)(0.240000,2.000000)
\qbezier(1.000000,0.000000)(0.940000,-0.500000)(0.940000,-1.000000)
\qbezier(1.000000,0.000000)(1.120000,1.000000)(1.480000,2.000000)
\qbezier(-1.500000,1.180000)(0.000000,1.000000)(1.500000,0.820000)
\thicklines
\qbezier(-1.000000,0.000000)(-1.029928,0.248202)(-1.059856,0.496404)
\qbezier(-1.059856,0.496404)(-1.069503,0.447344)(-1.079151,0.398283)
\qbezier(-1.059856,0.496404)(-1.038824,0.451043)(-1.017791,0.405682)
\qbezier(0.000000,0.000000)(0.000000,0.250000)(0.000000,0.500000)
\qbezier(0.000000,0.500000)(-0.015451,0.452447)(-0.030902,0.404894)
\qbezier(0.000000,0.500000)(0.015451,0.452447)(0.030902,0.404894)
\put(0.060000,0.070000){\makebox(0,0){$0$}}
\qbezier(1.000000,0.000000)(1.029928,0.248202)(1.059856,0.496404)
\qbezier(1.059856,0.496404)(1.038824,0.451043)(1.017791,0.405682)
\qbezier(1.059856,0.496404)(1.069503,0.447344)(1.079151,0.398283)
\put(1.060000,0.070000){\makebox(0,0){$0$}}
\qbezier(-1.060000,1.127200)(-1.060000,1.377200)(-1.060000,1.627200)
\qbezier(-1.060000,1.627200)(-1.075451,1.579647)(-1.090902,1.532094)
\qbezier(-1.060000,1.627200)(-1.044549,1.579647)(-1.029098,1.532094)
\qbezier(0.060000,0.992800)(0.089928,1.241002)(0.119856,1.489204)
\qbezier(0.119856,1.489204)(0.098824,1.443843)(0.077791,1.398482)
\qbezier(0.119856,1.489204)(0.129503,1.440144)(0.139151,1.391083)
\put(0.190000,1.062800){\makebox(0,0){${\Delta \lambda}_0$}}
\qbezier(0.060000,0.992800)(-0.690000,1.082800)(-1.500000,1.037800)
\qbezier(0.060000,0.992800)(0.810000,0.902800)(1.500000,0.677800)
\qbezier(1.151200,0.861856)(1.210626,1.104690)(1.270051,1.347525)
\qbezier(1.270051,1.347525)(1.243740,1.305008)(1.217428,1.262491)
\qbezier(1.270051,1.347525)(1.273756,1.297662)(1.277461,1.247800)
\put(1.251200,0.931856){\makebox(0,0){${\Delta \lambda}$}}
\put(0.000000,0.000000){\vector(1,0){1.000000}}
\put(0.750000,0.080000){\makebox(0,0){${\bf x}$}}
\put(0.200000,0.350000){\makebox(0,0){${\bf \dot r}_0 = \hat{\bf z}$}}
\put(1.200000,0.350000){\makebox(0,0){${\bf \dot r} = \hat{\bf z}$}}
\put(1.300000,0.240000){\makebox(0,0){$+ {\bf K}\cdot{\bf x}$}}
\put(0.200000,1.300000){\makebox(0,0){${\bf \dot r}_0'$}}
\put(1.300000,1.150000){\makebox(0,0){${\bf \dot r}'$}}
\end{picture}
\end{center}
\caption{Illustration of a bundle of rays (thin curves) and associated
wave-fronts (thick curves) and ray direction vectors $\dot \br = d \br / d \lambda$ (arrows).
The base of each arrow is labelled by distance (physical for lumpy glass, 
background conformal for perturbed FRW) along the path.
Close to the guiding ray the ray vectors will vary linearly
with transverse displacement.
The optical tensor $\bK$ is the derivative of the ray direction with
respect to coordinates $\bx$ on the plane that is tangent to the
wavefront at the location of the guiding ray.  The optical tensor
transport equation tells us how $\bK$ evolves as the bundle propagates
through any metric or refractive index fluctuations.
Since rays are perpendicular to the wave-fronts, the transverse components of the
direction of rays are the 2D gradient of the wave-front displacement from the
tangent plane.  It follows that the optical tensor is also the Hessian
(2nd spatial derivative) matrix for this displacement.}
\label{fig:raybundle}
\end{figure}
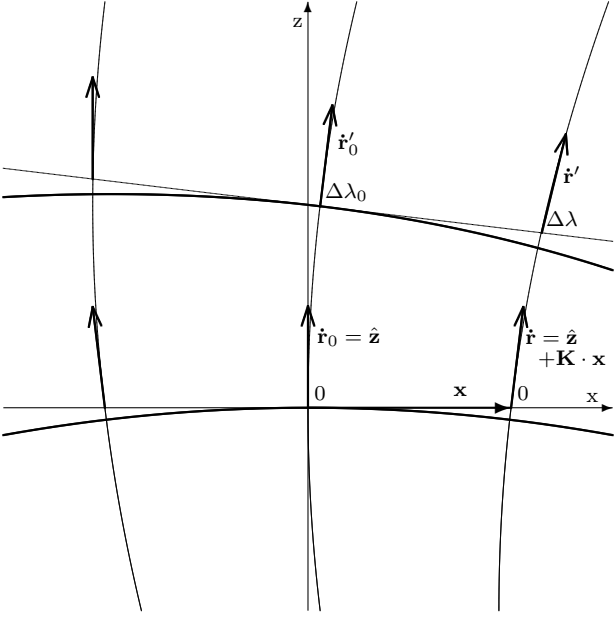

At some point $\lambda$ along the central (or `guiding') ray (which we denote by subscript 0),
and as illustrated in Figure \ref{fig:raybundle},
we can erect background spatial coordinates such that
the $z$-axis points along the direction of the central ray, i.e.\ $\brdot_0 = \bzhat$,
and define the 2-D orthogonal coordinates $\bx = \{x_1, x_2 \}$ on the plane orthogonal to 
be $\bx \equiv \br - \bzhat (\bzhat \cdot \br)$.  We set the origin of
coordinates at the location of the central ray: $\bx_0 = 0$.

Now consider a collection of neighbouring rays whose directions $\brdot$ vary smoothly
on the surface perpendicular to the central ray, so for infinitesimal displacements
$\bx$ they have orthogonal `velocity' $\bxdot = \brdot - \bzhat (\bzhat \cdot \brdot) = \bK \cdot \bx$
where $\bK$ is a $2\times 2$ matrix that we shall refer to as 
the `optical tensor', and which is the derivative of the orthogonal ray
velocity with respect to the orthogonal coordinates.  Our first goal is
to obtain a first order differential equation for how $\bK$ changes with path length along
the beam.

\subsubsection{The optical tensor transport equation}

At linear order in $\bx$ the ray directions are $\brdot = \bzhat + \bxdot$ and the
perpendicular gradient operator is 
\begin{equation}
\bnablaperp = \bnablax - \bxdot \partial_z - \bzhat (\bxdot \cdot \bnablax) .
\label{eq:os_bnablaperp}
\end{equation}
Let us now use this in the geodesic equation to propagate the guiding ray forward
by a path length corresponding to a given interval of optical path (or phase $\Delta \phi$ for
a monochromatic source): $\Delta \lambda_0 = \Delta \phi / n(\bzero)$.  To first order in
$\Delta \lambda_0$ the new position, which we denote by a prime, is 
\begin{equation}
\br_0' = \bzero + \brdot_0 \Delta \lambda_0 = \bzhat \Delta \lambda_0
\end{equation}
while the ray direction will be
\begin{equation}
\brdot_0' = \bzhat + \brddot_0 \Delta \lambda_0 = \bzhat + \bnablax \tilden(\bzero) \Delta \lambda_0 .
\end{equation}
As the direction has changed we have a new plane perpendicular
to the guiding ray, or equivalently tangent to the new wavefront,
that is tilted with respect to the plane $z = \Delta \lambda_0$.
The equation of this plane is 
\begin{equation}
z' - \Delta \lambda_0 = h(\bx) = - \Delta \lambda_0 \bnablax \tilden(\bzero) \cdot \bx .
\end{equation} 

Now consider a neighbouring ray that pierces the surface $z =  0$ at $\bx$
and propagate this to the new tangent plane.  To first order in $\Delta \lambda_0$ and $\bx$
this requires a path length $\Delta \lambda = \Delta \lambda_0 (1 - \bx \cdot \bnablax \tilden(\bx))$ -- this
can also be obtained from $\Delta \lambda = \Delta \phi / n(\bx)$.
The advanced position and direction will be
\begin{gather}
\br' = \bx + \brdot \Delta \lambda = \bx + (\bzhat + \bxdot) \Delta \lambda \\
\brdot' = \bzhat + \bxdot + 
(\bnablax - \bxdot \partial_z - \bzhat (\bxdot \cdot \bnablax)) \tilden \Delta \lambda
\end{gather}

One path forward at this point would be to apply rotations into
the local coordinate system defined by the new tangent plane to 
obtain the difference in direction between this ray and the guiding ray
$\bxdot'' = R(\brdot') - R(\brdot_0') = R(\brdot' - \brdot_0')$.  This will be a linear function
of the rotated displacement $\bx'' = R(\br' - \br_0')$ with tensorial coefficient
$\bK''$ such that $\bxdot'' = \bK'' \cdot \bx''$.  The rate of change with
path length of $\bK$ then being $\dot \bK = (\bK'' - \bK) / \Delta \lambda_0$. 

But this rotation is an unnecessary complication since
both of the vectors $\br' - \br_0'$ and $\brdot' - \brdot_0'$ are almost perpendicular to the
original (unrotated) $z$-axis, so they only change quadratically with the
angle. And the angle is first order in $\Delta \lambda_0$.
So the vectors $\bxdot''$ and $\bx''$
can be obtained at first order in $\Delta \lambda_0$ simply
by projecting $\br'$ and $\brdot'$ and $\brdot_0'$ onto the original $z = 0$ surface
to obtain $\bx' = \br' - \bzhat (\bzhat \cdot \br')$ and so on.

The transported transverse position and velocity are
\begin{gather}
\bx' = \bx + \bxdot \Delta \lambda = (\bI + \bK \Delta \lambda) \cdot \bx 
\label{eq:os_xprime} \\
\bxdot' - \bxdot_0'= \bxdot + (\bnablax - \bxdot \partial_z) \tilden(\bx)  \Delta \lambda
- \bnablax \tilden({\bf 0}) \Delta \lambda_0 .
\end{gather}
Making a first first order Taylor expansion 
$\bnablax \tilden(\bx) = \bnablax \tilden({\bf 0}) + (\bx \cdot \bnablax) \bnablax \tilden({\bf 0})$,
and realising that, at first order in displacement, $\bxdot \partial_z \tilden(\bx) = \bxdot \partial_z \tilden({\bf 0})$
since $\bxdot$ is of first order, this is
\begin{equation}
\bxdot' - \bxdot_0'= \bx \cdot 
[\bK + (\bnablax \bnablax \tilden - \bnablax \tilden \bnablax \tilden - \bK \partial_z \tilden ) \Delta \lambda]
\end{equation}
where the penultimate term, which like the last, is non-linear in the metric fluctuations,
comes from the first order (in $\bx$ and $\tilden$) difference between $\Delta \lambda$ and $\Delta \lambda_0$.

Writing the LHS as $\bxdot' - \bxdot_0' = \bK' \cdot \bx'$ 
and substituting $\bx = (\bI + \bK \Delta \lambda)^{-1} \cdot \bx'$ from (\ref{eq:os_xprime})
on the RHS and linearising in $\Delta \lambda$, gives
\begin{equation}
\bK' = \bK + [(\bnablax \bnablax - \bK \partial_z) \tilden - \bnablax \tilden \bnablax \tilden - \bK \cdot \bK] \Delta \lambda
\end{equation}
or equivalently, with $\bK' = \bK + {\dot \bK} \Delta \lambda$,
we have the optical tensor transport equation
\begin{equation}
{\dot \bK} = (\bnablax \bnablax - \bK \partial_z) \tilden - \bnablax \tilden \bnablax \tilden  - \bK \cdot \bK \; .
\label{eq:opticaltensortransporteq}
\end{equation}

The linear spatial derivative operator in the first term has a simple physical 
interpretation; it gives the second derivative of $\tilden$ on the curved wavefront 
with respect to the tangent plane coordinates.
The transport equation (\ref{eq:opticaltensortransporteq}) says that 
changes in $\bK$ are driven by any transverse gradients of the refractive
index on the wavefront surface that the beam encounters, which makes sense,
but there is also the non-linear term $- \bK \cdot \bK$ which `drives' changes in
$\bK$ even in the absence of refractive index variations.  This also has
a simple explanation; downstream of a refractive index fluctuation the ray directions are
unchanging, but their transverse positions evolve according to (\ref{eq:os_xprime}),
so the gradient of the fixed transverse velocity with respect to the evolving $\bx'$ coordinates
must change.

\subsubsection{Optical scalar transport equations}

The `optical scalar' transport equations (Sachs 1961) are obtained by
decomposing the optical tensor into the expansion rate $\theta = \Tr(\bK) / 2$
and the trace-free rate of shear
$\bSigma = \{\bK\}$ where the curly braces around a matrix indicates the
trace free projection: $\{\bM\} \equiv \bM - \bI \Tr(\bM) / 2$ (so $\bSigma  = \bK - \theta \bI$).  Now  
for any trace-free $2\times 2$ matrix $\bN = \{ \{ a,  b\}, \{c, -a\}\}$ it is easy to see that
$\bN \cdot \bN = - |\bN| \bI$, from which it follows that
$\bK \cdot \bK = (\theta \bI + \bSigma) \cdot (\theta \bI + \bSigma) = 
(\theta^2 + \Sigma^2) \bI + 2 \theta \bSigma$ where we have defined
$\Sigma^2 \equiv \Tr(\bSigma \cdot \bSigma) / 2 = - |\bSigma|$. 

Taking the trace and trace-free projections of (\ref{eq:opticaltensortransporteq}) yields
the coupled transport equations
\begin{gather}
\dot \theta = \left(\frac{\nablaperp^2}{2} - \theta \partial_\lambda\right) \tilden  
- |\bnablaperp \tilden |^2 / 2 - \theta^2 - \Sigma^2 
\label{eq:os_thetadot}\\
{\dot \bSigma} = (\{\bnablaperp \bnablaperp\} - \bSigma \partial_\lambda) \tilden 
- \{\bnablaperp \tilden \bnablaperp \tilden\}- 2 \theta \bSigma
\label{eq:os_sigmadot}
\end{gather}
where we are now using $\nablaperp^2$ to denote the transverse
Laplacian $\nablax^2$ on the guiding ray (this is not the same
as the dot product the operator in (\ref{eq:os_bnablaperp}) with itself which, containing $\bxdot$, is position
dependent) and $\partial_\lambda$ to denote derivative with respect to position along the
guiding ray.
The rate of shear tensor $\bSigma$ being trace-free has three independent components which
can be further decomposed to a 2-component shear that is sometimes represented as a complex number 
and a vorticity.  We shall not use that decomposition and will just work with $\bSigma$ as
a tensor.  But separating the expansion rate $\theta$ is useful, since unlike $\bSigma$ it
is non-vanishing in the unperturbed universe.

The form of (\ref{eq:os_thetadot}) \& (\ref{eq:os_sigmadot}) is a little different to
e.g.\ equations (6.6) of Blandford \& Narayan (1986) which have the 
linear 2nd derivative terms and the terms involving $\theta^2$, $\Sigma^2$ and $\theta \bSigma$,
but are missing the other non-linear derivative terms.  As we discuss shortly, these differences
arise in part because the spatial derivatives here are with respect to conformal
background coordinates rather than local proper coordinates; using the latter
eliminates the derivative along the line of sight $\partial_\lambda$, but we
are still left with the terms involving the square of the 
transverse gradient.  It is certainly the case that, for lensing by random 
structures, these terms are smaller than both the linear 2nd derivative terms
and the terms involving products of the cumulative rate of shear and expansion,
but they still need to be kept here.  If we ignore these terms we find that there
is a contribution to the mean fractional area perturbation on the order $\phi^2 (\lambda / L)^2$.
This is smaller than the claims by e.g.\ CUMD14, which are 
$\langle \Delta A \rangle / A_0 \sim \phi^2 (\lambda / L)^3$, but larger
than the correct result which is $\sim \phi^2 \lambda / L$.

Starting at some initial point on the central ray, and with some choice of orientation of the
initial orthogonal coordinate system, then for a given log refractive index field $\tilden(\br)$
one could integrate these equations, along with the geodesic equation to track the motion
of the guiding centre, to transport $\theta$ and $\bSigma$ along the ray.\footnote{There is a
slight subtlety here in that one needs to keep track of the rotation of the
perpendicular coordinate system as the central ray direction changes.  The coordinate system
we have used here is not tied to any neighbouring rays.  Instead, the new coordinate
axes $\{{\hat \bx}_1', {\hat \bx}'_2 \}$, viewed as 3-vectors in $\br$-space, 
are, after propagating a distance $\Delta \lambda$,
obtained from the unprimed ones by applying a rotation about the axis that is the cross product
$\brdot \times (\Delta \lambda \bnablaperp \tilden)$.  This will not concern us here, however.}

If the refractive index has no spatial gradients, equations (\ref{eq:os_thetadot}) \& (\ref{eq:os_sigmadot}) 
admit a solution $\theta = 1 / \lambda$ and $\bSigma = 0$.  This is the
appropriate initial condition for a narrow bundle of rays that leave the observer,
and is the zeroth order solution about which we will develop our perturbative analysis.
Note that in the case of an observer at the centre of a spherically
symmetric `lens' with $\tilden(\br) = \tilden(r)$ this will still be a
solution.  This is required by symmetry, and can be confirmed by calculation since 
for any spherically symmetric function 
$f(r = \sqrt{z^2 + |\bx|^2})$ it is
easily shown that $\nablax^2 f$ evaluated at $\bx = 0$ is just $2 (df/dr) / r$ so
the transverse Laplacian of $\tilden$ in (\ref{eq:os_thetadot}) is cancelled by
the longitudinal gradient term $- 2 \theta \partial_\lambda \tilden = - 2 \lambda^{-1} \partial_\lambda \tilden$.
 
The reason that these equations are of interest to us is that,
according to (\ref{eq:os_xprime}), the area of the bundle evolves as $A' 
= A |\bI + \bK \Delta \lambda | = A (1 + \Tr(\bK) \Delta \lambda + \ldots) 
= A (1 + 2 \theta \Delta \lambda + \ldots)$, where $\ldots$ indicates terms of
higher than 1st order in $\Delta \lambda$. Thus
$\theta = {\dot A} / 2 A = {\dot D} / D$  where $D \equiv \sqrt{A}$, 
which is why $\theta$ is called the expansion rate.
Note that we are justified in calculating the first order change
in the area using the projected, rather than rotated, coordinates here
since the difference in the areas is second order.

The solution of $\dot A / 2 A = \theta(\lambda) = \lambda^{-1} + \Delta \theta(\lambda)$ is
\begin{equation}
A = \Omega \lambda^2 \exp\left(2 \int\limits_0^\lambda d \lambda' \ispace \Delta \theta(\lambda')\right)
\label{eq:os_Ageneralsolution}
\end{equation}
where $\Omega$ is a constant of integration (which has an obvious interpretation
as the solid angle of the beam at the source or observer) and 
where $\Delta \theta$ must be obtained by solving 
(\ref{eq:os_thetadot}) \& (\ref{eq:os_sigmadot}).  We will presently
do this by means of expansion up to second
order in the assumed small refractive index fluctuations.
But first we make connection with the, arguably more elegant, relativistic
treatment and discuss the interpretation of the `focusing theorem'.

\subsection{The focusing theorem}

The rate of change with distance of ${\dot D} / D$ is $\dot \theta = \ddot D / D - (\dot D / D)^2 = \ddot D / D - \theta^2$ so,
according to (\ref{eq:os_thetadot}),
\begin{equation}
{\ddot D} / D  = \left(\frac{\nablaperp^2}{2} - \theta \partial_\lambda\right) \tilden - |\bnablaperp \tilden |^2 / 2 - \Sigma^2.
\label{eq:appendixddotD}
\end{equation}  
This appears to differ from the usual expression (e.g.\ Schneider, Ehlers \& Falco 1992)
\begin{equation}
{\ddot D} / D  = - R_{\alpha\beta} k^\alpha k^\beta / 2 - \Sigma^2
\label{eq:appendixddotD_GR}
\end{equation}  
where $R_{\alpha\beta}$ is the Ricci tensor and $k^\alpha$ is the guiding ray 4-vector.
In particular, the rate of expansion $\theta$ does not appear in (\ref{eq:appendixddotD_GR}).
The difference is partly because we are working in terms of the metric
fluctuations -- assumed to take the weak-field form -- and in part because our
$D$ is a distance in conformal background coordinate units whereas in (\ref{eq:appendixddotD_GR})
the distance is in proper distance units.  
In the weak-field approximation $g_{rr} = 1 - 2 \phi$, but $n^2 = (1 - 2 \phi) / (1 + 2 \phi)$ so at lowest
order in the metric fluctuations $g_{rr} = n$ and physical distances are related to background
distances by $d\lambda^* = n^{1/2} d \lambda$, so partial derivatives
with respect to physical coordinates are $\bnablaxstar = n^{-1/2} \bnablax$ and
$\partial_{\lambda^*} = n^{-1/2} \partial_\lambda$.  In terms of $D^* = n^{1/2} D$
(\ref{eq:appendixddotD}) becomes
\begin{equation}
\frac{\ddot D^*}{D^*}  = \frac{1}{2} \nabla_*^2 n - \frac{3}{4n} | \bnablaxstar n |^2 - \Sigma^2
\label{eq:appendixddotDstar}
\end{equation} 
where $\nabla_*^2 n$ is the 3D Laplacian operator in physical coordinates 
$\nabla_*^2 = \nablaxstar^2 + \partial^2_{\lambda^*}$ and
where, as in (\ref{eq:appendixddotD_GR}), the rate of expansion no longer appears.
Here dot denotes derivative with respect to background distance along the ray.

Equation (\ref{eq:appendixddotD_GR}) is the basis for the {\em focusing theorem\/} 
(Seitz, Schneider \& Ehlers 1994): since both terms on the RHS are
negative for any sensible equation of state for matter, 
then, rather generally, ${\ddot D} / D < 0$.  
The first term in (\ref{eq:appendixddotD_GR}) describes the local effect
of matter within the beam while the second term is the integrated effect of tidal
fields from matter outside the beam, or Weyl curvature, along the path of the
beam.  The focusing equation says that the latter can only act to enhance
the local focusing by positive density matter and that, as compared to
rays in Minkowski space-time where $\ddot D = 0$ beams are always focused (at least
up until caustic formation).  
This result seems also to be in accord with calculations based
on the lens equation (Schneider 1984; Ehlers \& Schneider 1986; Seitz \& Schneider 1992)
that any lens will give rise to at least one image that is magnified.
See Schneider, Ehlers \& Falco (1992) and Narlikar (2010) for further discussion.

In the cosmological context, the
width of an unperturbed beam in conformal (or co-moving) coordinate is $D = \sqrt{\Omega} \lambda$, so $\ddot D = 0$.  
The local tidal focusing, in this context, is caused by the density fluctuations around the
mean value, which averages to zero.  More interesting is the effect of the shear, which
is cumulative and always negative.  To get a sense of the size of the effect,
we note that in the perturbative regime the linearised version of (\ref{eq:os_sigmadot})
is $\dot \bSigma_1 = \{ \bnablaperp \bnablaperp \} \tilden$ where $\tilden \simeq -2 \phi$.
So, in a model of random positive or negative perturbations to the Newtonian potential 
with RMS value $\phi$ and characteristic scale $L$, $\bSigma_1$ will perform a
random walk and will have typical mean squared value $\langle \Sigma_1^2 \rangle \sim N \phi^2 / L^2$.
But $N \sim \lambda / L$, so $\langle \Sigma_1^2 \rangle \sim \phi^2 \lambda / L^3 \simeq \langle \kappa^2 \rangle / \lambda^2$.
There are other non-linear terms in (\ref{eq:appendixddotD}), but it is not difficult to show
that their expectation values are all much smaller than $\langle \Sigma_1^2 \rangle$.
If the change to the distance is small we can approximate $\ddot D / D$ by $\ddot D / D_0$
and it then follows that (\ref{eq:appendixddotD}) after integration implies a perturbation to 
the mean of $D = \sqrt{A}$, or equivalently the mean angular diameter distance, that is on the 
order of the mean squared convergence $\langle \Delta D \rangle / D_0 \simeq \langle \kappa^2 \rangle$
with a numerical coefficient that is negative.

Thus the optical scalar formalism shows, rather nicely, that structure
causes a negative bias in the mean (direction averaged) apparent distance.  
But that should come as no particular surprise.
As discussed in the Introduction and shown in \S\ref{subsec:diravgdist} we expect 
$\langle \Delta D \rangle / D_0 = - \langle \kappa^2 \rangle / 2$
when averaged over directions simply because $D$ is the square root of the
fluctuating area per unit solid angle.  The obvious question is whether,
as found by CUMD14, the optical scalar equations actually predict a decrease in the {\em area\/}
of a surface of constant redshift.
To answer this requires a more quantitative analysis.

In what follows we will show, using the optical scalar formalism,
and in the perturbative regime, that the direction averaged perturbation to the 
area is much smaller than the 
distance perturbation $\langle \Delta D \rangle/ D_0\sim \phi^2 \lambda^3 / L^3$. In fact $\langle \Delta A \rangle/ A_0$ 
is suppressed compared to $\langle \Delta D \rangle/ D_0$ by two powers
of $L / \lambda$ so $\langle \Delta A \rangle/ A_0 \sim \phi^2 \lambda / L$.
This means that the next order terms, which include post-Born corrections,
actually cancel (as was also seen in \S\ref{subsec:MSreview}), 
but it gives a result in accord with simple-minded consideration of
reduction in distance reached and wrinkling of surfaces, 
as presented in appendix \ref{sec:areacalculation}.
We then show how, consistent with this, in perturbation theory the
leading order distance bias is $\langle \Delta D \rangle/ D_0 = \alpha \langle \kappa^2 \rangle$
with numerical coefficient $\alpha = -1/2$ as one would expect if the distance bias
comes from statistical fluctuations.  But this seems to us to be a somewhat
backward step; the distance bias may well {\em not\/} be well described
by linear theory when small scale structure is taken into account.
The {\em un-focusing theorem\/} $\langle \Delta A \rangle/ A_0 = 0 + {\cal O}(\phi^2 \lambda / L)$
is, we will argue, more powerful.

\subsection{Perturbation analysis}

As already mentioned, in the absence of refractive index fluctuations $\bnabla \tilden = 0$ 
the solution of (\ref{eq:os_thetadot}) \& (\ref{eq:os_sigmadot}) is $\theta = 1 / \lambda$ and $\bSigma = 0$.
If we let $\theta = \theta_0 + \theta'$ with $\theta_0 = 1 / \lambda$ and drop the 
prime the optical scalar equations become
\begin{gather}
\frac{1}{\lambda^2}\frac{d \lambda^2 \theta}{d \lambda} = \left(\frac{\nablaperp^2}{2} 
- \frac{\partial_\lambda}{\lambda}\right) \tilden 
- \frac{|\bnablaperp\tilden|^2}{2}
 - \theta \partial_\lambda \tilden  - \theta^2 - \Sigma^2 
\label{eq:os_deltathetadot}\\
\frac{1}{\lambda^2}\frac{d \lambda^2 \bSigma}{d \lambda}  = (\{ \bnablaperp \bnablaperp \} - \bSigma \partial_\lambda) \tilden 
- \{\bnablaperp \tilden \bnablaperp \tilden\} - 2 \theta \bSigma\; .
\label{eq:os_deltasigmadot}
\end{gather}

We solve these in two steps.  Dropping all of the second order terms above yields the
first order solutions:
\begin{gather}
\theta_1 = \frac{1}{\lambda^2} \int\limits_0^\lambda d\lambda' \ispace \left(\frac{\lambda'^2 \nablax^2}{2} + 1 \right) \tilden'
- \frac{\tilden}{\lambda} 
\label{eq:os_theta1}\\
\bSigma_1  = \frac{1}{\lambda^2} \int\limits_0^\lambda d\lambda' \ispace \lambda'^2 \{\bnablax \bnablax\} \tilden'
\label{eq:os_sigma1}
\end{gather}
where the integrals are taken along the undeflected path, i.e.\ $\tilden' = \tilden(\br = \bzhat \lambda')$,
and where we have integrated by parts to eliminate the longitudinal derivative in $\theta_1$,
and where the spatial derivatives are with respect to the Cartesian coordinates.
From now on we will consider $\tilden$ to be the perturbation to the log refractive index.

We then insert these in the second order terms on the RHS of (\ref{eq:os_deltathetadot}) 
to obtain the second order solution
\begin{equation}
\begin{split}
\theta_{1+2} & = \frac{1}{\lambda^2} \int\limits_0^\lambda d\lambda' \lambda'^2 
\Biggl[\left(\frac{\nablaperp^2}{2} - \frac{\partial_\lambda}{\lambda'}\right) \tilden' \\
& \quad\quad - |\nablax\tilden|^2 / 2 - \theta_1 \partial_z \tilden' - \theta_1^2 - \Sigma_1^2\Biggr]
\end{split}
\label{eq:os_theta2}
\end{equation}
where $\theta_{1+2}$ includes the first order solution and where, 
to obtain 2nd order precision, we need to evaluate the first 
occurrence of $\tilden$ along the 1st order perturbed path
\begin{equation}
\br_1 = \bzhat \lambda + \int\limits_0^\lambda d\lambda' \ispace (\lambda - \lambda') \bnablax \tilden(\bzhat \lambda') 
\label{eq:os_r1}
\end{equation}
and we need to pay attention to first order perturbations to the derivatives $\bnablaperp$ and $\partial_\lambda$
in $\nablaperp^2 - 2 \lambda^{-1} \partial_\lambda$.

The beam area, after propagating a physical path length $\lambda$, 
is then from (\ref{eq:os_Ageneralsolution}) just $A = A_0 + \Delta A = \Omega \lambda^2 + \Delta A$
with
\begin{equation}
\frac{\Delta A}{A_0} = 2 \int\limits_0^{\lambda} d\lambda' \; \theta_{1+2}  
+ 2 \left(\int\limits_0^{\lambda} d \lambda' \; \theta_1 \right)^2 + \ldots
\label{eq:os_DeltaA}
\end{equation}
and our ultimate goal is to obtain the expectation value of this in terms of the
spatial autocorrelation function of the log refractive index fluctuations
$\xi(\br - \br') \equiv \langle \tilden(\br) \tilden(\br') \rangle$.

At linear order the area perturbation is $\Delta A / A_0 = - 2 \kappa + \ldots$
so from (\ref{eq:os_DeltaA}) we can identify the linear convergence $\kappa$ with
minus the integral of the first order expansion
$\kappa = - \int d\lambda' \ispace \theta_1(\lambda')$.
So the ensemble average of the second term here is, at leading order, just $2 \langle \kappa^2 \rangle$.
CUMD14 obtained an equation similar to (\ref{eq:os_DeltaA}) -- though they
have a different numerical coefficient for the second term -- and claimed
that the first term vanishes at all orders in the perturbation.  
We will now show that this is not the case and that not only does the
leading order contribution to $\Delta A / A_0$ from the second term in
(\ref{eq:os_DeltaA}) get cancelled by the first, but the next order terms -- in an expansion
of powers of the assumed small ratio $L / \lambda$ -- cancel also. 

The calculation is conceptually straightforward, but tedious in the number of terms
it generates (many of which cancel in the end).  To reduce the effort, it helps to 
define $\Delta_A \equiv \Delta A / A_0$
i.e.\ the fractional perturbation to the area at any point along the beam.
The rate of change of this with path length is, from (\ref{eq:os_DeltaA}), and to second order,
\begin{equation}
\dot{\Delta_A}  = 2 \theta_{1+2} + 4 \theta_1 \int\limits_0^\lambda d \lambda' \ispace \theta_1
\label{eq:os_avgDeltaAdot}
\end{equation}  
and differentiating once more, and using (\ref{eq:os_deltathetadot}), yields
\begin{equation}
\begin{split}
\frac{1}{2\lambda^2}\frac{d \lambda^2 \dot{\Delta_A}}{d\lambda} & = 
\left(\frac{\nablaperp^2}{2} - \frac{\partial_\lambda}{\lambda}\right) \tilden - \frac{|\nablax \tilden|^2}{2} 
- \theta_1 \partial_z \tilden \\
& \quad\quad + \theta_1^2 - \Sigma_1^2 + \frac{2}{\lambda^2}\frac{d \lambda^2 \theta_1}{d\lambda} 
\int\limits_0^\lambda d\lambda' \ispace \theta_1' \\
\end{split}
\label{eq:os_avgDeltaAddot}
\end{equation}
where we note that the large squared first order expansion and shear terms now appear
with opposite sign so their leading order effects cancel.
The next step is to work out the expectation values of each of the terms here.
In doing so all but the first term can be calculated using the Born approximation,
using (\ref{eq:os_theta1}) to express $\theta_1$ in terms of the refractive
index fluctuations.  The first term is then calculated in the post-Born
approximation and allowing for the first order perturbations to e.g.\ $\bnablaperp$.
This gives $d (\lambda^2 \dot{\langle \Delta_A \rangle}) / d \lambda$
in terms of $\xi$ and shows how the large terms cancel.  We then integrate to obtain $\langle \Delta_A \rangle$ 
and show that this agrees with what we found in the simpler calculation in the main text.
The next few sub-sections give the details, term by term; 
\S\ref{subsec:osfinalareapert} takes stock.

\subsubsection{The last term}

From (\ref{eq:os_theta1}) the two factors in the last term above are
\begin{gather}
\frac{2}{\lambda^2}\frac{d \lambda^2 \theta_1}{d\lambda} = (\nablax^2 - 2 \lambda^{-1} \partial_z) \tilden \\
\int\limits_0^\lambda d\lambda' \ispace \theta_1' = 
\frac{1}{2\lambda} \int\limits_0^\lambda d\lambda' \ispace [\lambda' (\lambda - \lambda') \nablax^2 - 2] \tilden(\lambda')
\end{gather} 
and the expectation value of their product is 
\begin{equation}
\begin{split}
\left\langle \frac{2}{\lambda^2}\frac{d \lambda^2 \theta_1}{d\lambda} \int\limits_0^\lambda d\lambda' \ispace \theta_1' \right\rangle
= - 2 \nablax^2 \xi_0 \\ 
+ \frac{2}{\lambda^2}  \int\limits_0^\lambda d\lambda' \ispace \lambda' \nablax^2 \xi_{\lambda - \lambda'}
- \frac{2[\xi]_\lambda^0}{\lambda^2} 
\end{split}
\label{eq:os_last_term}
\end{equation}
where we have used $\langle (\partial_\lambda \tilden(\lambda)) \tilden(\lambda')\rangle = - \partial_{\lambda'} \xi(\lambda - \lambda')$ and
have also made use of the identities $\nablax^2 \xi(y) = 2 \xi' / y$, where $\xi' = d\xi/dy$, and $\nablax^4 \xi(y) = 8 (\xi' / y)'/y = 4 (\nablax^2 \xi)' / y$.

The first term here is relatively large, scaling as $1/L^2$.
This is one of a number of similar terms that cancel collectively.

\subsubsection{Expectation value of $\Sigma_1^2$}

We now turn to the three second order terms on the first line of (\ref{eq:os_avgDeltaAddot}).  
Starting with $\Sigma_1^2$, this is 
\begin{equation}
\begin{split}
\langle \Sigma_1^2 \rangle & = \Tr(\langle \bSigma_1 \cdot \bSigma_1 \rangle) / 2 \\ 
& = \frac{1}{2 \lambda^4} 
\int\limits_0^\lambda d\lambda' \ispace \lambda'^2 
\int\limits_0^\lambda d\lambda'' \ispace  \lambda''^2 \\
& \quad\quad \times \Tr(\langle \{\bnablax \bnablax\} \tilden \cdot \{\bnablax \bnablax\} \tilden' \rangle)
\end{split}
\end{equation}
Writing $\tilden(\br) = (2 \pi)^{-3} \int d^3k \; \tilden_\bk \exp(i \bk \cdot \br)$ we have
\begin{equation}
\bSigma_1 = - \int \frac{d^3 k}{(2 \pi)^3} \tilden_\bk
\begin{bmatrix}
(k_1^2 - k_2^2) / 2 & k_1 k_2 \\
k_2 k_1 & (k_2^2 - k_1^2) / 2
\end{bmatrix}
e^{i\bk\cdot\br}
\end{equation}
from which we find, on invoking statistical translational invariance: 
$\langle \tilden_\bk \tilden^*_{\bk'} \rangle = (2 \pi)^3 \delta(\bk - \bk') P_\tilden(\bk)$,
\begin{equation}
\langle \Sigma_1^2 \rangle = \frac{1}{4\lambda^4} 
\int\limits_0^\lambda d\lambda' \ispace \lambda'^2
\int\limits_0^\lambda d\lambda'' \ispace  \lambda''^2 \nablax^4 \xi_{\lambda' - \lambda''}
\end{equation}
where $\xi(\br) = (2 \pi)^{-3} \int d^3k \; P_\tilden(\bk) \exp(i \bk \cdot \br)$. 
The integral here scales as $L^{-3}$ so this is a very large term, but it 
is cancelled by an identical leading order term in $\langle \theta_1^2 \rangle$.

\subsubsection{Expectation value of $\theta_1^2 - \Sigma_1^2$}

Using (\ref{eq:os_theta1}) we find that the average of the term involving two Laplacians in $\theta_1^2$
is identical to $\Sigma_1^2$ and we have, for the residual net effect,
\begin{equation}
\begin{split}
\theta_1^2 - \Sigma_1^2  &  = 
- \frac{1}{\lambda^4} \left[\tilden \lambda - \int\limits_0^\lambda d\lambda' \ispace \tilden'\right]
\int\limits_0^\lambda d\lambda'' \ispace  \lambda''^2 \nablax^2 \tilden'' \\
& + \frac{1}{\lambda^4} \left[\tilden \lambda - \int\limits_0^\lambda d\lambda' \ispace \tilden'\right]^2
\end{split}
\end{equation}
which has expectation value
\begin{equation}
\begin{split}
\langle \theta_1^2 & - \Sigma_1^2 \rangle  = 
\frac{1}{\lambda^4} 
\int\limits_0^\lambda d\lambda'  \int\limits_0^\lambda d\lambda'' \ispace  \lambda''^2 \nablax^2 \xi_{\lambda' - \lambda''} \\
& \quad - \frac{1}{\lambda^3}\int\limits_0^\lambda d\lambda' \ispace \lambda'^2 \nablax^2 \xi_{\lambda - \lambda'} 
+ \frac{\xi_0}{\lambda^2} \\
& - \frac{2}{\lambda^3} \int\limits_0^\lambda d\lambda' \ispace \xi_{\lambda - \lambda'}
+ \frac{1}{\lambda^4} \int\limits_0^\lambda d\lambda'  \int\limits_0^\lambda d\lambda'' \ispace \xi_{\lambda' - \lambda''} \; .
\end{split}
\label{eq:os_thetaminusSigma}
\end{equation}
The third term scales as $L^0$ and the last two scale as $L$ so all three
are ignorable.  The first two terms both scale as $L^{-1}$.

\subsubsection{Expectation value of $- \theta_1 \partial_z \tilden$} 

Next we consider
\begin{equation}
- \theta_1 \partial_z \tilden = 
- \partial_z \tilden \left[\frac{1}{\lambda^2} \int\limits_0^\lambda d\lambda' \ispace 
\left(\frac{\lambda'^2 \nablax^2}{2} + 1 \right) \tilden' - \frac{\tilden}{\lambda}\right] 
\end{equation}
which has expectation value
\begin{equation}
- \langle \theta_1 \partial_z \tilden \rangle = 
\frac{\nablax^2 \xi_0}{2} 
- \frac{1}{\lambda^2} \int\limits_0^\lambda d\lambda' \ispace \lambda' \nablax^2 \xi_{\lambda - \lambda'} 
+ \frac{[\xi]_\lambda^0}{\lambda^2} \; .
\label{eq:os_minusthetadzn}
\end{equation}
These scale as $L^{-2}$, $L^{-1}$ and $L^0$ respectively, so the last
is ignorable.

\subsubsection{Expectation value of $(\nablaperp^2 / 2 - \lambda^{-1} \partial_\lambda) \tilden - |\bnablax \tilden|^2 / 2$}

The first term on the RHS of (\ref{eq:os_avgDeltaAddot})
is of first order in the refractive index fluctuation, but acquires a non-zero average 
at second order because $\tilden$ must be evaluated on the 1st order perturbed beam path (\ref{eq:os_r1}) and
the spatial derivatives are also perturbed.

The spatial derivative operators are, according to (\ref{eq:os_bnablaperp}),
perturbed at first order relative to the derivatives with respect to the Cartesian coordinates:
\begin{gather}
\nablaperp^2 = \nablax^2 - 2 \bxdot_1 \cdot \bnablax \partial_z \\
\partial_\lambda = \partial_z + \bxdot_1 \cdot \bnablax 
\end{gather}
where the first order perturbation to the beam direction is
\begin{equation}
\bxdot_1 =  \int\limits_0^\lambda d\lambda' \ispace \bnablax \tilden' .
\label{eq:os_bxdot1}
\end{equation}

Using these and making a Taylor expansion to obtain $\tilden$
along the 1st order perturbed path and discarding terms that are higher than second order in $\tilden$ gives
\begin{equation}
\begin{split}
& \left(\frac{\nablaperp^2}{2} - \frac{\partial_\lambda}{\lambda}\right) \tilden 
= \left(\frac{\nablax^2}{2} - \frac{\partial_z}{\lambda}\right) \tilden \\
& \quad\quad + \left(\frac{\nablax^2}{2} - \frac{\partial_z}{\lambda}\right) \bnablax \tilden \cdot 
\int\limits_0^\lambda d\lambda' \ispace (\lambda - \lambda') \bnablax \tilden' \\
& \quad\quad - (\partial_z + 1 / \lambda) \bnablax \tilden \cdot \int\limits_0^\lambda d\lambda' \ispace \bnablax \tilden'
\end{split} 
\end{equation}
Including the expectation value of $- |\bnablax \tilden|^2 / 2$ which is just $\nablax^2 \xi_0 / 2$ we obtain
the expectation value for these final terms:
\begin{equation}
\left\langle \left(\frac{\nablaperp^2}{2} - \frac{\partial_\lambda}{\lambda}\right) \tilden 
- \frac{|\bnablax \tilden |^2}{2} \right\rangle = \frac{3}{2} \nablax^2 \xi_0 \; .
\label{eq:os_lineartermsetc}
\end{equation}

\subsection{The final result for the mean area perturbation}
\label{subsec:osfinalareapert}

Putting all the above pieces together 
-- equations \ref{eq:os_last_term}, \ref{eq:os_thetaminusSigma}, \ref{eq:os_minusthetadzn} and \ref{eq:os_lineartermsetc}
-- the first thing we note is that the
`large' terms involving $\nablax^2 \xi_0$ and which scale as $L^{-2}$
all cancel.  So not only is there no very large Born-level 
contribution from $\Sigma^2$ and $\theta^2$ and scaling as $L^{-3}$, the next order $L^{-2}$ terms, including first post-Born
approximation corrections, also cancel out in the end.
At leading order -- i.e.\ at level $\phi^2 \lambda / L$ -- the surviving net effect is equal to 
the first term in equation (\ref{eq:os_thetaminusSigma}) for $\langle \theta_1^2 - \Sigma_1^2 \rangle$:
\begin{equation}
\frac{1}{2\lambda^2}\frac{d \lambda^2 \langle \dot{\Delta_A} \rangle}{d\lambda} 
= \frac{1}{\lambda^4} 
\int\limits_0^\lambda d\lambda'  \int\limits_0^\lambda d\lambda'' \ispace  \lambda''^2 \nablax^2 \xi_{\lambda' - \lambda''} .
\end{equation}
The Laplacian is a narrow function of width $\sim L$ so for all $\lambda'$ except for
within $\sim L$ of the observer or the source sphere the second integral will have
converged and will be well approximated by $\lambda'^2 J$.  We have assumed here
that the metric fluctuations are not evolving -- this was quite hard enough -- so the first integral can
be performed too to give
\begin{equation}
\frac{1}{2\lambda^2}\frac{d \lambda^2 \langle \dot{\Delta_A} \rangle }{d\lambda} 
= - \frac{2 J}{3 \lambda}
\end{equation}
where we have defined
\begin{equation}
J = - \int\limits_{-\infty}^0 dy \ispace \nablax^2 \xi(y)
\end{equation}
which is the same as the definition (\ref{eq:Jdefinition_maintext}) 
since here $\xi = 4 \xi_\phi$ and $\nablax^2 \xi(y) = 2 \xi'(y) / y$.

Finally, integrating this gives
\begin{equation}
\langle \Delta_A \rangle = - 2 J \lambda / 3 .
\label{eq:os_finalDeltaA}
\end{equation}
This is the ensemble average of the fractional perturbation to the area at the end of
a beam of path length (physical for glass, conformal background in cosmology) $\lambda$.  It is not the same as the perturbation to the
area of the surface of constant path $2 \langle \Delta r \rangle / r$ in (\ref{eq:distancereached}) as elements of area on that surface
are not perpendicular to the beam as is the case here.  So (\ref{eq:os_finalDeltaA})  should be
compared to sum of (\ref{eq:distancereached}) and (\ref{eq:deltaxdotterm}) 
$2 \langle \Delta r\rangle/ r - \langle \Delta \bxdot^2 \rangle/ 2 $.  They agree.
This does not provide the full expression for the perturbation to a surface of constant redshift, or the
cosmic photosphere.  For that it is necessary to add the contributions arising from the
fluctuating path length coming from time delays.  This is done in the main part of the paper.

\subsection{The focusing equation in the perturbative regime}
\label{subsec:osperturbativefocusingequation}

We can obtain the mean distance perturbation from the focusing equation
in the perturbative regime in much the same way.  The 
solution of $\dot D / D = \lambda^{-1} + \theta$ is
$D = \sqrt{\Omega} \lambda \exp(\int d\lambda \ispace \theta)$,
and expanding this up to second
order, and defining $\Delta_D = (D - D_0) / D_0$, gives
\begin{equation}
\Delta_D = \int\limits_0^{\lambda} d\lambda' \; \theta_{1+2}  
+ \left(\int\limits_0^{\lambda} d \lambda' \ispace \theta_1 \right)^2 + \ldots
\end{equation}
rather like (\ref{eq:os_DeltaA}). Differentiating this gives
\begin{equation}
\begin{split}
\frac{1}{\lambda^2}\frac{d \lambda^2 \dot{\Delta_D}}{d\lambda} & = 
\left(\frac{\nablaperp^2}{2} - \frac{\partial_\lambda}{\lambda}\right) \tilden - \frac{|\nablax \tilden|^2}{2} 
- \theta_1 \partial_z \tilden \\
& \quad\quad - \Sigma_1^2 + \frac{1}{\lambda^2}\frac{d \lambda^2 \theta_1}{d\lambda} 
\int\limits_0^\lambda d\lambda' \ispace \theta_1' \\
\end{split}
\end{equation}
which is very similar to (\ref{eq:os_avgDeltaAddot}) but
there is no longer cancellation of the large leading order contribution from
$\Sigma_1^2$ by $\theta_1^2$.  The results from the previous section show
that the other terms are relatively negligible (i.e.\ of higher order in $L / \lambda$) so
\begin{equation}
\frac{1}{\lambda^2}\frac{d \lambda^2 \dot{\Delta_D}}{d\lambda} = - \Sigma_1^2
\end{equation}
at leading order.

On the other hand, the linear convergence is $\kappa(\lambda) = \int d\lambda' \theta_1(\lambda')$
so $\dot{\kappa^2} = 2 \theta_1 \int d\lambda' \theta_1(\lambda')$
from which we find
\begin{equation}
\frac{1}{\lambda^2}\frac{d \lambda^2 \dot{\kappa^2}}{d\lambda} = 2 \theta_1^2 +
\frac{1}{\lambda^2}\frac{d \lambda^2 \theta_1}{d\lambda} \int\limits_0^{\lambda} d \lambda' \ispace \theta_1\; .
\end{equation}
But again the results from the previous section show that, in the ensemble average sense, the second
term is negligible compared to the first and that, in the ensemble average,
$\langle \theta_1^2 \rangle = \langle \Sigma_1^2 \rangle$ to leading order. 
Thus we have
\begin{equation}
\frac{1}{\lambda^2}\frac{d \lambda^2 \langle \dot{\Delta_D} \rangle}{d\lambda} = - \langle \Sigma_1^2 \rangle = -\langle \theta_1^2 \rangle =
- \frac{1}{2\lambda^2}\frac{d \lambda^2 \langle \dot{\kappa^2} \rangle}{d\lambda} \; .
\end{equation}
This can be integrated twice, with appropriate boundary conditions at the observer to give,
for the average of $\Delta_D = \Delta D / D_0$,
\begin{equation}
\langle \Delta D \rangle / D_0 = - \langle \kappa^2 \rangle / 2
\end{equation}
consistent with conservation of area.
At the end of this journey, we therefore have a conclusion consistent with the one
obtained previously by more elementary means: cosmological inhomogeneities have
no tendency to focus beams of light in terms of changing their area. 
\section{Source averaged convergence}
\label{sec:sourceaveragedkappa}

The mean magnification of sources is almost precisely unity, but we
have shown that
the mean inverse magnification, averaged over sources, is non-zero:
\begin{equation}
\langle \mu^{-1} \rangle_A = 1 + \langle (\Delta \mu)^2 \rangle + \ldots = 1 + 4 \langle \kappa^2 \rangle + \ldots
\label{eq:sourceavginvmu}
\end{equation}
This effect can be understood qualitatively as being a consequence of extremal paths to sources tending to avoid over-densities
and therefore sampling paths for which the convergence, on average, is negative.  
We have invoked this in e.g.\ \S\ref{sec:areabias}.  Here we expand on this
and compute the bias in the the mean convergence, or in the column density of matter, 
in the perturbative regime.  This may be of some
relevance to cosmic gas abundance measurements from absorption line studies.

It is well known that images of sources behind
clusters of galaxies appear to be `repelled' by the cluster, being biased
against high-density  regions.
This makes sense: rather than going through the centre of a cluster,
light rays can minimise their total travel time by taking a longer path
to one side of the cluster in order  to reduce gravitational time delay
near the centre. Another viewpoint on this bias
is to consider a thin screen populated with weak lensing regions 
scattered around the sky; some with enhanced surface density and an equal number of negative lenses,
all the lenses being of the same area.  It is easy to see that the light paths that pass through the negative
lenses will diverge and will map to a larger area of the source sphere than those which pass through
the positive lenses.  Thus the observer will see more sources 
through the negative lenses than through the positive lenses (and if the observer
can resolve the sources would see the former to be shrunken relative to the
latter).  Averaged over the sources then the mean surface density fluctuation and
convergence will be biased negative.
We now show how this works out with 3-dimensional metric fluctuations rather
than a thin screen.

The convergence, for a bundle of rays that arrives at the observer along the $z$-axis from distance $\lambda_0$, is 
\begin{equation}
\kappa = \frac{1}{\lambda_0} \int\limits_0^{\lambda_0} d \lambda \ispace \lambda (\lambda_0 - \lambda) \nablaperp^2 \phi(\lambda) .
\label{eq:kappaavg_kappa}
\end{equation}
At linear order, the integral can be taken along the $z$-axis and we can ignore the
difference between $\nablaperp^2$ and $\nablax^2$.  The ensemble average of this vanishes.

Going beyond first order we must allow for first order change to spatial derivatives: 
$\nablaperp^2 = \nablax^2 - \bxdot \cdot \bnablax \partial_z$ and also for the 
displacement of the path.  
Using (\ref{eq:transversexdot}) for $\bxdot$ and taking the ensemble average of the
extra derivative term  gives non-zero contribution to $\langle \kappa \rangle$ given at leading order by 
\begin{equation}
\delta \langle \kappa \rangle = \frac{2}{\lambda_0}
\int\limits_0^{\lambda_0} d \lambda \ispace \lambda (\lambda_0 - \lambda) 
\langle |\bnablax \phi|^2 \rangle
\end{equation}
This scales (with the lens properties) as $\phi^2 / L^2$, which is large compared to the
effect on the area, but sub-dominant here, where the leading order effect scales,
like $\langle \kappa^2 \rangle$ as $L^{-3}$. So we can ignore the
distinction between $\nablaperp^2$ and $\nablax^2$ in (\ref{eq:kappaavg_kappa}).

The first order displacement, also for a ray that arrives at the observer along the $z$-axis, is, at the source plane,
\begin{equation}
\bx(\lambda_0) = -2 \int\limits_0^{\lambda_0} d \lambda \;  (\lambda_0 - \lambda) \bnablax \phi(\lambda).
\label{eq:kappaavg_xi}
\end{equation} 
If there were a source sitting on the $z$-axis then the ray that we need to
fire in order to reach that source would have to arrive at the observer with
direction $\bTheta = - \bx(\lambda_0) / \lambda_0$.  The displacement from the $z$-axis for that ray,
at some distance $\lambda$ along the ray, again to 1st order, is
\begin{equation}
\begin{split}
\bx(\lambda) &= -2 \int\limits_0^{\lambda} d \lambda' \;  (\lambda - \lambda') \bnablax \phi(\lambda') \\
& \quad\quad + 2 \frac{\lambda}{\lambda_0}  
\int\limits_0^{\lambda_0} d \lambda' \;  (\lambda_0 - \lambda') \bnablax \phi(\lambda') 
\end{split}
\label{eq:kappaavg_xj}
\end{equation}
which vanishes at both ends of the ray.

To compute the ensemble average of $\kappa$ in (\ref{eq:kappaavg_kappa}), 
correct to 2nd order, for the ray that reaches the source we
simply need to replace $\nablaperp^2\phi$ in (\ref{eq:kappaavg_kappa}) by 
$\nablax^2\phi + \bx(\lambda) \cdot \bnablax \nablax^2\phi$ with $\bx(\lambda)$ as in (\ref{eq:kappaavg_xj}).
As mentioned, the average of $\nablaperp^2 \phi$, understood to be along the unperturbed path
vanishes.  This gives a double integral involving $\langle \bnablax \phi(\lambda') \cdot \bnablax \nablax^2 \phi(\lambda) \rangle$.
But under the assumption that $\phi$ is a statistically homogeneous random process this is
just minus $\langle \nablax^2 \phi(\lambda') \nablax^2 \phi(\lambda) \rangle$ (since each time we move an
index we pick up a factor $i^2 = -1$) which is clearly a symmetric, positive
function of $\lambda - \lambda'$ and  
$\langle \bnablax \phi(\lambda') \cdot \bnablax \nablax^2 \phi(\lambda) \rangle = - \nablax^4 \xi_\phi(\lambda - \lambda')$.

The result, in gory detail except for suppressing the argument of $\nablax^4 \xi_\phi(\lambda - \lambda')$, is
\begin{equation}
\begin{split}
\langle \kappa \rangle_A &= 
- \frac{2}{\lambda_0^2} \int\limits_0^{\lambda_0} d \lambda \; \lambda^2 (\lambda_0 - \lambda) 
\int\limits_0^{\lambda_0} d \lambda' \;  (\lambda_0 - \lambda') \nablax^4 \xi_\phi \\
& \quad\quad + \frac{2}{\lambda_0} \int\limits_0^{\lambda_0} d \lambda \; \lambda (\lambda_0 - \lambda) 
\int\limits_0^{\lambda} d \lambda' \;  (\lambda - \lambda') \nablax^4 \xi_\phi .
\end{split}
\end{equation}

These two double integrals are very similar looking, but are quite different.  When we
apply the condition that the correlation length is much less than $\lambda \sim c / H$ 
we can effectively replace the factor $\lambda_0 - \lambda'$ by $\lambda_0 - \lambda$
in the first expression and take it outside the second integral and replace that with
an unrestricted integral of $\int dy\; \nablax^4 \xi_\phi(y)$ (which does not vanish as the integrand is even).
In the second line however, when we change variables in the inner integral from $\lambda'$ to $y = \lambda' - \lambda'$
we only have a one-sided integral.  That is not particularly significant, but instead of $\lambda_0 - \lambda'$
we have  $\lambda - \lambda'$, which is very small
whenever $\nablax^4 \xi_\phi(\lambda - \lambda')$ is not negligible.  The result is that the
second line is negligible compared to the first and, on dropping it, we have
\begin{equation}
\langle \kappa \rangle_A = 
- \frac{2}{\lambda_0^2} \int\limits_0^{\lambda_0} d \lambda \; \lambda^2 (\lambda_0 - \lambda)^2 
\int\limits_{-\infty}^{\infty} d y \;  \nablax^4 \xi_\phi(y) .
\end{equation}
However, if we were calculating the sky-direction weighted mean convergence 
then we would not have the second line in (\ref{eq:kappaavg_xj})
and we would only have the much smaller term we are discarding here.

The integral appearing here is superficially similar to the definition of $J$ but
involves $\nablax^4 \xi_\phi$ rather than $\nablax^2 \xi_\phi$.  By the same reasoning
that led us to (\ref{eq:JfromP}) one can express this integral in terms of the power
spectrum:
\begin{equation}
\int\limits_{-\infty}^{\infty} d y \;  \nablax^4 \xi_\phi(y) = \pi \int d \ln k \ispace k^3 \Delta_\phi^2(k) .
\end{equation}
The extra two powers of $k$ in the integrand as compared to (\ref{eq:JfromP}) mean that
this integral is dominated by small-scale structure.  Indeed, if the mass auto-correlation function
is similar to that of galaxies: $\xi \propto r^{-\gamma}$ with $\gamma \simeq 1.8$ then
$\Delta_\phi^2 \propto k^{\gamma - 4}$ and the integral here is $\sim \int d \ln k \ispace k^{\gamma - 1}$
which diverges for large $k$ provided $\gamma > 1$. 

The same line of argument gives the expectation for $\langle \kappa^2 \rangle$ with $\kappa$
given by (\ref{eq:kappaavg_kappa}) which is almost identical but with $+1$ in place of the factor $-2$ so 
\begin{equation}
\langle \kappa \rangle_A = -2 \langle \kappa^2 \rangle
\end{equation}
so the mean column density along paths to sources is lower than on average.  
Despite this, the flux density
of sources is not biased, but the inverse magnification is biased positive.
This result ignores selection effects, however.  If sources are selected according
to luminosity there will be a magnification bias and other effects
such as extinction by dust may be important.  These effects have been discussed
in the context of estimation of the neutral HI density from damped Ly-$\alpha$ systems
by Bartelmann \& Loeb (1996) using single lenses modelled as isothermal spheres.

\end{document}